%% file: dune-review-paper.tex
\documentclass{elsarticle}
\pdfoutput=1

\usepackage{url}
\usepackage{dune}
\usepackage{graphicx}
\usepackage{bold-extra}

\usepackage[english]{babel}
\usepackage{amsthm,amssymb,amsmath}
\usepackage{setspace}
\usepackage[verbose,section]{placeins}

\usepackage[inline]{enumitem}
\setlist[enumerate]{label=\roman*.}
\setlist[enumerate,2]{label=\roman{enumi}.\alph*.}

\usepackage{tikz}
\usetikzlibrary{patterns}
\usetikzlibrary{decorations.pathreplacing}
\usetikzlibrary{arrows,positioning}
\usepackage{tikz-qtree}

\usepackage{comment}
\usepackage{color}
\usepackage{xspace}
\usepackage{graphicx,epstopdf}
\usepackage{array}
\usepackage{multirow}
\usepackage{todonotes}
\usepackage{subcaption}
\usepackage{colonequals}

\newcommand{\enablelinenumbers}{}

\journal{Computers \& Mathematics with Applications}

\usepackage{hyperref}
\hypersetup{pdftitle={The DUNE Framework: Basic Concepts and Recent Developments},
            pdfborderstyle={/S/U/W 1}}

\input{macros}

\graphicspath{{gfx/}}

\begin{document}

\begin{frontmatter}

\title{The \dune Framework: Basic Concepts and Recent Developments}

\author[1]{Peter Bastian}
\author[2]{Markus Blatt}
\author[3]{Andreas Dedner}
\author[4]{Nils-Arne Dreier}
\author[4]{Christian Engwer}
\author[4]{Ren\'e Fritze}
\author[5]{Carsten Gr{\"a}ser}
\author{Christoph Gr{\"u}ninger}
\author[1]{Dominic Kempf}
\author[7]{Robert Kl{\"o}fkorn}
\author[4]{Mario Ohlberger}
\author[8]{Oliver Sander}
\address[1]{Heidelberg University}
\address[2]{Dr.~Blatt HPC-Simulation-Software \& Services}
\address[3]{University of Warwick}
\address[4]{Applied Mathematics: Institute of Analysis and Numerics, University of M\"unster}
\address[5]{Freie Universit\"at Berlin}
\address[7]{NORCE Norwegian Research Centre AS}
\address[8]{Technische Universit\"at Dresden}


\begin{abstract}
This paper presents the basic concepts and the module structure of the 
\textit{Distributed and Unified Numerics Environment} and reflects on 
recent developments and general changes that happened since the release of the first \dune version in 2007
 and the main papers describing that state \cite{dunegridpaperI, dunegridpaperII}. 
This discussion is accompanied with a description of various advanced
features, such as coupling of domains and cut cells, grid modifications such as
adaptation and moving domains, high order discretizations and node level
  performance, non-smooth multigrid methods, and multiscale
methods. A brief discussion on current and future development directions of the framework concludes the paper.   
\end{abstract}

\end{frontmatter}

\enablelinenumbers



\input{introduction}

\section{\texorpdfstring{\dune}{Dune} core modules and re-usable concepts}
\label{sec:coremodules}
In this section we want to give an overview of the central components
of \dune{}, as they are offered through the core modules.
These modules are described as they are
in \dune version 2.7, released in January 2020.
Focus of the core modules is on those components modeling mathematical abstractions needed in a finite element method.
We will discuss in detail the \dunegrid{} and
\duneistl{} modules, explain the basic ideas of the
\dunelocalfunctions{} and \dunefunctions{} module,
and discuss how the recently added
Python support provided by the \dunepython{} module works.
While
\dunecommon{} offers central infrastructure and foundation classes, its
main purpose is that of a generic C++ toolbox and we will only briefly
introduce it, when discussing the build system and infrastructure in general.

\input{dunegrid}
\input{duneistl}

\input{dunelocalfunc}

\input{dunepython}

\input{infrastructure}



\section{Selected advanced features with applications}\label{sec:advancedfeatures}

After having discussed the central components of \dune{} and their
recent changes, we now want to highlight some advanced features.
The following
examples all showcase features of \dune extension modules in conjunctions with
the core modules.

\input{adaptation}
\input{complexdomains}
\input{solvers}
\input{multiscale}
\input{hpc}

\section{Development trends in \texorpdfstring{\dune}{Dune}}
\label{sec:trends}
\input{trends}

\section*{Acknowledgements}
We thank the \dune users and contributors for their continuous support,
as only a vivid community unfolds the power of open source.
We would like to point out the essential contributions of those \dune
core developers that are not authors of this paper:
Ansgar Burchardt, Jorrit Fahlke, Christoph Gersbacher,
Steffen M\"uthing, and Martin Nolte. The file LICENSE.md
within every \dune module attributes the work of numerous
more contributors.

Robert Kl\"ofkorn acknowledges the support of the
Research Council of Norway through the INTPART project INSPIRE (274883).
Peter Bastian, Nils-Arne Dreier, Christian Engwer, Ren\'e Fritze,
and Mario Ohlberger acknowledge funding 
by the Deutsche Forschungsgemeinschaft (DFG, German Research Foundation) 
under SPP 1648: Software for Exascale Computing through the project
EXA-DUNE - Flexible PDE Solvers, Numerical Methods, and Applications
under contract numbers Ba1498/10-2, EN 1042/2-2, and OH98/5-2.
Christian Engwer, Dominic Kempf, and Peter Bastian also acknowledge funding through the
BMBF project HPC$^2$SE under reference number 01H16003A.
Nils-Arne Dreier, Christian Engwer, Ren\'e Fritze, and Mario Ohlberger
acknowledge funding by the Deutsche
Forschungsgemeinschaft under Germany's Excellence 
Strategy EXC 2044-390685587, Mathematics M\"unster: Dynamics -- Geometry -- Structure.

\bibliographystyle{elsarticle-num}
\bibliography{dune-review-paper}

\end{document}

%% file: macros.tex
\newcommand{\R}{\mathbb{R}}   

\newcommand{\T}{\ensuremath{\mathcal{T}}\xspace}

\newcommand{\be}{\mathbf{e}}
\newcommand{\bu}{\mathbf{u}}

\DeclareMathOperator*{\argmin}{arg\,min}

\providecommand{\jump}[1]{[\![ #1 ]\!]}


%% file: introduction.tex

\section{Introduction}

The \textit{Distributed and Unified Numerics Environment}
\dune{}\footnote{\url{www.dune-project.org}} \cite{dunegridpaperI,dunegridpaperII} is a free and open source
software framework for the grid-based 
numerical solution of partial differential equations (PDEs) that has been
developed for more than 15 years as a collaborative effort of several universities and research
institutes. In its name, the term ``distributed'' refers to distributed development as well as distributed computing. 
The enormous importance of numerical methods for PDEs in applications has lead to the development of a large number of 
general (i.e. not restricted to a particular application) PDE software projects. Many of them will be presented
in this special issue and an incomplete list includes
AMDIS \cite{amdis:07},
deal.II \cite{dealII91}, 
FEniCS \cite{Logg:2012:ASD:2331176},
FreeFEM \cite{MR3043640},
HiFlow \cite{emclpp42879},
Jaumin \cite{Liu2019},
MFEM \cite{mfem-library},
Netgen/NGSolve \cite{ngsolve},
PETSc \cite{petsc-web-page}, and
UG4 \cite{Vogel2013}.

The distinguishing feature of \dune{} in this arena is its flexibility combined with efficiency. 
\textit{The main goal of \dune{} is to
provide well-defined interfaces for the various components of a PDE solver for which then
specialized implementations can be provided}. \dune{} is \textit{not} build upon one single grid data structure
or one sparse linear algebra implementation nor is the intention to focus on
one specific discretization method only. 
All these components are meant to be exchangeable. This philosophy
is based on a quote from \textit{The Mythical Man-Month: Essays on Software Engineering} by Frederick Brooks
\cite[p. 102]{Brooks1975}:
\begin{quote}
Sometimes the strategic breakthrough will be a new algorithm, \ldots\\
Much more often, strategic breakthrough will come from redoing
the representation of the data or tables. This is where the heart of a
program lies.
\end{quote}
This observation has lead to the design principles of \dune{} stated in \cite{dunegridpaperII}:
\begin{enumerate}
\item Keep a clear separation of data structures and algorithms by providing abstract interfaces algorithms
can be built on. Provide different, special purpose implementations of these data structures.
\item Employ generic programming using templates in C++ to remove any overhead of these abstract interfaces
at compile-time. This is very similar to the approach taken by the C++ standard template library (STL).
\item Do not reinvent the wheel. This approach allows us also to reuse existing legacy code from our own and other projects in one
common platform.
\end{enumerate}
Another aspect of flexibility in \dune{} is to structure code as much
as possible into separate modules with a clear dependence.

This paper is organized as follows. In Section \ref{sec:ecosystem} we describe the modular structure of \dune{}
and the ecosystem it provides. Section \ref{sec:coremodules} describes the core modules and their concepts
while Section \ref{sec:advancedfeatures} describes selected advanced features and illustrates the concepts with
applications. The latter section is intended for selective reading depending on the interests of the reader.
Section \ref{sec:trends} concludes the paper with current development trends in \dune{}.

\section{The Dune ecosystem}\label{sec:ecosystem}

The modular structure of \dune{} is implemented by conceptually splitting
the code into separate, interdependent libraries.
These libraries are referred to as \dune{}-modules
(not to be confused with C++-modules and translation units).
The \dune{} project offers a common infrastructure
for hosting, developing, building, and testing these modules.
However, modules can also be maintained independently
of the official \dune{} infrastructure.

The so-called \emph{\dune{} core
modules} are maintained by the \dune{} developers and each stable release provides consistent releases of all core modules.
These core modules are:
\begin{description}[nosep]
  \item[\dunecommon:] Basic infrastructure for all \dune{} modules such as build scripts or dense linear algebra.
  \item[\dunegeometry:] Reference elements, element transformations, and quadrature rules.
  \item[\dunegrid:] Abstract interface for general grids featuring any dimension, various element types, 
    conforming and nonconforming, local hierarchical refinement, and
    parallel data decomposition. Two example implementations are provided. 
  \item[\duneistl:] Abstract interfaces and implementations for sparse
    linear algebra and iterative solvers on matrices with small dense
    blocks of sizes often known at compile time.
  \item[\dunelocalfunctions:] Finite elements on the reference element.
\end{description}

While the core modules build a common, agreed-upon foundation for the \dune{} framework,
higher level functionality based on the core modules is developed by independent groups,
and concurrent implementations of some features focusing on different aspects exist.
These additional modules can be grouped into the following categories:
\begin{description}[nosep]
  \item[Grid modules] provide additional implementations of the grid interface including so-called meta grids, implementing additional
    functionality based on another grid implementation.
  \item[Discretization modules]  provide full-fledged implementations of finite element, finite volume, or other grid based 
    discretization methods using the core modules.
    The most important ones are \dunefem{}, \dunefufem{}, and \dunepdelab{}.
  \item[Extension modules] provide additional functionality interesting for all \dune{} users which are not yet core modules.
    Examples currently are Python bindings, a generic implementation of grid functions, and support for system testing.
  \item[Application modules] provide frameworks for applications. They make heavy
    use of the features provided by the \dune ecosystem but do not intend to merge
    upstream as they provide application-specific physical laws, include third-party
    libraries, or implement methods outside of \dune's scope. Examples are the porous
    media simulators OPM \cite{opm} and DuMu$^x$ \cite{dumux3}, the FEM toolbox
    \textsc{KASCADE7} \cite{kaskade7}, and the
    reduced basis method module \dune{}-RB \cite{dune-rb}.
  \item[User modules] are all other \dune modules that usually 
    provide applications implementations for specific research projects and also new development features that 
    are not yet used by a large number of other \dune users but over time may
    become extension modules. 
\end{description}
Some of the modules that are not part of the \dune core are designated
as so-called \emph{staging modules}.
These are considered to be of wider interest and may be proposed to
become part of the core in the future.

The development of the \dune software takes place on the \dune gitlab instance
(\url{https://gitlab.dune-project.org}) where users can download and clone all
openly available \dune git repositories. They can create their own new projects,
discuss issues, and open merge requests to contribute to the code base.
The merge requests are reviewed by \dune developers and others who want to
contribute to the development. For each commit to the core and extension modules
continuous integration tests are run to ensure code stability.

%


%% file: dunegrid.tex
\subsection{The \texorpdfstring{\dune}{Dune} grid interface -- \dunegrid}
\label{sec:DuneGrid}

The primary object of interest when solving partial differential equations (PDEs) are functions
\begin{equation*}
f : \Omega \to R,
\end{equation*}
where the domain $\Omega$ is a (piecewise) differentiable $d-$manifold embedded in $\mathbb{R}^w$, $w\geq d$,
and $R=\mathbb{R}^m$ or $R=\mathbb{C}^m$ is the range. 
In grid-based numerical methods for the solution of PDEs the domain $\Omega$ is partitioned into a finite number of
open, bounded, connected, and nonoverlapping subdomains $\Omega_e$, $e\in E$, $E$ the set of elements, satisfying
\begin{equation*}
\bigcup_{e\in E} \overline{\Omega}_e = \overline{\Omega} \qquad\text{and}\qquad 
\Omega_e\cap\Omega_{e'}=\emptyset\,\text{ for $e\neq e'$}.
\end{equation*}
This partitioning serves three separate but related tasks:
\begin{enumerate}
\item \textit{Description of the manifold.}
Each element $e$ provides a diffeomorphism (invertible and differentiable map) $\mu_e : \hat\Omega_e \to \Omega_e$ from
its reference domain $\hat\Omega_e\subset\mathbb{R}^d$ to the subdomain $\Omega_e\subset\mathbb{R}^w$. 
It is assumed that the maps $\mu_e$ are
continuous and invertible up to the boundary $\partial\hat\Omega_e$. 
Together these maps give a piecewise description of the manifold.
\item \textit{Computation of integrals.} Integrals can be computed by partitioning and transformation of integrals 
$\int_\Omega f(x)\,dx = \sum_{e\in E} \int_{\hat\Omega_e} f(\mu_e(\hat x)) \,d\mu_e(\hat x)$. 
Typically, the reference domains $\hat\Omega_e$ have simple shape that is amenable to numerical quadrature.
\item \label{it:task_approximation} \textit{Approximation of functions.} Complicated functions can be approximated subdomain by subdomain for 
example by multivariate polynomials $p_e(x)$ on each subdomain $\Omega_e$.
\end{enumerate}
The goal of \dunegrid{} is to provide a C++ interface to describe such subdivisions, from now on called a ``grid'', in a generic way.
Additionally, approximation of functions (Task \ref{it:task_approximation}) requires further information to associate data with the constituents of a grid.
The grid interface can handle arbitrary dimension $d$ (although naive grid-based methods become 
inefficient for larger $d$), arbitrary world dimension $w$ as well as
different types of elements, local grid refinement, and parallel processing.

\subsubsection{Grid entities and topological properties}
\label{sec:basicgridconcepts}

Our aim is to separate the description of grids into a \textit{geometrical part},
mainly the maps $\mu_e$ introduced above, 
and a \textit{topological part} describing how the elements of the grid
are constructed hierarchically from lower-dimensional objects and
how the grid elements are glued together to form the grid.

The topological description can be understood recursively over the dimension $d$.
In a one-dimensional grid, the elements are edges connecting two
vertices and two neighboring elements share a common vertex. In the combinatorial description
of a grid the position of a vertex is not important but the fact that two edges share a vertex is.
In a two-dimensional grid the elements might be triangles and quadrilaterals
which are made up of three or four edges, respectively. Elements could also be polygons with any number of edges.  
If the grid is \textit{conforming}, neighboring elements share a common edge with two
vertices or at least one vertex if adjacent.

In a three-dimensional grid elements might be tetrahedra or hexahedra consisting of triangular
or quadrilateral faces, or other types up to very general polyhedra.

In order to facilitate a dimension-independent description of a grid we call its constituents \textit{entities}. An entity $e$
has a dimension $\dim(e)$, where the dimension of a vertex is 0, the dimension of an edge is 1, and so on.
In a $d$-dimensional grid the highest dimension of any entity is $d$ and we define the \textit{codimension}
of an entity as  $$ \text{codim}(e) = d - \dim(e).$$ 
We introduce the \textit{subentity relation} $\subseteq$ with
$e'\subseteq e$ if $e'=e$ or $e'$ is an entity contained in $e$, e.g. a face of a hexahedron.
The set $U(e) = \{e' : e' \subseteq e\}$ denotes all subentities of $e$. The \textit{type} of an entity
$\text{type}(e)$ is characterized by the graph $(U(e),\subseteq)$ being
isomorphic to a specific reference entity $\hat e\in \hat E$ (the set of all reference entities).

A $d$-dimensional grid is now given by all its entities $E^c$ of codimension $0\leq c \leq d$.
Entities of each set $E^c$ are represented by a different C++ type depending on the codimension $c$ as a template parameter.
In particular we call $E^d$ the set of vertices, $E^{d-1}$ the set of edges, $E^{1}$ the set of facets, and $E^0$ the
set of elements. Grids of mixed dimension are not allowed, i.e. for every $e'\in E^c$, $c>0$ there exists $e\in E^0$
such that $e'\subseteq e$.
We refer to \cite{dunegridpaperI,casc2019} for more details on formal properties of a grid.

\dune{} provides several implementations of grids all implementing the \dunegrid{} interface.
Algorithms can be written generically to operate on different grid implementations.
We now provide some code snippets to illustrate the \dunegrid{} interface.
First we instantiate a grid:
\begin{c++}
const int dim = 4;
using Grid = Dune::YaspGrid<dim>; 
Dune::FieldVector<double,dim> length; for (auto& l : length) l=1.0;
std::array<int,dim> nCells; for (auto& c : nCells) c=4;
Grid grid(length,nCells);
\end{c++}
Here we selected the \cpp{YaspGrid} implementation providing a $d$-dimensional structured, parallel grid.
The dimension is set to 4 and given as a template parameter to the \cpp{YaspGrid} class. Then arguments
for the constructor are prepared, which are the length of the domain per coordinate direction 
and the number of elements per direction.
Finally, a grid object is instantiated.
Construction is implementation specific. Other grid implementations might
read a coarse grid from a file. 

Grids can be refined in a hierarchic manner, meaning that elements are subdivided into several smaller elements.
The element to be refined is kept in the grid and remains accessible. More details on local grid refinement
are provided in Section \ref{sec:gridadaptation} below. The following
code snippet refines all elements once and then provides access to the most refined elements in a so-called 
\cpp{GridView}:
\begin{c++}
grid.globalRefine(1);
auto gv = grid.leafGridView();
\end{c++}
A \cpp{GridView} object provides read-only access to the entities of all codimensions in the view.
Iterating over entities of a certain codimension is done by the following snippet using a range-based
for loop:
\begin{c++}
const int codim = 2;
for (const auto& e : entities(gv,Dune::Codim<codim>{}))
   if (!e.type().isCube()) std::cout << "no cube" << std::endl;
\end{c++}
In the loop body the type of the entity is accessed and tested for being a cube (here of dimension 2=4-2).
Access via more traditional names is also possible:
\begin{c++}
for (const auto& e : elements(gv)) assert(e.codim()==0); 
for (const auto& e : vertices(gv)) assert(e.codim()==dim);
for (const auto& e : edges(gv)) assert(e.codim()==dim-1);
for (const auto& e : facets(gv)) assert(e.codim()==1);
\end{c++}
Range-based for loops for iterating over entities have been introduced with release 2.4 in 2015.
Entities of codimension 0, also called elements, provide an extended range of methods.
For example it is possible to access subentities of all codimensions that are contained
in a given element:
\begin{c++}
for (const auto& e : elements(gv))
   for (unsigned int i=0; i<e.subEntities(codim); ++i)
      auto v = e.template subEntity<codim>(i);
\end{c++}
This corresponds to iterating over $U(e)\cap E^c$ for a given $e\in E^0$.

\subsubsection{Geometric aspects}
\label{sec:dune_grid:geometric_aspects}

Geometric information is provided for $e\in E^c$ by a map $\mu_e : \hat\Omega_e \to \Omega_e$, where 
$\hat\Omega_e$ is the domain associated with the reference entity $\hat e$ of $e$ and 
$\Omega_e$ is its image on the manifold $\Omega$.
Usually $\hat\Omega_e$ is one of the usual shapes (simplex, cube, prism, pyramid) 
where numerical quadrature formulae are available.
However, the grid interface also supports arbitrary polygonal elements. In that case no maps $\mu_e$ are provided
and only the measure and the barycenter of each entity is available.
Additionally, the geometry of \textit{intersections} $\Omega_e\cap\Omega_{e'}$ with $d-1$-dimensional measure for $e, e'\in E^0$
is provided as well.

Working with geometric aspects of a grid requires working with positions, e.g. $x\in \hat\Omega_e$,
functions, such as $\mu_e$, or matrices. In \dune{} these follow the \cpp{DenseVector} and
\cpp{DenseMatrix} interface and the most common implementations are the class templates
\cpp{FieldVector} and \cpp{FieldMatrix} providing vectors and matrices with compile-time
known size built on any data type having the operations of a field. Here are some examples (using dimension 3):
\begin{c++}
Dune::FieldVector<double,3> x({1.0,2.0,3.0}); // construct a vector
Dune::FieldVector<double,3> y(x);
y *= 1.0/3.0; // scale by scalar value
double s = x*y; // scalar product
double norm = x.two_norm(); // compute Euclidean norm
Dune::FieldMatrix<double,3,3> A({{1,0,0},{0,1,0},{0,0,1}});
A.mv(x,y); // y = Ax
A.usmv(0.5,x,y); // y += 0.5*Ax
\end{c++}
An entity $e$ (of any codimension) offers the method \cpp{geometry()} returning (a reference to) a geometry object
which provides, among other things, the map $\mu_e: \hat\Omega_e\to\Omega_e$ mapping a local
coordinate in its reference domain $\hat\Omega_e$ to a global coordinate in $\Omega_e$.
Additional methods provide the barycenter of $\Omega_e$ and the volume of $\Omega_e$, for example.
They are used in the following code snipped to approximate the integral over a given function using the
midpoint rule:
\begin{c++}
auto u = [](const auto& x){return std::exp(x.two_norm());};
double integral=0.0;
for (const auto& e : elements(gv))
   integral += u(e.geometry().center())*e.geometry().volume();
\end{c++}
For more accurate integration \dune{} provides a variety of quadrature rules which can be
selected depending on the reference element and quadrature order. Each rule is a container of quadrature
points having a position and a weight. The code snippet below computes the integral over a given function
with fifth order quadrature rule on any grid in any dimension. It illustrates the use of
the \cpp{global()} method on the geometry which evaluates the map $\mu_e$ for a given (quadrature) point.
The method \cpp{integrationElement()} on the geometry provides the measure arising in the transformation formula
of the integral.
\begin{c++}
double integral = 0.0;
using QR = Dune::QuadratureRules<Grid::ctype,Grid::dimension>;
for (const auto& e : elements(gv)) {
   auto geo = e.geometry();
   const auto& quadrature = QR::rule(geo.type(),5);
   for (const auto& qp : quadrature)
      integral += u(geo.global(qp.position()))
          *geo.integrationElement(qp.position())*qp.weight();
}
\end{c++}

\begin{figure}
\begin{center}
\begin{tikzpicture}[font=\footnotesize,rotate=270,xscale=1.0,yscale=0.9]
\draw(2,0) -- (3,0) -- (3,1) -- cycle;
\draw[thick,red] (1.5,0) -- (1.5,1);
\draw (0,0) rectangle (1,1);
\draw[black!20!white] (1.3,4) -- (2.5,4.2) -- (2,5.2) -- (0.8,5) -- cycle;
\draw (1.8,3) -- (1.3,4) -- (2.5,4.2) -- cycle;
\draw (1.8,3) -- (0.8,5) -- (0,4.7) -- (0.3,2.8) -- cycle;
\draw (0.95,0.5) -- (1.05,0.5);
\draw[thick,red]
  (1,0) -- (1,0.5)
  (2,0) -- (3,1)
  (1.8,3) -- (1.3,4);
\path
  (0.5,0.5) coordinate(e1r)
  (2.7,0.3) coordinate(e2r)
  (1.5,0.5) coordinate(ir)
  (0.725,3.875) coordinate(e1w)
  (1.867,3.733) coordinate(e2w)
  (1.55,3.5) coordinate(iw)
  (1,0.25) coordinate(ie1)
  (2.5,0.5) coordinate(ie2);
\draw[dashed]
  (ir)  edge[->,bend left]  node[above] {$\mu_I$}       (iw)
  (e1r) edge[->,bend left]  node[above] {$\mu_{i(I)}$}   (e1w)
  (e2r) edge[->,bend right] node[below]  {$\mu_{o(I)}$}   (e2w)
  (ir)  edge[->,bend left]  coordinate(ie1c)  (ie1)
  (ir)  edge[->,bend right] node(ie2c) [near start] {}  (ie2);
\node[left] (ie1label) at(1.1,-0.8) {$\eta_{I,i(I)}$};
\node[left] (ie2label) at(1.9,-0.8) {$\eta_{I,o(I)}$};
\node (ilabel) at (1.7,2.6) {$\Omega_I$};
\draw[densely dotted]
  (ie1label)+(0,+1ex) edge (ie1c)
  (ie2label)+(0,+1ex) edge (ie2c.center)
  (ilabel)+(0,+1ex) edge (iw);
\node at(1.5,-0.3) {$\hat{\Omega}_I$};
\node at(0.5,-0.4) {$\hat{\Omega}_{i(I)}$};
\node at(2.5,-0.4) {$\hat{\Omega}_{o(I)}$};
\node at(0.7,4.25) {$\Omega_i$};
\node at(2.5,3.7) {$\Omega_o$};
\end{tikzpicture}
\end{center}
\caption{Maps related to an interior intersection.}
\label{fig:Intersections}
\end{figure}
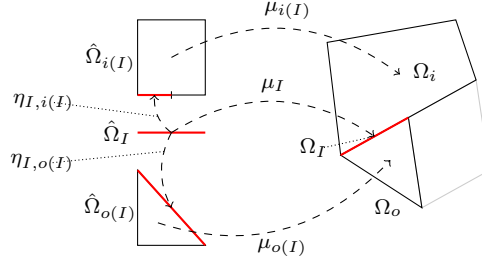

An \textit{intersection} $I$ describes the intersection  $\Omega_I = \partial\Omega_{i(I)} \cap \partial\Omega_{o(I)}$ of two elements
$i(I)$ and $o(I)$ in $E^0$. 
Intersections can be visited from each of the two elements involved. The element
from which $I$ is visited is called the \textit{inside} element $i(I)$ and the other one is called
the \textit{outside} element $o(I)$. Note that $I$ is \textit{not} necessarily an
entity of codimension $1$ in the grid. In this way \dunegrid{} allows for nonconforming grids. In a conforming grid, however, every intersection
corresponds to a codimension 1 entity. For an intersection three maps are provided:
\begin{equation*}
\mu_I : \hat\Omega_I \to \Omega_I, \qquad
\eta_{I,i(I)} = \hat\Omega_I \to \hat\Omega_{i(I)}, \qquad
\eta_{I,o(I)} = \hat\Omega_I \to \hat\Omega_{o(I)} .
\end{equation*}
The first map describes the domain $\Omega_I$ by a map from a corresponding reference element. The second two maps
describe the embedding of the intersection into the reference elements of the inside and outside element, see Figure 
\ref{fig:Intersections}, such that
\begin{equation*}
\mu_I(\hat x) =  \mu_{i(I)}( \eta_{I,i(I)}(\hat x) ) = \mu_{o(I)}( \eta_{I,o(I)}(\hat x) ) .
\end{equation*}
Intersections $\Omega_I=\partial\Omega_{i(I)}\cap\partial\Omega$ with the domain boundary are treated in the same
way except that the outside element is omitted.

As an example consider the approximative computation of the elementwise divergence of a vector field
$\text{div}_e = \int_{\Omega_e} \nabla\cdot f(x) dx = \int_{\partial\Omega_e} f\cdot n ds$ for all elements $e\in E^0$.
Using again the midpoint rule for simplicity this is achieved by the following snippet:
\begin{c++}
auto f = [](const auto& x){return x;};
for (const auto& e : elements(gv)) {
   double divergence=0.0;
   for (const auto& I : intersections(gv,e)) {
      auto geo = I.geometry();
      divergence += f(geo.center())*I.centerUnitOuterNormal()
                       *geo.volume();
   }
}
\end{c++}

\subsubsection{Attaching data to a grid}

In grid-based methods data, such as degrees of freedom in the finite element method,
is associated with geometric entities and stored in containers, such as vectors, external to the grid.
To that end, the grid provides an index for each entity that can be used to access random-access containers.
Often there is only data for entities of a certain codimension and geometrical type (identified by 
its reference entity). Therefore we consider subsets of entities
having the same codimension and reference entity
$$E^{c,\hat e} = \{ e\in E^c : \text{$e$ has reference entity $\hat e$} \}.$$
The grid provides bijective maps
$$\text{index}_{c,\hat{e}} : E^{c,\hat e} \to \{0,\ldots,|E^{c,\hat e}|-1\}$$
enumerating all the entities in $E^{c,\hat e}$ consecutively and starting with zero.
In simple cases where only one data item is to be stored for each entity of  a given
codimension and geometric type this can be used directly to store data in a vector as shown in the following
example:
\begin{c++}
auto& indexset = gv.indexSet();
Dune::GeometryType gt(Dune::GeometryTypes::cube(dim)); // encodes $(c,\hat e)$
std::vector<double> volumes(indexset.size(gt)); // allocate container
for (const auto& e : elements(gv))
   volumes[indexset.index(e)] = e.geometry().volume();
\end{c++}
Here, the volumes of the elements in a single element type grid are stored in a vector.
Note that a \cpp{GeometryType} object encodes both, the dimension and the geometric type,
e.g. simplex or cube.
In more complicated situations an index map for entities of different codimensions and/or geometry
types needs to be composed of several of the simple maps. This leaves the layout of degrees of
freedom in a vector under user control and allows realization of different blocking strategies.
\dunegrid offers several classes for this purpose, such as
\cpp{MCMGMapper} which can map entities of multiple codimensions and multiple geometry types to a consecutive index.

When a grid is modified through adaptive refinement, coarsening, or load balancing in the parallel case,
the index maps may change as they are required to be consecutive and zero-starting.  
In order to store and access data reliably when the grid is modified each geometric entity is equipped
with a global id:
$$\text{globalid} : \bigcup_{c=0}^d E^c \to \mathbb{I}$$ 
where $\mathbb{I}$ is a set of unique identifiers.
The map globalid is injective and persistent, i.e. globalid($e$) does not change under grid modification
when entity $e$ is in the old \textit{and} the new grid, and globalid($e$)
is not used when $e$ was in the old grid and
is not in the new grid (note that global ids may be used again after the next grid modification).
There are very weak assuptions on the ids provided by globalid. They
don't need to be consecutive, actually they don't even need to be
numbers.
Here is how element volumes would be stored in a map:
\begin{c++}
const auto& globalidset = gv.grid().globalIdSet();
using GlobalId = Grid::GlobalIdSet::IdType;
std::map<GlobalId,double> volumes2;
for (const auto& e : elements(gv))
   volumes2[globalidset.id(e)] = e.geometry().volume();
\end{c++}
The type \cpp{GlobalId} represents $\mathbb{I}$ and must be
sortable and hashable. This requirement is necessary to be able to
store data for example in an \cpp{std::map} or \cpp{std::unordered_map}. For
example, \cpp{YaspGrid} uses the \cpp{bigunsignedint}
class from \dunecommon that implements arbitrarily large unsigned integers,
while \dune{-UGGrid} uses a \cpp{std::uint_least64_t} which is stored in each
entity. In \dunealugrid the  \cpp{GlobalId} of an element is computed 
from the macro element's unique vertex ids, codimension, and refinement
information.

The typical use case would be to store data in vectors and use an \cpp{indexset} while the grid is in \textit{read-only} state 
and to copy only the necessary data to a map using \cpp{globalidset} when the grid is being modified.
Since using a \cpp{std::map} may not be the most efficient way to store data, 
a utility class \cpp{PersistentContainer<Grid, T>} exists, that implements
the strategy outlined above 
for arbitrary types \cpp{T}.
 To allow for optimization, this class can be specialized by the grid implementation
 using structural information to optimize performance.

\subsubsection{Grid refinement and coarsening}
\label{sec:dunegridrefinement}

Adaptive mesh refinement using a posteriori error estimation is an established
and powerful technique to reduce the computational effort in the numerical solution
of PDEs, see e.g.\cite{AINSWORTH19971}. \dunegrid{} supports the typical \textit{estimate-mark-refine} paradigm
as illustrated by the following code example:
\begin{c++}
const int dim = 2;
using Grid = Dune::UGGrid<dim>;
auto pgrid = std::shared_ptr<Grid>(
                Dune::GmshReader<Grid>::read("circle.msh"));
auto h = [](const auto& x) 
           {auto d=x.two_norm(); return 1E-6*(1-d)+0.01*d;};
for (int i=0; i<15; ++i) {
  auto gv = pgrid->leafGridView();
  for (const auto& e : elements(gv)) {
    auto diameter=std::sqrt(e.geometry().volume()/M_PI);
    if (diameter>h(e.geometry().center())) pgrid->mark(1,e);
  }
  pgrid->preAdapt();
  pgrid->adapt();
  pgrid->postAdapt();
}
\end{c++}
Here the \cpp{UGGrid} implementation is used in dimension 2.
In each refinement iteration those elements with a diameter larger than a desired
value given by the function \cpp{h} are marked for refinement. The method
\cpp{adapt()} actually modifies the grid, while \cpp{preAdapt()} determines grid entities which might be deleted
and \cpp{postAdapt()} clears the information about new grid entities. In between
\cpp{preAdapt()} and \cpp{adapt()} data from the old grid needs to be stored using persistent global ids
and in between \cpp{adapt()} and \cpp{postAdapt()} this data is transferred to the new grid.
In order to identify elements that may be
affected by grid coarsening and refinement the element offers two
methods. The method \cpp{mightVanish()}, typically used between \cpp{preAdapt()} and 
\cpp{adapt()}, returns \cpp{true} if the entity might vanish during the grid modifications carried out in \cpp{adapt()}.
Afterwards, the method \cpp{isNew()} returns \cpp{true} if an element was newly created during the previous 
\cpp{adapt()} call.
\textit{How} an element is refined when it is marked for refinement is specific to the implementation. Some implementations
offer several ways to refine an element.
Furthermore some grids may refine non-marked elements in an implementation specific way
to ensure certain mesh properties like, e.g., conformity.
For implementation of data restriction
and prolongation a \cpp{geometryInFather()} method provides geometrical mapping between parent and children elements.

\begin{figure}
\begin{center}
\begin{tikzpicture}[scale=1]
\draw[very thick] (0.0,0.0) -- (8.0,0.0); 
\filldraw (0,0) circle (1mm);
\filldraw (4,0) circle (1mm);
\filldraw (8,0) circle (1mm);
\node[left] at (-0.3,0.0) {level 0};
\draw[very thick] (0.0,0.5) -- (8.0,0.5); 
\filldraw (0,0.5) circle (1mm);
\filldraw (2,0.5) circle (1mm);
\filldraw (6,0.5) circle (1mm);
\filldraw (4,0.5) circle (1mm);
\filldraw (8,0.5) circle (1mm);
\node[left] at (-0.3,0.5) {level 1};
\draw[very thick] (0.0,1.0) -- (4.0,1.0); 
\draw[very thick] (6.0,1.0) -- (8.0,1.0); 
\filldraw (0,1.0) circle (1mm);
\filldraw (1,1.0) circle (1mm);
\filldraw (3,1.0) circle (1mm);
\filldraw (7,1.0) circle (1mm);
\filldraw (2,1.0) circle (1mm);
\filldraw (6,1.0) circle (1mm);
\filldraw (4,1.0) circle (1mm);
\filldraw (8,1.0) circle (1mm);
\node[left] at (-0.3,1.0) {level 2};
\draw[very thick] (1.0,1.5) -- (3.0,1.5); 
\filldraw (1.0,1.5) circle (1mm);
\filldraw (1.5,1.5) circle (1mm);
\filldraw (2.0,1.5) circle (1mm);
\filldraw (2.5,1.5) circle (1mm);
\filldraw (3.0,1.5) circle (1mm);
\node[left] at (-0.3,1.5) {level 3};
\draw[very thick,blue] (0.0,2.5) -- (8.0,2.5); 
\node[left] at (-0.3,2.5) {\color{blue}leaf};
\filldraw[blue] (0.0,2.5) circle (1mm);
\filldraw[blue] (1.0,2.5) circle (1mm);
\filldraw[blue] (1.5,2.5) circle (1mm);
\filldraw[blue] (2.0,2.5) circle (1mm);
\filldraw[blue] (2.5,2.5) circle (1mm);
\filldraw[blue] (3.0,2.5) circle (1mm);
\filldraw[blue] (4.0,2.5) circle (1mm);
\filldraw[blue] (6.0,2.5) circle (1mm);
\filldraw[blue] (7.0,2.5) circle (1mm);
\filldraw[blue] (8.0,2.5) circle (1mm);
\draw[->,very thick,blue] (0.5,1.2) -- (0.5,2.3); 
\draw[->,very thick,blue] (1.25,1.7) -- (1.25,2.3); 
\draw[->,very thick,blue] (1.75,1.7) -- (1.75,2.3); 
\draw[->,very thick,blue] (2.25,1.7) -- (2.25,2.3); 
\draw[->,very thick,blue] (2.75,1.7) -- (2.75,2.3); 
\draw[->,very thick,blue] (3.5,1.2) -- (3.5,2.3); 
\draw[->,very thick,blue] (5,0.7) -- (5,2.3); 
\draw[->,very thick,blue] (6.5,1.2) -- (6.5,2.3); 
\draw[->,very thick,blue] (7.5,1.2) -- (7.5,2.3); 
\end{tikzpicture}
\end{center}
\caption{Level and leaf grid views.}
\label{fig:GridViews}
\end{figure}
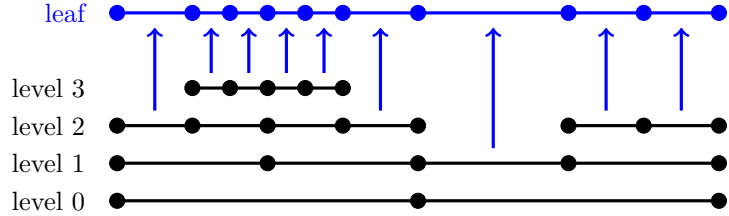

Grid refinement is hierarchic in all currently available \dunegrid{} implementations. 
Each entity is associated with a \textit{grid level}. After construction of a grid object all its entities are on level 0. When an
entity is refined the entities resulting from this refinement, also called its direct children, are added on the next higher level.
Each level-0-element and all its descendants form a tree. All entities on level $l$ are the entities of the level $l$ grid view.
All entities that are not refined are the entities of the leaf grid view. This is illustrated in Figure \ref{fig:GridViews}.
The following code snippet traverses all vertices on all levels of the grid using a
\cpp{levelGridView}:
\begin{c++}
for (int l=0; l<=pgrid->maxLevel(); l++) 
   for (const auto& v : vertices(pgrid->levelGridView(l)))
      assert( v.level() == l); // check level consistency 
\end{c++}
Each \cpp{GridView} provides its own \cpp{IndexSet} and so allows to associate data with
entities of a single level or with all entities in the leaf view.

\subsubsection{Parallelization}

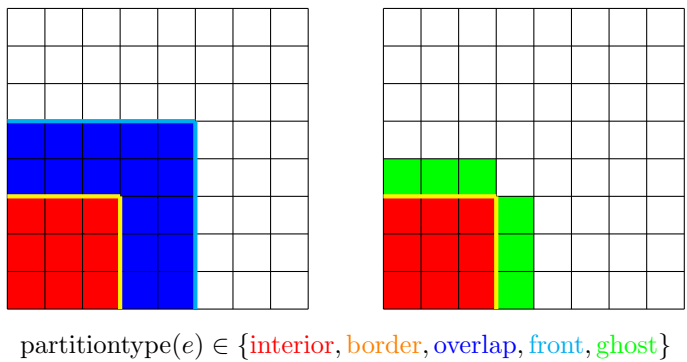
\begin{figure}
\begin{center}
\begin{tikzpicture}[scale=1]
\fill[fill=red] (0,0) rectangle (1.5,1.5);
\fill[fill=blue] (1.5,0) rectangle (2.5,2.5);
\fill[fill=blue] (0.0,1.5) rectangle (1.5,2.5);
\draw (0.0,0.0) -- (0.0,4.0); 
\draw (0.5,0.0) -- (0.5,4.0); 
\draw (1.0,0.0) -- (1.0,4.0); 
\draw (1.5,0.0) -- (1.5,4.0); 
\draw (2.0,0.0) -- (2.0,4.0); 
\draw (2.5,0.0) -- (2.5,4.0); 
\draw (3.0,0.0) -- (3.0,4.0); 
\draw (3.5,0.0) -- (3.5,4.0); 
\draw (4.0,0.0) -- (4.0,4.0); 
\draw (0.0,0.0) -- (4.0,0.0); 
\draw (0.0,0.5) -- (4.0,0.5); 
\draw (0.0,1.0) -- (4.0,1.0); 
\draw (0.0,1.5) -- (4.0,1.5); 
\draw (0.0,2.0) -- (4.0,2.0); 
\draw (0.0,2.5) -- (4.0,2.5); 
\draw (0.0,3.0) -- (4.0,3.0); 
\draw (0.0,3.5) -- (4.0,3.5); 
\draw (0.0,4.0) -- (4.0,4.0); 
\draw[ultra thick,yellow] (0.0,1.5) -- (1.5,1.5); 
\draw[ultra thick,yellow] (1.5,0.0) -- (1.5,1.5); 
\draw[ultra thick,cyan] (0.0,2.5) -- (2.5,2.5); 
\draw[ultra thick,cyan] (2.5,0.0) -- (2.5,2.5); 
\fill[fill=red] (5,0) rectangle (6.5,1.5);
\fill[fill=green] (6.5,0) rectangle (7,1.5);
\fill[fill=green] (5,1.5) rectangle (6.5,2);
\draw (5.0,0.0) -- (5.0,4.0); 
\draw (5.5,0.0) -- (5.5,4.0); 
\draw (6.0,0.0) -- (6.0,4.0); 
\draw (6.5,0.0) -- (6.5,4.0); 
\draw (7.0,0.0) -- (7.0,4.0); 
\draw (7.5,0.0) -- (7.5,4.0); 
\draw (8.0,0.0) -- (8.0,4.0); 
\draw (8.5,0.0) -- (8.5,4.0); 
\draw (9.0,0.0) -- (9.0,4.0); 
\draw (5.0,0.0) -- (9.0,0.0); 
\draw (5.0,0.5) -- (9.0,0.5); 
\draw (5.0,1.0) -- (9.0,1.0); 
\draw (5.0,1.5) -- (9.0,1.5); 
\draw (5.0,2.0) -- (9.0,2.0); 
\draw (5.0,2.5) -- (9.0,2.5); 
\draw (5.0,3.0) -- (9.0,3.0); 
\draw (5.0,3.5) -- (9.0,3.5); 
\draw (5.0,4.0) -- (9.0,4.0); 
\draw[ultra thick,yellow] (5.0,1.5) -- (6.5,1.5); 
\draw[ultra thick,yellow] (6.5,0.0) -- (6.5,1.5); 
\node at (4.5,-0.5) {$\text{partitiontype}(e) \in \{\text{\color{red}interior}, \text{\color{orange}border}, \text{\color{blue}overlap}, 
\text{\color{cyan}front}, \text{\color{green}ghost} \}$};
\end{tikzpicture}
\end{center}
\caption{Two types of data decomposition implemented by \cpp{YaspGrid} (left) and \cpp{UGGrid}/\cpp{ALUGrid} (right).
The colored entities show the entities of one rank.
Sub-entities of elements assume the partition type of the element unless those
sub-entities are located on a border between different partition types of \cpp{interior,overlap} or \cpp{ghost}.
  }
\label{fig:data_decomposition}
\end{figure}

Parallelization in \dunegrid{} is based on three concepts:
\begin{enumerate*}
\item data decomposition,
\item message passing paradigm and
\item single-program-multiple-data (SPMD) style programming.
\end{enumerate*}
As for the refinement rules in grid adaptation the 
data decomposition is implementation specific but must adhere to certain rules:
\begin{enumerate}
\item The decomposition of codimension 0 entities $E^0$ into sets $E^{0,r}$ assigned to process rank $r$ form a
(possibly overlapping) partitioning $\bigcup_{i=0}^{p-1} E^{c,r} = E^c$.
\item When process $r$ has a codimension 0 entity $e$ then it also stores all its subentities, i.e.
$e\in E^{0,r} \wedge E^c\ni f\subseteq e \Rightarrow f\in E^{c,r}$ for $c>0$.
\item Each entity is assigned a partition type attribute
$$\text{partitiontype}(e) \in \{\text{interior}, \text{border}, \text{overlap}, \text{front}, \text{ghost} \}$$
with the following semantics:
\begin{enumerate}
\item Codimension 0 entities may only have the partition types $\text{interior}, \text{overlap}$, or $\text{ghost}$.
The interior codimension 0 entities $E^{0,r,\text{interior}} = \{e\in E^{0,r} : \text{partitiontype}(e)=\text{interior}\} $
form a nonoverlapping partitioning of $E^0$.
Codimension 0 entities with partition type overlap can be used like regular entities whereas those with partition type ghost
only provide a limited functionality (e.g. intersections may not be provided).
\item The partition type of entities with codimension $c>0$ is derived from the codimension 0 entities they are contained in.
For any entity $f\in E^{c}$, $c>0$, set $\Sigma^{0}(f)=\{e\in E^{0} : f\subseteq e \}$ and $\Sigma^{0,r}(f) = \Sigma^{0}(f) \cap E^{0,r}$. 
If $\Sigma^{0,r}(f) \subseteq E^{0,r,\text{interior}}$ and $\Sigma^{0,r}(f) = \Sigma^0(f)$ then $f$ is interior, 
else if $\Sigma^{0,r}(f) \cap E^{0,r,\text{interior}} \neq\emptyset$ then $f$ is border, else
if $\Sigma^{0,r}(f)$ contains only overlap entities and $\Sigma^{0,r}(f) = \Sigma^0(f)$ then $f$ is overlap, else 
if $\Sigma^{0,r}(f)$ contains overlap entities then $f$ is front, else $f$ is ghost. 
\end{enumerate}
\end{enumerate}
Two examples of typical data decomposition models are shown in Figure \ref{fig:data_decomposition}. Variant a) on the left with
interior/overlap codimension 0 entities is implemented by \cpp{YaspGrid}, variant b) on the right with the interior/ghost model is implemented
by \cpp{UGGrid} and \cpp{ALUGrid}.

To illustrate SPMD style programming we consider a simple example.
Grid instantiation is done by all processes $r$ with identical arguments and each stores its respective grid partition $E^{c,r}$.
\begin{c++}
const int dim = 2;
using Grid = Dune::YaspGrid<dim>; 
Dune::FieldVector<double,dim> length; for (auto& l:length) l=1.0; 
std::array<int,dim> nCells; for (auto& c : nCells) c=20;
Grid grid(length,nCells,std::bitset<dim>(0ULL),1);
auto gv = grid.leafGridView();
\end{c++}
Here, the third constructor argument of the grid
controls periodic boundary conditions and the last argument sets the amount of overlap in codimension 0 entities.

Parallel computation of an integral over a function using the midpoint rule is illustrated by the following code snippet:
\begin{c++}
auto u = [](const auto& x){ return std::exp(x.two_norm()); };
double integral=0.0;
for (const auto& e : elements(gv,Dune::Partitions::interior))
  integral += u(e.geometry().center())*e.geometry().volume();
integral = gv.comm().sum(integral);
\end{c++}
In the range-based for loop we specify in addition that iteration is restricted to interior elements.
Thus, each element of the grid is visited exactly once. After each process has computed the integral 
on its elements a global sum (allreduce) produces the result which is now known by
each process.

Data on overlapping entities $E^{c,r}\cap E^{c,s}$ stored by two processes $r\neq s$ can be communicated
with the abstraction \cpp{CommDataHandleIF} describing which information is sent for each entity and how it is processed.
Communication is then initiated by the method \cpp{communicate()} on a \cpp{GridView}.
Although all current parallel grid implementation use the message
passing interface (MPI) in their implementation, nowhere the user has to make explicit MPI calls. Thus, an implementation could 
also use shared memory access to implement the \dunegrid{} parallel functionality. Alternatively, multithreading can be used
within a single process by iterating over grid entities in parallel. This has been implemented in the \exadune{} project
\cite{BEGIIOTFKMR14,exa:highlevel2} or in \dunefem \cite{dgimpl:12} but a common interface concept 
is not yet part of \dune{} core functionality.

\subsubsection{List of grid implementations}
\label{sec:gridimplemenations}

The following list gives an overview of existing implementations of the \dunegrid{} interface and their properties and the \dune module these are implemented in.
Where noted, the implementation wraps access to an external library. A complete
list can be found on the \dune web page \url{https://dune-project.org/doc/grids/}.
\begin{description}[font=\ttfamily\mdseries]
  \item[AlbertaGrid (dune-grid)] Provides simplicial grids in two and three dimensions with bisection refinement based
  on the ALBERTA software \cite{alberta:05}.
\item[ALUGrid (dune-alugrid)] Provides a parallel unstructured grid in two and three dimensions
using either simplices or cubes. Refinement is nonconforming for simplices and cubes. Conforming refinement based on bisection is 
    supported for simplices only\cite{alugrid:16}.
\item[CurvilinearGrid (dune-curvilineargrid)] Provides a parallel simplicial grid \cite{2016arXiv161202967F} supporting curvilinear grids read from Gmsh
\cite{doi:10.1002/nme.2579} input.
\item[CpGrid (opm-grid)] Provides an implementation of a corner point grid, a nonconforming hexahedral grid,
which is the standard in the oil industry \url{https://opm-project.org/}.
\item[FoamGrid (dune-foamgrid)] Provides one and two-dimensional grids embedded in three-dimensional space including 
  non-manifold grids with branches \cite{foamgrid}.
\item[OneDGrid (dune-grid)] Provides an adaptive one-dimensional grid.
\item[PolygonGrid (dune-polygongrid)] A grid with polygonal cells (2d only).
\item[UGGrid (dune-grid)] Provides a parallel, unstructured grid with mixed element types (triangles and
quadrilaterals in two, tetrahedra, pyramids, prisms, and hexahedra in three dimensions) and local
refinement. Based on the UG library \cite{Bastian1997}.
\item[YaspGrid (dune-grid)] A parallel, structured grid in arbitrary dimension using cubes. Supports non-equidistant mesh spacing and
periodic boundaries.
\end{description}

Metagrids use one or more implementations of the \dunegrid{} interface to provide
either a new implementation of the \dunegrid{} interface or new functionality all together. Here are examples:
\begin{description}[font=\ttfamily\mdseries]
  \item[GeometryGrid (dune-grid)] Takes any grid and replaces the geometries of all entities $e$ by the concatenation
$\mu_{geo}\circ\mu_e$ where $\mu_{geo}$ is a user-defined mapping, see Section  \ref{sec:moving_grids}.
\item[PrismGrid (dune-metagrid)] Takes any grid of dimension $d$ and extends it by a structured grid
in direction $d+1$ \cite{dune-prismgrid:12}.
\item[GridGlue (dune-grid-glue)] Takes two grids and provides a projection of one on
the other as a set of intersections \cite{gridglue,engwer2016concepts}, see Section \ref{sec:dunegridglue}.
\item[MultiDomainGrid (dune-multidomaingrid)] Takes a grid and provides possibly overlapping sets of elements
as individual grids \cite{Steffen2015}, see Section~\ref{sec:multidomaingrid}.
\item[SubGrid (dune-subgrid)] Takes a grid and provides a subset of its entities as a new grid \cite{Graeser2009}.
\item[IdentityGrid (dune-grid)] Wraps all classes of one grid implementation in new classes that delegate to the existing
implementation. This can serve as an ideal base to write a new metagrid.
\item[CartesianGrid (dune-metagrid)] Takes a unstructured quadrilateral or hexahedral grid (e.g. \cpp{ALUGrid} or \cpp{UGGrid}) and replaces the
  geometry implementation with a strictly Cartesian geometry implementation for
    performance improvements. 
  \item[FilteredGrid (dune-metagrid)] Takes any grid and applies a binary filter to the entity sets for codimension $0$ 
  provided by the grid view of the given grid.
\item[SphereGrid (dune-metagrid)] A meta grid that provides the correct
  spherical mapping for geometries and normals for underlying spherical grids. 
\end{description}

\subsubsection{Major developments in the \dunegrid interface}
Version 1.0 of the \dunegrid{}  module was released on December 20, 2007.
Since then a number of improvements were introduced, including the following:
\begin{itemize}
\item Methods \cpp{center()}, \cpp{volume()}, \cpp{centerUnitOuterNormal()} on \cpp{Geometry} and \cpp{Intersection} 
were introduced to support FV methods on polygonal and polyhedral grids.
\item \cpp{GridFactory} provides an interface for portably creating initial meshes.
\cpp{GmshReader} uses that to import grids generated with \cpp{gmsh}.
\item \cpp{EntitySeed} replaced \cpp{EntityPointer}; this allows the grid to free the memory occupied
by the entity and to recreate the entity from the seed.
\item The \dunegeometry{} module was introduced as a separate module to provide reference
elements, geometry mappings, and quadrature formulae independent of \dunegrid{}.
\item The automatic type deduction using \cpp{auto} makes using heavily template-based libraries
such as \dune{} more convenient to use.
\item Initially, the \dunegrid{} interface tried to avoid copying objects for performance reasons.
Many methods returned const references to internal data and disallowed copying.
With copy elision becoming standard, copyable lightweight entities and intersections were introduced.
Given an \cpp{Entity} with codimension $c$ to obtain the geometry one would write: 
\begin{c++}
  using Geometry = typename Grid::template Codim< c >::Geometry;
  const Geometry& geo = entity.geometry();
\end{c++}
whereas in newer \dune versions one can simply write:
\begin{c++}
  const auto geo = entity.geometry();
\end{c++}
using both, the automatic type deduction and the fact that objects are
copyable. 
A performance comparison discussing the references vs. copyable grid objects 
can be found in \cite{dune24:16,perf:11}.
In order to save on memory when storing entities the entity seed concept was introduced. 
\item Range-based for loops for entities and intersections made iteration over grid entities very convenient.
  With newer \dune versions this is simply:
\begin{c++}
  for( const auto& element : Dune::elements( gridView ) )
    const auto geo = element.geometry();
\end{c++}
\item \cpp{UGGrid} and \cpp{ALUGrid} became dune modules instead of being external libraries. This way they can 
be downloaded and installed like other dune modules.
\end{itemize}

As the \dune grid interface has been adapted only slightly, it proved to work for a wide audience.
Looking back, the separation of both topology and geometry, and mesh and data were good principles.
Further, having entities as a view was a successful choice.
Having \dune split up in modules helped to keep the grid interface separated from other concerns.
Some interface changes resulted from extensive advancements of C++11 and the subsequent standards;
many are described in \cite{dune24:16}.
Other changes turned out to make the interface easier to use or to enable different methods on top
of the grid interface.

%% file: duneistl.tex
\subsection{The template library for iterative solvers -- \duneistl} 
\label{sec:duneistl} 
\duneistl is the linear algebra library of \dune. It consists of two
main components. First it offers a collection of different vector and
matrix classes. Second it features different solvers and
preconditioners. While the grid interface consists of fine grained interfaces and
relies heavily on static polymorphism, the abstraction in \duneistl
uses a combination of dynamic and static polymorphism.

\subsubsection{Concepts behind the \duneistl{} interfaces}

A major design decision in \duneistl was influenced by the
observation, that linear solvers can significantly benefit from
inherent structure of PDE discretizations. For example a discontinuous Galerkin (DG)
discretization leads to a block structured matrix for certain
orderings of the unknowns, the same holds for
coupled diffusion reaction systems with many components. Making use of
this structure often allows to improve convergence of the linear solver,
reduce memory consumption and improve memory throughput.

\duneistl offers different vector and matrix implementations and many
of these can be nested. The whole interface is fully templatized
w.r.t. the underlying scalar data type (double, float, etc.), also
called field type. Examples of such nested types are:
\begin{descclasses}
\item[Dune::BlockVector<Dune::FieldVector<std::complex<double>,2>>] a
  dynamic block vector, which consists of $N$ blocks of vectors with static size
  2 over the field of the complex numbers, i.e. a vector in $(\mathbb{C}^2)^N$.
\item[\normalfont\texttt{Dune::BCRSMatrix<Dune::FieldMatrix<float,27,27>>}]
  a sparse block matrix with dense $27\times 27$ matrices as its
  entries. The dense matrices  use a 
  low precision \cpp{float} representation. The whole matrix
  represents a linear mapping  $(\mathbb{R}^{27\times 27})^{N\times M}$.
\item[\normalfont\texttt{Dune::BCRSMatrix<Dune::BCRSMatrix<double>>}]
  a (sparse) matrix whose entries are sparse matrices with scalar entries. These might for example arise from a Taylor--Hood  \cite{Wathen2014} discretization of the
  Navier--Stokes equations, where we obtain a $4 \times 4$ block matrix of sparse matrices. 
\end{descclasses}
It is not necessary to use the same field type for
matrices and vectors, as the library allows for mixed-precision
setups with automatic conversions and determination of the correct
return type of numeric operations. In order to allow for an efficient access to individual
entries in matrices or vectors, these matrix/vector interfaces are static and
make use of compile-time polymorphism, mainly by using duck
typing\footnote{Used characteristics rather than actual type for algorithms: ``if it walks like a duck and quacks like a duck, it is
  a duck''}
like in the STL\footnote{standard template library}.

Solvers on the other hand can be formulated at a very high level of
abstraction and are a perfect candidates for dynamic polymorphism.
\duneistl defines abstract interfaces for operators, scalar products,
solvers, and preconditioners. A solver, like \cpp{LoopSolver},
\cpp{CGSolver}, and similar is parameterized with the operator,
possibly a preconditioner, and usually the standard euclidean scalar
product. The particular implementations, as well as the interface, are
strongly typed on the underlying vector types, but it is
possible to mix and shuffle different solvers and preconditioners
dynamically at runtime.  While linear operators are most often stored as
matrices, the interface only requires that an operator can be applied
to a vector
and thus also allows for implementing on-the-fly operators for implicit
methods; this drastically reduces the memory consumption and
allows for increased arithmetic intensity
and thus overcomes performance
limitations due to slow memory access. The benefit of on-the-fly
operators is highlighted in section~\ref{sec:hpc}.
It should be noted that many strong preconditioners, like the
\cpp{Dune::Amg::AMG} have stricter
requirements on the operator and need access to the full matrix.

The interface design also
offers a simple way to introduce parallel solvers.
The parallelization of \duneistl's data structures and solvers differs
significantly from that of other libraries like PETSc. While many
libraries rely on a globally consecutive numbering of unknowns, we only require a
locally consecutive numbering, which allows for fast access into
local containers. Global consistency is then achieved by choosing
appropriate parallel operators, preconditioners, and scalar products.
Note that the linear
solvers do
not need any information about parallel data distribution, as they
only rely on operator (and preconditioner) applications and scalar
product computations, which are hidden under the afore introduced high
level abstractions. This allows for a fully transparent switch between
sequential and parallel computations.

\subsubsection{A brief history}

The development of \duneistl began nearly in parallel with
\dunegrid around the year 2004 and it was included in the 1.0 release
of \dune in
2007. 
The serial implementation was presented in
\cite{blatt_bastian_ISTL:2007}. One goal was to make \duneistl usable
standalone without \dunegrid. Hence a powerful abstraction of parallel
iterative solvers based on the concept of parallel index sets was
developed as described in \cite{blatt_bastian_ISTLParallel:2008}. As a
first showcase of it an aggregation based parallel algebraic multigrid method
for continuous and discontinuous Galerkin
discretizations of heterogeneous elliptic problems was added to the
library, see \cite{blatt_parallel:2010,bastian_algebraic:2012}. It was one
of the first solvers scaling to nearly 295,000 processors on a problem
with 150 billion unknowns, see \cite{ippisch_scalability:2011}.

\subsubsection{Feature overview and recent developments}

The afore outlined concepts are implemented using several templatized
C++ structures. Linear operators, that do not do their computations on
the fly, will often use an underlying matrix representation. \duneistl
offers dense matrices, both either of dynamic size or size known
already at compile time, as well as several sparse (block) matrices. For
a comprehensive list see Table \ref{tab:istl_matrices}. The
corresponding vector classes can be found in Table \ref{tab:istl_vectors}.

\begin{table}[ht]
  \renewcommand{\arraystretch}{1.05}
  \centering
  \caption{Matrix types in \duneistl, the first three matrix types
    can not be used as a block matrices.  }
  \label{tab:istl_matrices}
  \begin{tabular}{l|p{7cm}}
    class & implements \\\hline\hline
    \texttt{FieldMatrix} & (small) matrix with size known at compile
                           time\\
    \texttt{DiagonalMatrix} & storage optimal representation of a diagonal matrix with size known at compile time\\
    \texttt{ScaledIdentityMatrix} & storage optimal representation of a scaled identity matrix with size known at compile time\\\hline
    \texttt{BCRSMatrix} & (block) compressed row storage matrix \\
    \texttt{BDMatrix} & (block) diagonal matrix \\
    \texttt{BTDMatrix} & (block) tri-diagonal matrix \\
    \texttt{Matrix} & generic dynamic dense (block) matrix\\
    \texttt{MultiTypeBlockMatrix} & dense block matrix with differing block
                           type known at compile time
  \end{tabular}
\end{table}
\begin{table}[ht]
  \renewcommand{\arraystretch}{1.05}
  \centering
  \caption{Vector types in \duneistl, the first vector type can not
    be used as a block vector.}
  \label{tab:istl_vectors}
  \begin{tabular}{l|p{7cm}}
    class & implements \\\hline\hline
    \texttt{FieldVector} & (small) vector with size known at compile time\\\hline
    \texttt{BVector} & (block) vector, blocks have same size \\
    \texttt{VariableBlockVector} & block vector where each block may vary in size \\
    \texttt{MultiTypeBlockVector} & block vector with differing block
                           type known at compile time
  \end{tabular}
\end{table}

The most important building blocks of the iterative solvers in \duneistl are the
preconditioners. Together with the scalar product and linear operator
they govern whether a solver will be serial/sequential only or capable of running
in parallel. To mark sequential solvers the convention is that their
name starts with \cpp{Seq}.
Using the idea of inexact block Jacobi methods or Schwarz type methods, the
\cpp{BlockPreconditioner} allows to turn any sequential 
into a parallel preconditioner, given information about the parallel data decomposition.
Such
so-called hybrid preconditioners are commonly used
in parallel (algebraic) multigrid methods, see \cite{yang_hybrid:2004}.
A list of preconditioners provided by \duneistl is in Table
\ref{tab:istl_preconditioners}. The third column indicates whether a
preconditioner is sequential (s), parallel (p), or both. For simple
preconditioners, that do not need to store a decomposition, a
recursion level can be given to the class. Those are marked with ``yes''
in the last column. The level given to the class indicates where the
inversion on the matrix block happens. For a
\cpp{BCRSMatrix<FieldMatrix<double,n,m> >} a level of $0$ will lead to
the inversion of
the scalar values inside of the small dense matrices whereas a level of
$1$ would invert the \cpp{FieldMatrix}. The latter variant, which leads to a
block preconditioner, is the default. All of the listed preconditioners
can be used in the iterative solvers provided by \duneistl. Table
\ref{tab:istl_solvers} contains a list of these together with the direct
solvers. The latter are only wrappers to existing well established libraries.
\begin{table}[ht]
  \renewcommand{\arraystretch}{1.05}
  \centering
  \caption{Preconditioners in \duneistl}
  \label{tab:istl_preconditioners}
  \begin{tabular}{l|p{5cm}|l|l}
    class & implements & s/p & recursive\\\hline\hline
    \texttt{Richardson} & Richardson (multiplicate with a scalar) & s & no\\
    \texttt{SeqJac} & Jacobi method & s & yes\\
    \texttt{SeqSOR} & successive overrelaxation (SOR) & s & yes\\
    \texttt{SeqSSOR} & symmetric SOR & s & yes\\
    \texttt{SeqOverlappingSchwarz} & overlapping Schwarz for arbitrary
                            subdomains &s & no \\
    \texttt{SeqILU} & incomplete LU decomposition & s & no\\
    \texttt{SeqILDL} & incomplete LDL decomposition & s & no\\
    \texttt{Pamg::AMG} & algebraic multigrid solver based on aggregation & s/p
                             & no\\
    \texttt{BlockPreconditioner} & wraps sequential preconditioner to
                          parallel hybrid one 
                          & p & no
  \end{tabular}
\end{table}

\begin{table}[ht]
  \renewcommand{\arraystretch}{1.05}
  \centering
  \caption{Iterative and direct solvers in \duneistl. Some of these
    solvers can handle non-static preconditioner, i.e. the
    preconditioner might change from iteration to iteration.}
  \label{tab:istl_solvers}
  \begin{tabular}{l|p{5cm}|l}
    class & implements & direct \\\hline\hline
    \texttt{LoopSolver} & simply applies preconditioner in each step & no\\
    \texttt{GradientSolver} & simple gradient solver & no\\
    \texttt{CGSolver} & conjugate gradient method & no\\
    \texttt{BiCGSTABSolver} & biconjugate gradient stabilized method & no\\
    \texttt{MINRESSolver} & minimal residual method & no\\
    \texttt{RestartedGMResSolver} & restarted GMRes solver & no \\
    \texttt{RestartedFlexibleGMResSolver} & flexible restarted GMRes solver (for
                                   non-static preconditioners) & no \\
    \texttt{GeneralizedPCGSolver} & flexible conjugate gradient solver (for non-static
                           preconditioners) & no \\
    \texttt{RestartedFCGSolver} & flexible conjugate gradient solver proposed
                         by Notay (for non-static preconditioners) & no\\
    \texttt{CompleteFCGSolver} & flexible conjugate gradient method reusing
                        old orthogonalizations when restarting& no\\\hline
    \texttt{SuperLU} & Wrapper for SuperLU library & yes\\
    \texttt{UMFPack} & Wrapper for UMFPack direct solver library & yes\\
  \end{tabular}
\end{table}

  In recent time \duneistl has seen a lot of new development. Both,
  regarding usability, as well as feature-wise. We now briefly
  discuss some noteworthy improvements.
%
\begin{enumerate}
    \item From the start, \duneistl was designed to support nested vector and
      matrix structures. However, the nesting recursion always had to end in
      \cpp{FieldVector} and \cpp{FieldMatrix}, respectively.  Scalar entries
      had to be written as vectors of length~1 or matrices of size $1 \times 1$.
      Exploiting modern C++ idioms now allows to support scalar values directly to end the
      recursion. In other words, it is now possible to write
      \begin{c++}
Dune::BCRSMatrix<double>
\end{c++}
      instead of the lengthy
      \begin{c++}
Dune::BCRSMatrix<FieldMatrix<double,1,1>>
\end{c++}
      Internally, this is implemented using the \cpp{Dune::IsNumber} traits class
      to recognize scalar types. Note that the indirections needed internally
      to implement the transparent use of scalar and blocked entries
      is completely optimized away by the compiler.

  \item As discussed in the concepts section, operators, solvers, preconditioners,
    and scalar products offer only coarse grained interfaces. This
    allows to use dynamic polymorphism. To enable full
    exchangeability of these classes at runtime
    we introduced abstract base
    classes and now store shared pointers to these base
    classes. With this change it is now possible to configure the
    solvers at runtime. Additionally, most solvers can now be
    configured using a \cpp{Dune::ParameterTree} object, which holds
    configuration parameters for the whole program. A convenient
    solver factory is currently under development, which will complete
    these latest changes.
    For example the restarted GMRes solver was
    constructed as
    \begin{c++}
Dune::RestartedGMResSolver<V> solver(op, preconditioner, reduction, restart, maxit, verbose);
\end{c++}
    where reduction, restart, maxit, and verbose are just scalar
    parameters, which the user usually wants to change often to tweak
    the solvers. Now these parameters can be specified in a section of
    an INI-style file like:
    \begin{c++}
[GMRES]
reduction = 1e-8
maxit = 500
restart = 10
verbose = 0
\end{c++}
    This configuration is parsed into a \cpp{ParameterTree} object,
    which is passed to the constructor:
    \begin{c++}
Dune::RestartedGMResSolver<V> solver(op, preconditioner, parametertree);
\end{c++}
  \item From a conceptual point of view \duneistl was designed to
    support vectors and matrices with varying block structure since the
    very first release. In practice, it took a very long time to
    actually fully support such constructs. Only since the new
    language features of C++11 are available it was possible to
    implement the classes
    \cpp{MultiTypeBlockVector} and
    \cpp{MultiTypeBlockMatrix} in a fully featured way.
    These classes implement
    dense block matrices and vectors with different block
    types in different entries.
    The user
    now can easily define matrix structures like
    \begin{c++}
using namespace Dune;
using Row0 = MultiTypeBlockVector<
    Matrix<FieldMatrix<double,3,3>>, 
    Matrix<FieldMatrix<double,3,1>> >;
using Row1 = MultiTypeBlockVector<
    Matrix<FieldMatrix<double,1,3>>, 
    Matrix<double> >;

MultiTypeBlockMatrix<Row0,Row1> A;
    \end{c++}
    Such a matrix type would be natural, e.g., for a Taylor--Hood discretization of
    a three-dimensional Stokes or Navier--Stokes problem, combining a velocity vector field
    with a scalar pressure.

    \item With the \dune 2.6 release an abstraction layer for
      SIMD-vectorized data types
      was introduced. This abstraction layer provides functions for
      transparently handling
      SIMD data types, as provided by libraries, e.g.
      Vc\footnote{note that an interface similar to Vc is
        part of the C++ \emph{Parallelism TS 2} standard}~\cite{kretz2015extending,kretz2012vc} or
      VectorClass~\cite{fog2013c++}, and scalar data types, like
      \cpp{double} or \cpp{std::complex}.
The
layer consists of free-standing functions, for example
\cpp{Simd::lane(int l, VT& v)}, where \cpp{v} is of vector-type
\cpp{VT} and \cpp{Simd::lane}
gives access to the \cpp{l}-th entry of the vector.
Operators like \cpp{+} or \cpp{*} are overloaded and applied
component-wise. The result of boolean expressions are also
vectorized and return
data types with \cpp{bool} as scalar type.
To handle these values \dune
offers functions like \cpp{Simd::cond}, \cpp{Simd::allTrue}, or
\cpp{Simd::anyTrue} for testing them.
The \cpp{Simd::cond} function has the semantics of the ternary operator, which
cannot be overloaded. This operator is necessary, as \cpp{if-else}
expressions might lead to different branches in the different lanes,
which contradicts the principle of vectorization.
\par
Using this abstraction layer it is possible to construct a solver in \duneistl
supporting multiple right hand sides. This is achieved by using the vectorized type as
\cpp{field_type} in the data structures. For example using the type
\cpp{Vec4d}, provided by VectorClass, the Vector type is constructed as:
\begin{c++}
  Dune::BlockVector<Dune::FieldVector<Vec4d, 1>>
\end{c++}
It can be interpreted as a tall-skinny matrix
in $\mathbb{R}^{N\times 4}$. Using
these data types has multiple advantages:
\begin{itemize}
\item \emph{Explicit use of vectorization instructions} - Modern CPUs provide
  SIMD-vectorization instructions, that can perform the same instruction on
  multiple data simultaneously. It is difficult for the compiler to make use of these
  instructions automatically. With the above approach we can make explicit
  use of the vectorization instructions.\\
\item \emph{Better utilization of memory bandwidth} - The application
  of the operator or the preconditioner is in most cases limited by
    the available memory bandwidth. This means the runtime of these operations 
    depends on the amount of data that must be transferred from or to the memory. 
    With our vectorization approach the matrix has to be loaded from memory 
    only once for calculating $k$ matrix--vector products.\\
\item \emph{Reduction of communication overhead} - On distributed
  systems the cost for sending a message is calculated as $\alpha D +
  \beta$, where $D$ is the amount of data, $\alpha$ is the bandwidth,
  and $\beta$ is the latency. When using vectorized solvers, $k$ messages are fused 
    to a single message. Therefore the costs are reduced from $k(\alpha D + \beta)$ to $k \alpha D + \beta$.\\
\item \emph{Block Krylov methods} - Block Krylov methods are Krylov methods for
  systems with multiple right hand sides. In every iteration the energy error is
  minimized in all search directions of all lanes. This improves the number of
  iterations, that are needed to achieve a certain residual reduction.
\end{itemize}
\end{enumerate}

%% file: dunelocalfunc.tex
\subsection{Finite element spaces on discretization grids}

While \dune focuses on grid-based discretization methods for PDEs,
its modular design explicitly avoids any reference to ansatz functions
for discretizations in the interfaces of grids and linear algebra of
the modules discussed so far. Instead of this the corresponding interfaces 
and implementations are contained in separate modules.
However, the \dune infrastructure is not limited to finite element
discretizations and, for example, a number of applications based on the finite volume 
method exist, for example \dumux \cite{dumux3} and the Open Porous Media
Initiative \cite{opm}, or higher order finite volume schemes on 
polyhedral grids \cite{reconpoly:17} as well as a tensor product multigrid approach for
grids with high aspect ratios in atmospheric flows \cite{mueller-tpmg:16}.

\subsubsection{Local functions spaces}
\label{sec:local_fe}
The \dunelocalfunctions core module contains interfaces and implementations
for ansatz functions on local reference domains. In terms of finite element
discretizations, this corresponds to the finite elements defined on reference
geometries. Again following the modular paradigm, this is done independently
of any global structures like grids or linear algebra, such that the
\dunelocalfunctions module does not depend on the \dunegrid and \duneistl
module.
The central concept of the \dunelocalfunctions module is the
\cpp{LocalFiniteElement} which is defined along the lines
of a \emph{finite element} in terms of \cite{Ciarlet2002}.
There, a finite element is a triple $(\mathcal{D}, \Pi, \Sigma)$
of the local domain $\mathcal{D}$, a local function space $\Pi$,
and a finite set of functionals $\Sigma=\{\sigma_1,\dots,\sigma_n\}$
which induces a basis $\lambda_1,\dots,\lambda_n$ on the local ansatz space by
$\sigma_i(\lambda_j) = \delta_{ij}$.

Each \cpp{LocalFiniteElement} provides access to its polyhedral
domain $\mathcal{D}$ by exporting a \cpp{GeometryType}.
The exact geometry of the type is defined in the \dunegeometry module.
The local basis functions $\lambda_i$ and the functionals $\sigma_i$
are provided by each \cpp{LocalFiniteElement} by exporting
a \cpp{LocalBasis} and \cpp{LocalInterpolation} object, respectively.
Finally, a \cpp{LocalFiniteElement} provides a \cpp{LocalCoefficients}
object.
The latter maps each basis function/functional to a triple
$(c,i,j)$ which identifies the basis function as the $j$-th
one tied to the $i$-th codimension-$c$ face of $\mathcal{D}$.

\subsubsection{Global functions spaces}

In contrast to local shape functions provided by \dunelocalfunctions,
a related infrastructure for global function spaces on a grid view%
---denoted \emph{global finite element spaces} in the following---%
is not contained in the \dune core so far.
Instead several concurrent/complementary discretization modules, like
\dunefem, \dunepdelab, \dunefufem, used to provide their own implementations.
To improve interoperability, an interface for global function spaces
was developed as a joint effort in the staging module \dunefunctions.
This intended to be a common foundation for higher level discretization modules.

It can often be useful to use different bases of the same global
finite element space for different applications. For example, a discretization
in the space $P_2$ will often use a classical Lagrange basis of the second
order polynomials on the elements, whereas hierarchical error estimators make
use of the so called hierarchical $P_2$ basis.
As a consequence the \dunefunctions module does not use the global finite element
space itself but its basis as the central abstraction. This is represented
by the concept of a \cpp{GlobalBasis} in the interface.

Inspired by the \dunepdelab module, \dunefunctions provides a flexible
framework for global bases of hierarchically structured
finite element product spaces.
Such spaces arise, e.g. in non-isothermal phase field models,
primal plasticity, mixed formulations of higher order PDEs, multi-physics problems,
and many more applications.
The central feature is a generic mechanism for the construction of bases for
arbitrarily structured product spaces from existing elementary spaces.
Within this construction, the global DOF indices
can easily be customized according to the matrix/vector data structures
suitable for the problem at hand, which may be flat, nested, or even multi-type
containers.

In the following we will illustrate this using the $k$-th order Taylor--Hood
space $P_{k+1}^3 \times P_k$ in $\mathbb{R}^3$ for $k\geq 1$ as example.
Here $P_j$ denotes the space of $j$-th order continuous finite elements.
Notice that the Taylor--Hood space provides a natural hierarchical structure:
It is the product of the space $P_{k+1}^3$ for the
velocity with the space $P_k$ for the pressure, where the former is again the 3-fold
product of the space $P_{k+1}$ for the individual velocity components.
Any such product space can be viewed as a tree of spaces,
where the root denotes the full space (e.g. $P_{k+1}^3 \times P_k$)
inner nodes denote intermediate product spaces (e.g. $P_{k+1}^3$),
and the leaf nodes represent \emph{elementary} spaces that are not
considered as products themselves (e.g. $P_{k+1}$ and $P_k$).


The \dunefunctions module on the one hand defines an interface for
such nested spaces and on the other hand provides implementations
for a set of elementary spaces together with a mechanism for
the convenient construction of product spaces. For example, the
first order Taylor--Hood space on a given grid view can be
constructed using
\begin{c++}
using namespace Dune::Functions;
using namespace Dune::Functions::BasisFactory;
auto basis = makeBasis(gridView,
    composite( power<3>( lagrange<2>()), lagrange<1>()));
\end{c++}
Here, \cpp{lagrange<k>()} creates a descriptor of the
Lagrange basis of $P_k$ which is one of the pre-implemented elementary bases,
\cpp{power<m>(...)} creates a descriptor of an $m$-fold product of a given basis,
and \cpp{composite(...)} creates a descriptor of the product
of an arbitrary family of (possibly different) bases. Finally, \cpp{makeBasis(...)}
creates the global basis of the desired space on the given grid view.
For other applications, \cpp{composite} and \cpp{power} can be nested
in an arbitrary way, and the mechanism can be extended by implementing
further elementary spaces providing a certain implementers interface.

The interface of a \cpp{GlobalBasis} is split into to several parts.
All functionality that is related to the basis as a whole is
directly provided by the basis, whereas all functionality that can be
localized to grid elements is accessible by a so called \cpp{LocalView}
obtained using \cpp{basis.localView()}.
Binding a local view to a grid element using \cpp{localView.bind(element)}
will then initialize (and possibly pre-compute and cache) the local properties.
To avoid costly reallocation of internal data,
one can rebind an existing \cpp{LocalView} to another element.

Once a \cpp{LocalView} is bound, it gives access to
all non-vanishing basis functions on the bound-to element.
Similar to the global basis, the localized basis forms a local tree
which is accessible using \cpp{localView.tree()}.
Its children can either be directly obtained using the \cpp{child(...)} method
or traversed in a generic loop.
Finally, shape functions can be accessed on each local leaf node
in terms of a \cpp{LocalFiniteElement} (cf. Section~\ref{sec:local_fe}).

The mapping of the shape functions to global indices is done in several stages.
First, the shape functions of each leaf node have
unique indices within their \cpp{LocalFiniteElement}.
Next, the per-\cpp{LocalFiniteElement} indices of each leaf node can be
mapped to per-\cpp{LocalView} indices using \cpp{leafNode.localIndex(i)}.
The resulting local indices enumerate all basis functions on the bound-to element uniquely.
Finally, the local per-\cpp{LocalView} indices can be mapped to
globally unique indices using \cpp{localView.index(j)}.
To give a full example, the global index of the $i$-th shape function
for the $d$-th velocity component of the Taylor--Hood basis on a given element
can be obtained using
\begin{c++}
using namespace Dune::Indices;                 // Use compile-time indices
auto localView = basis.localView();            // Create a LocalView
localView.bind(element);                       // Bind to a grid element
auto&& velocityNode = localView.child(_0, d);  // Obtain leaf node of the
                                               // d-th velocity component
auto localIndex = velocityNode.localIndex(i);  // Obtain local index of
                                               // i-th shape function
auto globalIndex = localView.index(localIndex);// Obtain global index
\end{c++}
Here, we made use of the compile time index \cpp{Dune::Indices::\_0}
because direct children in
a \cpp{composite} construction may have different types.

While all local indices are flat and zero-based, global indices
can in general be multi-indices which allows to efficiently access hierarchically structured
containers.
The global multi-indices do in general form an index tree.
The latter can be explored using \cpp{basis.size(prefix)} with a given prefix multi-index.
This provides the size of the next digit following the prefix,
or, equivalently, the number of direct children of the (possibly interior)
node denoted by the prefix. Consistently, \cpp{basis.size()} provides
the number of entries appearing in the first digit.
In case of flat indices, this corresponds to the total number of basis functions.

The way shape functions are associated to indices can be influenced according
to the needs of the used discretization, algebraic data structures, and algebraic solvers.
In principle an elementary basis provides a pre-defined set of global indices.
When defining more complex product space
bases using \cpp{composite} and \cpp{power},
the indices provided by the respective direct children
are combined in a customizable way.
Possible strategies are, for example, to prepend or append the number of the child
to the global index within the child, or to increment the global indices to get consecutive
flat indices.

Additionally to the interfaces and implementations of global finite element
function space bases, the \dunefunctions module provides utility functions
for working with global bases.
The most basic utilities are
\cpp{subspaceBasis(basis, childIndices)},
which constructs a view of only those basis functions corresponding to a certain
child in the ansatz tree,
\cpp{makeDiscreteGlobalBasisFunction<Range>( basis, vector)},
which allows to construct the finite element function (with given range type) obtained by
weighting the basis functions with the coefficients stored in a suitable vector,
and \cpp{interpolate(basis, vector, function)},
which computes the interpolation
of a given function storing the result as coefficient vector with respect to a basis.

The following example interpolates a function into the pressure degrees
of freedom only and later construct the velocity vector field as a function.
The latter can e.g. be used to write a subsampled representation in the VTK format.
\begin{c++}
// Interpolate f into vector x
auto f = [] (auto x) { return sin(x[0]) * sin(x[1]); };
interpolate(subspaceBasis(basis, _1), x, f);
// [ Do something to compute x here ]
using Range = FieldVector<double,dim>;
auto velocityFunction =
  makeDiscreteGlobalBasisFunction<Range>(subspaceBasis(basis, _0), x);
\end{c++}

A detailed description of the \cpp{GlobalBasis} interface,
the available elementary basis implementations,
the mechanism to construct product spaces,
the rule-based combination of global indices,
and the basis-related utilities
can be found in \cite{EngwerEtAl2018}.
The type-erasure based polymorphic interface of
global functions and localizable grid functions
as e.g. implemented by \cpp{makeDiscreteGlobalBasisFunction}
is described in \cite{EngwerEtAl2017}.


%% file: dunepython.tex
\subsection{Python interfaces for \texorpdfstring{\dune}{Dune}}
\label{sec:dunepython}

Combining easy to use scripting languages with state-of-the-art
numerical software has been a continuous effort in scientific computing
for a long time.
For solution of PDEs
the pioneering work of the FEniCS team \cite{Logg:2012:ASD:2331176}
inspired many others, e.g. \cite{petsc4py:11,firedrake} to also provide
Python scripting for high performance PDE solvers usually coded in C++.

Starting with the 2.6 release in 2018, \dune can also be used
from within the Python scripting environment.
The \dunepython staging module provides
\begin{enumerate*}
\item a general concept for exporting realizations of polymorphic interfaces
  as used in many \dune{} modules and
\item Python bindings for the central
  interfaces of the \dune{} core modules described in this section.
\end{enumerate*}
These bindings make rapid prototyping of new numerical algorithms easy
since they can be implemented and tested within a scripting environment.
Our aim was to keep the Python interfaces as close as possible to their C++
counterparts so that translating the resulting Python algorithms to C++ to
maximize efficiency of production code is as painless as possible. Bindings
are provided using \cite{pybind11}.

We start with an example demonstrating these concepts.
We revisit the examples given in Section~\ref{sec:basicgridconcepts}
starting with the construction of a simple Cartesian grid
in four space dimensions and the approximation of the integral
of a function over that grid. 
The corresponding Python code is
\begin{python}
from dune.grid import yaspGrid, cartesianDomain
from math import exp
dim     = 4
lower   = 4*[0.]
upper   = 4*[1.]
nCells = 4*[4]
gv = yaspGrid(constructor=cartesianDomain(lower,upper,nCells))
u = lambda x: exp(x.two_norm)
integral = 0.0
for e in gv.elements:
   integral += u(e.geometry.center)*e.geometry.volume
\end{python}
A few changes were made to make the resulting code more Pythonic, i.e., the
use of class attributes instead of class methods, but the only major change
is that the function returning the grid object in fact returns the leaf
grid view and not the hierarchic grid structure.
Notice that the life time of this underlying grid is managed automatically by Python's internal
reference counting mechanism. It can be
obtained using a class attribute, i.e., to refine the grid globally
\begin{python}
  gv.hierarchicalGrid.globalRefine(1);
\end{python}
Other interface classes and their realizations have also been exported so
that for example the more advanced quadrature rules used in the previous
sections can also be used in Python:
\begin{python}
from dune.geometry import quadratureRule
integral = 0.0
for e in gv.elements:
   geo = e.geometry
   quadrature = quadratureRule(e.type,5)
   for qp in quadrature:
      integral += u( geo.toGlobal(qp.position) )\
          *geo.integrationElement(qp.position)*qp.weight
\end{python}
Again, the changes to the C++ code is mostly cosmetics or due to the
restrictions imposed by the Python language.

While staying close to the original C++ interface facilitates rapid prototyping,
it also can lead to a significant loss of efficiency. A very high level of efficiency
was never a central target during the design of the Python bindings
to \dune{}---to achieve this, a straightforward mechanism is provided to call
\dune{} algorithms written in C++. Nevertheless,
we made some changes to the interface and added a few extra features to
improve the overall efficiency of the code. The two main strategies are to
reduce the number of calls from Python to C++ by, for example, not returning
single objects for a given index but iterable structures instead.
The second strategy is to introduce an extended interface taking a vector of its
arguments to allow for vectorization.

Consider, for example the C++ interface methods on the
\pyth{Geometry} class
\pyth{geometry.corners()} and \pyth{geometry.corner(i)} which return the
number of corners of the elements and their coordinates in physical space,
respectively.
Using these methods, loops would read as follows:
\begin{c++}
auto center = geometry.corner(0);
for (std::size_t i=1;i<geometry.corners();++i)
    center += geometry.corner(i);
center /= geometry.corners();
\end{c++}
To reduce the number of calls into C++,
we decided to slightly change the semantics of method
pairs of this type:
the plural version now returns an iterable object, while
the singular version still exists in its original form.
So in Python the above code snippet can be written as follows:
\begin{python}
corners = geometry.corners
center = corners[0]
for c in corners[1:]:
    center += c
center /= len(corners)
\end{python}
As discussed above, quadrature loops are an important ingredient of most
grid based numerical schemes. As the code snippet given at the beginning of
this section shows, this requires calling methods on the geometry for each
point of the quadrature rule which again can lead to a significant
performance penalty. To overcome this issue we provide vectorized versions
of the methods on the geometry class so that the above example can be more
efficiently implemented
\begin{python}
import numpy
from dune.geometry import quadratureRule
u = lambda x: numpy.exp( numpy.sqrt( sum(x*x) ) )
integral = 0.0
for e in gv.elements:
    hatxs, hatws = quadratureRule(e.type, 5).get()
    weights = hatws * e.geometry.integrationElement(hatxs)
    integral += numpy.sum(u(hatxs) * weights, axis=-1)
\end{python}

The following list gives a short overview of changes and extensions we made
to the \dune\ interface while exporting it to Python:

\begin{itemize}
\item
  Since \pyth{global} is a keyword in Python we cannot export the
  \cpp{global} method on the \cpp{Geometry} directly. So we have
  exported \pyth{global} as \pyth{toGlobal} and for symmetry reasons 
  \pyth{local} as \pyth{toLocal}.
\item
  Some methods take compile-time static arguments, e.g., the codimension
  argument for
  \cpp{entity.subEntity< c >( i )}.
  These had to be turned into dynamic arguments, so in Python the subEntity
  is obtained via \pyth{entity.subEntity(i, c)}.
\item
  In many places we replaced methods with properties, i.e.,
  \pyth{entity.geometry} instead of \cpp{entity.geometry()}.
\item
  Methods returning a \pyth{bool} specifying that other interface methods
  will return valid results are not exported (e.g. \pyth{neighbor}
  on the intersection class). Instead \pyth{None} is returned to specify
  a non valid call (e.g. to \pyth{outside}).
\item
  Some of the C++ interfaces contain pairs of methods where the
  method with the \emph{plural name} returns an integer (the \emph{number of})
  and the singular version takes an integer and returns the \emph{ith}
  element.
  The plural version was turned to a range-returning method in Python as discussed above.
\item
  In C++, free-standing functions can be found via argument-dependent lookup.
  As Python does not have such a concept, we converted those free-standing
  functions to methods or properties.
  Examples are \cpp{elements}, \cpp{entities}, \cpp{intersections}, or
  \cpp{localFunction}.
\item
  A \emph{grid} in \dunepython is always the \cpp{LeafGridView} of the
  hierarchical grid. To work with the actual hierarchy, i.e., to refine the
  grid, use the \pyth{hierarchicalGrid} property. Level grid views can
  also be obtained from that hierarchical grid.
\item
  In contrast to C++, partitions are exported as objects of their own.
  The interior partition, for example, can be accessed by
\begin{python}
partition = grid.interiorPartition
\end{python}
  The partition, in turn, also exports the method \pyth{entities} and the properties
  \pyth{elements}, \pyth{facets}, \pyth{edges}, and \pyth{vertices}.
\item
  An \pyth{MCMGMapper} can be constructed using the \pyth{mapper}
  method on the \pyth{GridView} class passing in the \pyth{Layout}
  as argument.
  The mapper class has an additional call method taking an entity,
    which returns an array with the indices of all DoFs (degrees of freedom) attached to that
  entity.
  A list of DoF vectors based on the same mapper can be communicated using
  methods defined on the mapper itself and without having to define a
  \cpp{DataHandle}.
\end{itemize}

A big advantage of using the Python bindings for \dune{} is that non
performance critical tasks, e.g., pre- and postprocessing can be carried
out within Python while the time critical code parts can be easily carried
out in C++. To make it easy to call functions written in C++ from within a
Python script, \dunepython provides a simple mechanism. Let us assume for
example that the above quadrature for $e^{|x|}$ was implemented in a C++
function \pyth{integral} contained in the header file
\pyth{integral.hh} using the \dune{} interface as described in
Section~\ref{sec:basicgridconcepts}:
\begin{c++}
template <class GridView>
double integral(const GridView &gv) {
  auto u = [](const auto& x){return std::exp(x.two_norm());};
  double integral=0.0;
  for (const auto& e : elements(gv))
     integral += u(e.geometry().center())*e.geometry().volume();
  return integral;
}
\end{c++}
We can now call this function from within Python using
\begin{python}
integral = algorithm.run('integral', 'integral.hh', gv)
\end{python}
Note that the correct version of the template function
\pyth{integral} will be exported using the C++ type of the
\pyth{gv} argument, i.e., \pyth{YaspGrid<4>}.

With the mechanism provided in the \dunepython module,
numerical schemes can first be implemented and tested within
Python and can then be translated to C++ to achieve a high level of
efficiency. The resulting C++ functions can be easily called from
within Python making it straightforward to simply replace parts of the
Python code with their C++ counterparts.

In addition to the features described so far,
the \dunepython module provides general infrastructure for
adding bindings to other \dune{} modules.
Details are given in \cite{dune-python-paper}.
We will demonstrate the power
of this feature in Section~\ref{sec:gridadaptation} where we also use the
domain specific language UFL \cite{ufl:14} to describe PDE models.


%% file: infrastructure.tex
\subsection{Build system and testing}
Starting with the 2.4 release, \dune{} has transitioned from its Autotools build system to a new, CMake-based build system.
This follows the general trend in the open source software community to use CMake.
The framework is split into separate modules; each module is treated as a CMake project in itself, with the build system
managing inter-module dependencies and propagation of configuration results. In order to simplify the inter-module
management, there is a shell script called \pyth{dunecontrol} (part of \dunecommon) that resolves dependencies and controls the build order.

In the CMake implementation of the \dune{} build system, special emphasis has been put on testing.
Testing has become increasingly important with the development model of the \dune{} core modules being heavily based on Continuous Integration.
In order to lower the entrance barrier for adding tests to a minimum, a one-line solution in the form of a CMake convenience function \pyth{dune_add_test} has been implemented.
Further testing infrastructure has been provided in the module \dunetesttools{} \cite{kempf2017system}, which allows the definition of system tests.
These system tests describe samples of framework variability covering both compile-time and run-time variations.

More information on the \dune{} CMake build system can be found in the
Sphinx-generated documentation, which is available on the \dune{} website\footnote{\url{https://dune-project.org/buildsystem/}}.


%% file: adaptation.tex
\subsection{Grid modification}
\label{sec:gridadaptation}

In this section we discuss two mechanisms of modifying grids within the DUNE
framework: \textit{dynamic local grid adaptation} and \textit{moving domains}. 
In particular, dynamic local grid adaptation is of interest
for many scientific and engineering applications due to the potential high
computational cost savings.
However, especially parallel dynamic local grid
adaptation is technically challenging and not many PDE frameworks offer a
seamless approach. We will demonstrate in this section how the general
concepts described in Section~\ref{sec:DuneGrid}
for grid views and adaptivity provided by the core modules are used to
solve PDE problems on grids with dynamic refinement and coarsening.
Especially important for these concepts is the separation of topology,
geometry, and user data provided by the grid interface.

To support grid modification the \dunefem{} module provides
two specialized \cpp{GridViews}:
\cpp{AdaptiveGridView} and \cpp{GeometryGridView}. Both are based on a
given grid view, i.e., the \cpp{LeafGridView}, and replace certain aspects
of the implementation. In the first case, the index set is replaced by an
implementation that provides additional information that can be used to
simplify data transfer during grid refinement and coarsening. In the second
case the geometry of each element is replaced by a given grid function,
e.g., by an analytic function or by some discrete function over the
underlying grid view. The advantage of this \emph{meta grid view}
approach is that any algorithm based on a \dune{} grid view can be
used without change while for example the data
transfer during grid modification can be transparently handled by 
specialized algorithms using features of the \emph{meta} grid view.

\subsubsection{Dynamic local grid adaptation}

A vast number of structured or Cartesian grid managers are available
which support adaptive
refinement\footnote{See \url{http://math.boisestate.edu/~calhoun/www_personal/research/amr_software/}.}.
There exist far fewer open source unstructured grid managers, supporting
adaptivity, for example,
deal.II~\citep{dealII91} which is build on top of p4est~\citep{burstedde:11} for
parallel computations, or another recent development the FEMPAR~\citep{fempar:18} package.
Both provide hexahedral grids with non-conforming refinement.
Other very capable unstructured grid managers providing tetrahedral elements are, for
example, AMDIS~\citep{amdis:07}, FEniCS \cite{Logg:2012:ASD:2331176}, HiFlow \cite{emclpp42879}, or
the "Flexible Distributed Mesh Database (FMDB)"~\citep{fmdb:12}, \texttt{libMesh}~\citep{libmesh:06}, and others.

As previously described in Section \ref{sec:dunegridrefinement} the \dune{} grid interface
offers the possibility to dynamically refine and coarsen grids if the underlying
grid implementation offers these capabilities. 
Currently, there are two implementations that support parallel dynamic grid adaptation
including load balancing, \cpp{UGGrid} and \cpp{ALUGrid}. \cpp{AlbertaGrid}
supports grid adaptation but cannot be used for parallel computations.

A variety of applications make use of the \dune{} grid interface for
adaptive computations. For example, adaptive discontinuous Galerkin computations of compressible flow, e.g. 
Euler equations \cite{limiter:11} or atmospheric flow \cite{cosmodg:14}. 
A number of applications focus on hp-adaptive schemes, e.g. 
for continuous Galerkin approximations of Poisson 
type problems \cite{dglagrange:14}, or 
discontinuous Galerkin approximations of two-phase flow 
in porous media \cite{twophase:18,fvca:17,kane:18,fuel3d:08} or conservation laws \cite{gersbacher:17}. 
Other works consider, for example, the adaptive solution of the
Cahn--Larch{\'e} system using finite elements \cite{graeser:14}.

In this section we demonstrate the capabilities of the \dune{} grid interface
and its realizations making use of the Python bindings for the Dune
module \dunefem. We show only small parts of the
Python code here, the full scripts are part of the tutorial \cite{dunefem:tutorial}.

To this end we solve the Barkley model, which is a system of reaction-diffusion equations modeling 
excitable media and oscillatory media.
The model is often used as a 
qualitative model in pattern forming systems like the 
Belousov--Zhabotinsky reaction and other systems that are well 
described by the interaction of an activator and an inhibitor component
\cite{Barkley:91}.

In its simplest form the Barkley model is given by
\begin{gather*}
  \frac{\partial u}{\partial t}
       = \frac{1}{\varepsilon}f(u,v) + D\Delta u, \qquad  \frac{\partial v}{\partial t} = h(u,v),
\end{gather*}
with 
  $f(u,v)=u\Big(1-u\Big)\Big(u-\frac{v+b}{a}\Big)$
and
  $h(u,v) = u - v$.
Finally, $\varepsilon = 0.02$, $a = 0.75$, $b = 0.02$, and $D = 0.01$ are chosen according to the 
web page \url{http://www.scholarpedia.org/article/Barkley_model} and
\cite{Barkley:91}. To evolve the equations in time, we employ the
carefully constructed linear time stepping scheme for this model described
in the literature:
let $u^n,v^n$ be approximations of the solution at a time $t^n$.
To compute approximations $u^{n+1},v^{n+1}$ at a later time
$t^{n+1}=t^n+\Delta t$ we replace the nonlinear function $f(u,v)$ by
$-m(u^n,v^n)u^{n+1} + f_E(u^n,v^n)$ where using
$U^*(V)\colonequals\frac{V+b}{a}$ 
\begin{align*}
  m(U,V) &\colonequals \begin{cases}
    (U-1)\;(\;U-U^*(V)\;) & U < U^*(V) \\
    U\;(\;U-U^*(V)\;)    & U \geq U^*(V),
  \end{cases} \\
    f_E(U,V) &\colonequals \begin{cases}
    0 & U < U^*(V) \\
    U\;(\;U-U^*(V)\;)    & U \geq U^*(V).
  \end{cases} 
\end{align*}
Note that $u,v$ are assumed to be between zero and one so
$m(u^n,v^n) > 0$.
We end up with a linear, positive definite elliptic operator
defining the solution $u^{n+1}$ given $u^n,v^n$.
In the following we will use a conforming Lagrange approximation with
quadratic basis functions. To handle possible nonconforming grids we add
interior penalty DG terms as discussed in \cite{dglagrange:14}.
The slow reaction $h(u,v)$ can be solved explicitly leading to a purely algebraic
equation for $v^{n+1}$.
The initial data is piecewise constant chosen in such a way that a spiral
wave develops.

The model and initial conditions are easily provided using the Unified Form Language
(UFL) \cite{ufl:14}. First the problem data needs to be provided
\begin{python}
dt,t = 0.1,0
spiral_a,spiral_b,spiral_eps,spiral_D = 0.75, 0.02, 0.02, 0.01
def spiral_h(u,v): return u - v
\end{python}
and the discrete space and functions constructed
\begin{python}
space = lagrange( gridView, order=2 )
x = ufl.SpatialCoordinate(space)
iu = lambda s: ufl.conditional(s > 1.25, 1, 0 )
top = ufl.conditional( x[2] > 1.25,1,0)
initial_u = iu(x[1])*top + iu(2.5-x[1])*(1.0 - top)
initial_v = ufl.conditional(x[0]<1.25,0.5,0)
uh   = space.interpolate( initial_u, name="u" )
vh   = space.interpolate( initial_v, name="v" )
uh_n, vh_n = uh.copy(), vh.copy()
\end{python}
Now we use UFL to describe the PDE for the function $u$
adding DG skeleton terms to take care of possible conforming intersections
caused by local grid modification \cite{dglagrange:14}:
\begin{python}
u, phi = ufl.TrialFunction(space), ufl.TestFunction(space)
hT  = ufl.MaxCellEdgeLength(space.cell())
hS  = ufl.avg( ufl.MaxFacetEdgeLength(space.cell()) )
hs =  ufl.MaxFacetEdgeLength(space.cell())('+')
n   = ufl.FacetNormal(space.cell())
penalty = 5 * (order * (order+1)) * spiral_D
ustar   = lambda v: (v+spiral_b)/spiral_a
source  = lambda u1,u2,u3,v: -1/spiral_eps * u1*(1-u2)*(u3-ustar(v))
# main terms
xForm  = inner(D_spiral*grad(u), grad(phi)) * dx
xForm += ufl.conditional(uh_n<ustar(vh_n),
           source(u,uh_n,uh_n,vh_n), source(uh_n,u,uh_n,vh_n)) * phi * dx
# dg terms
xForm -= ( inner( outer(jump(u), n('+')), avg(spiral_D*grad(phi))) +\
           inner( avg(spiral_D*grad(u)), outer(jump(phi), n('+'))) ) * dS
xForm += penalty/hS * inner(jump(u), jump(phi)) * dS
# adding time discretization
form   = ( inner(u,phi) - inner(uh_n, phi) ) * dx + dt*xForm
\end{python}


For adaptation we use a residual based error estimator derived in
\cite{dglagrange:14} for a Discontinuous Galerkin (DG) approximation for the Poisson
problem. The error estimator for an element E at a given time step is given by
$\int_E \eta_E(u^{n+1}) + \frac{1}{2}\int_{\partial E}\eta_{\partial E}(u^{n+1})$
with 
\begin{align*}
  \eta_E(u_h) &= h_E^2 \Big(\frac{u_h-u^n}{\Delta t}
                 - f(u_h,v^{n+1}) + \nabla \cdot D\nabla u_h\Big)^2~, \\
  \eta_{\partial E}(u_h) &= 
              h_E \jump{D\nabla u_h\cdot \nu}^2 + \jump{u_h}^2
\end{align*}
where $\jump{\cdot}$ is the jump of the given quantity over the boundary of
$E$ and $\nu$ denotes the outward unit normal.
To describe the estimator using UFL we rewrite it in the form
\begin{equation*}
  R(u_h,\varphi_h) \colonequals \int_\Omega \eta_h(u_h)\varphi_h +
                      \int_{\partial\Omega} \eta_{\partial\Omega}(u_h)\{\varphi_h\}
\end{equation*}
such that computing
$R(u^{n+1},\chi_E) = \int_E \eta_E(u^{n+1}) +
            \frac{1}{2}\int_{\partial E} \eta_{\partial E}(u^{n+1})$
provides the estimator on each element where $\chi_E$ is the characteristic
function on $E$. The characteristic functions are the basis of the
finite-volume space provided by \dunefem so that $R(\cdot,\cdot)$ can be
defined using UFL as bilinear form over the solution space of $u^{n+1}$ and
the scalar finite volume space:
\begin{python}
fvspace = dune.fem.space.finiteVolume(uh.space.grid)
estimate = fvspace.interpolate([0], name="estimate")
chi = ufl.TestFunction(fvspace)
residual = (u-uh_n)/dt - div(spiral_D*grad(u)) + source(u,u,u,vh)
estimator_ufl = hT**2 * residual**2 * chi * dx +\
       hS * inner( jump(spiral_D*grad(u)), n('+'))**2 * avg(chi) * dS +\
     1/hS * jump(u)**2 * avg(chi) * dS
estimator = dune.fem.operator.galerkin(estimator_ufl)
\end{python}
Now the grid can be modified according to the estimator within the time
loop by
\begin{enumerate*}
\item applying the operator constructed above to the discrete
solution $u^{n+1}$
\item marking all elements where the error indicator exceeds
a given tolerance for refinement and marking elements for coarsening with
an indicator below a given threshold and finally
\item modifying the grid
prolonging/restricting the data for $u_h,v_h$ to the new element:
\newline
\begin{python}
estimator(uh, estimate)
dune.fem.mark(estimate, maxTol,0.1*maxTol, 0,maxLevel)
dune.fem.adapt([uh,vh])
\end{python}
\newline
where \pyth{maxTol} is some prescribed tolerance (in the following set to
$10^{-4}$).
\end{enumerate*}
Note that only the data for $u_h,v_h$ is retained during the adaptation
process, the underlying storage for other discrete functions used in the
simulation is resized as required but the values of the functions are not
maintained.

\begin{figure}[!ht]
  \begin{subfigure}[b]{0.19\linewidth}
    \centering{
      \includegraphics[width=0.9\linewidth]{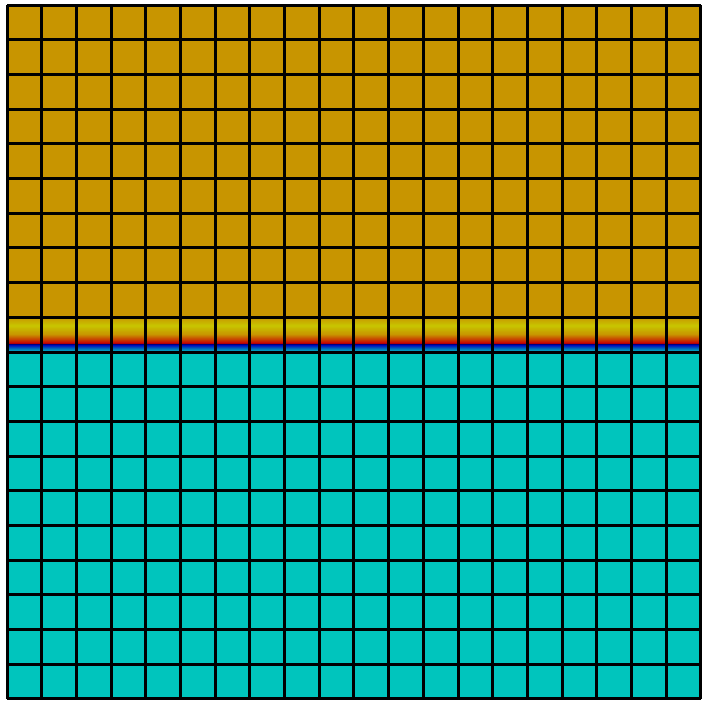}
    }
    \caption{$t=0$}
  \end{subfigure} 
  \begin{subfigure}[b]{0.19\linewidth}
    \centering{
      \includegraphics[width=0.9\linewidth]{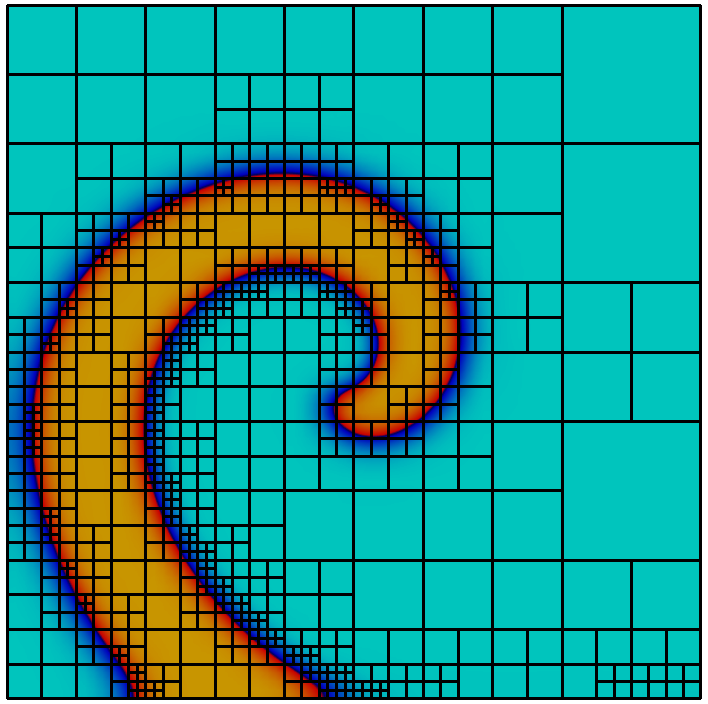}
    }
    \caption{$t=6$}
  \end{subfigure}
  \begin{subfigure}[b]{0.19\linewidth}
    \centering{
      \includegraphics[width=0.9\linewidth]{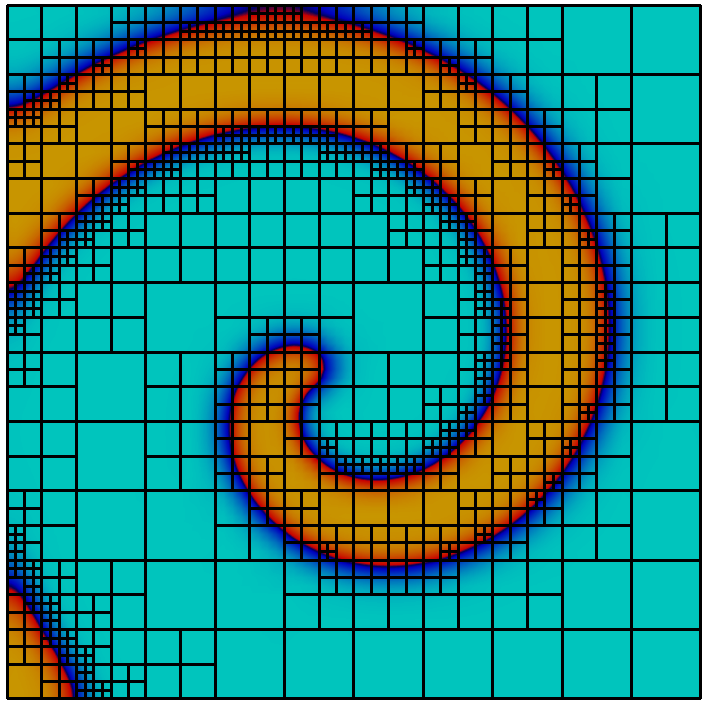}
    }
    \caption{$t=8$}
  \end{subfigure}
  \begin{subfigure}[b]{0.19\linewidth}
    \centering{
      \includegraphics[width=0.9\linewidth]{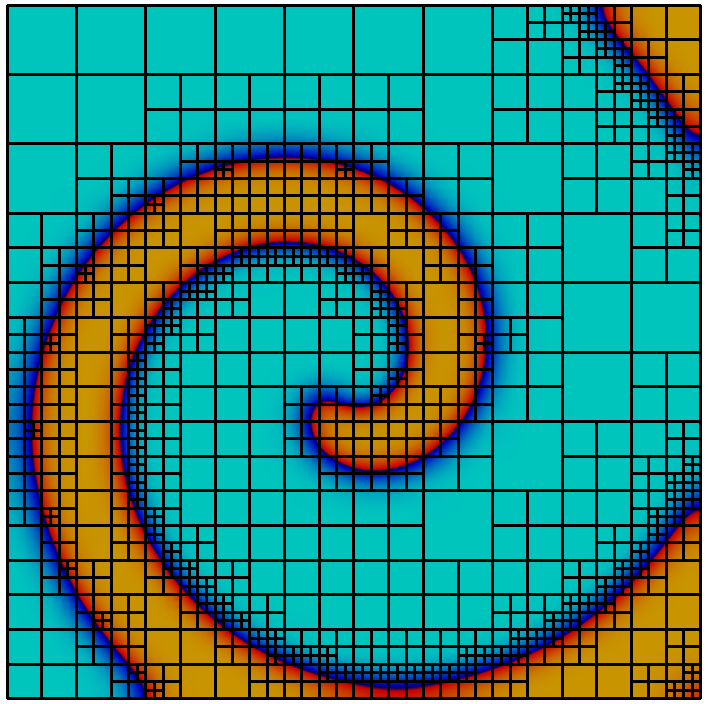}
    }
    \caption{$t=11$}
  \end{subfigure}
  \begin{subfigure}[b]{0.19\linewidth}
    \centering{
      \includegraphics[width=0.9\linewidth]{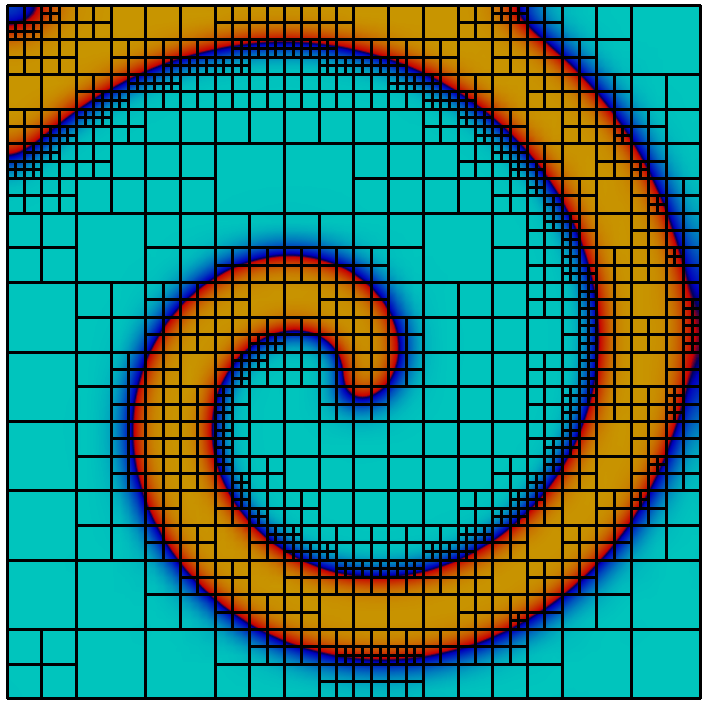}
    }
    \caption{$t=14$}
  \end{subfigure}
  \caption{The evolution of $u$ for different times using non-conforming 
  grid adaptation in 2d with quadrilaterals.}
  \label{fig:spiral-nonconf}
\end{figure}

Figure \ref{fig:spiral-nonconf} shows results 
for different times using a 2d quadrilateral grid with conforming, 
quadratic Lagrange basis functions.
When using a conforming discrete space, the additional 
terms in the DG formulation vanish whenever basis functions are smooth across
element intersection while these terms lead to a stabilization for nonconforming
refinement/coarsening with hanging nodes. On a conforming mesh without hanging
nodes the residual error estimator coincides with standard results known from the
literature \cite[and references therein]{alberta:05}. 

\begin{figure}[!ht]
  \begin{subfigure}[b]{0.19\linewidth}
    \centering{
      \includegraphics[width=0.9\linewidth]{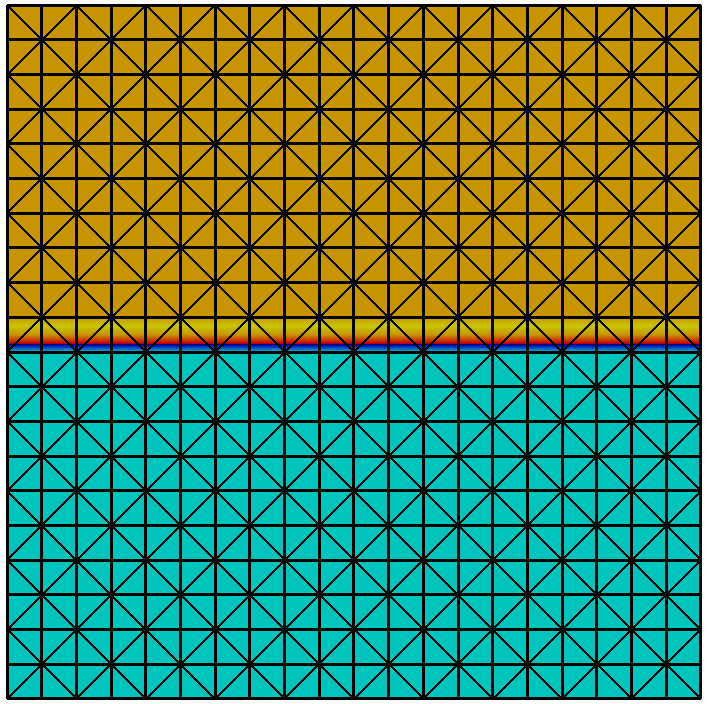}
    }
    \caption{$t=0$}
  \end{subfigure} 
  \begin{subfigure}[b]{0.19\linewidth}
    \centering{
      \includegraphics[width=0.9\linewidth]{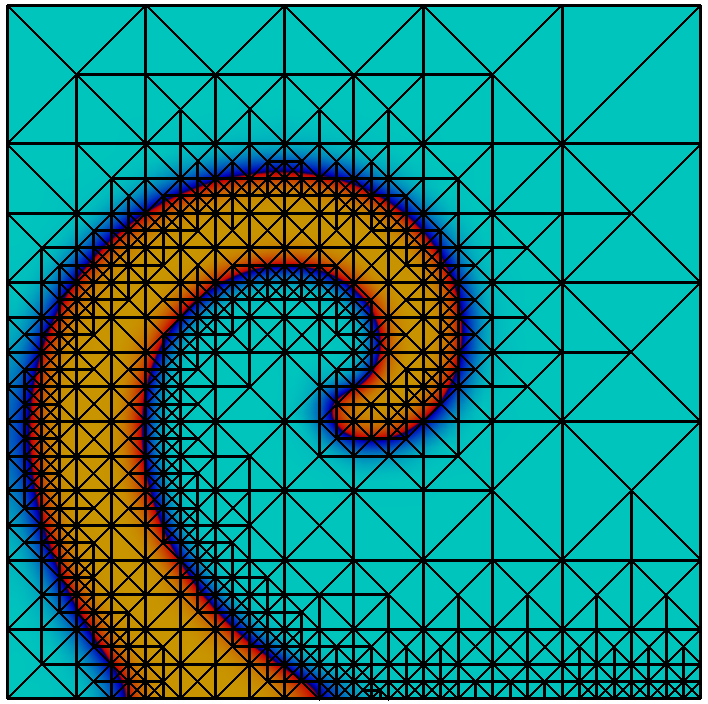}
    }
    \caption{$t=6$}
  \end{subfigure}
  \begin{subfigure}[b]{0.19\linewidth}
    \centering{
      \includegraphics[width=0.9\linewidth]{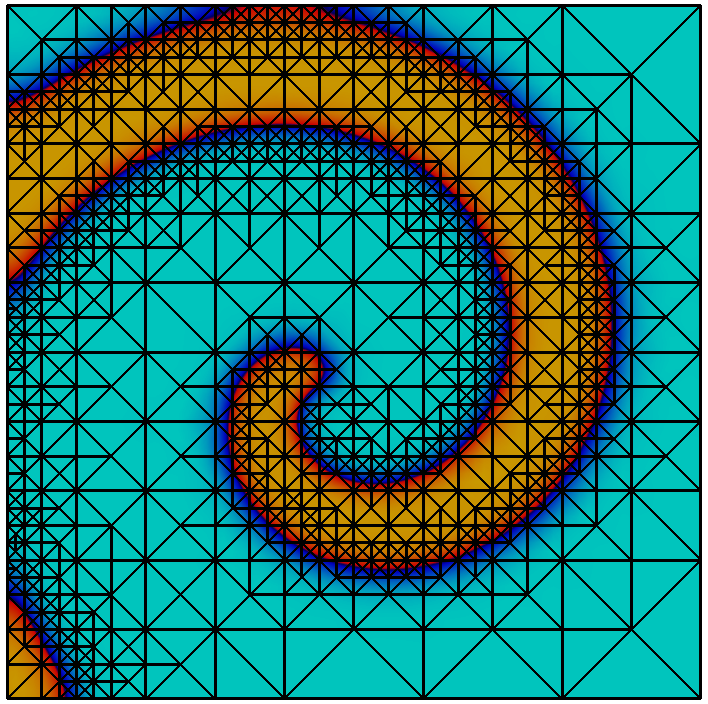}
    }
    \caption{$t=8$}
  \end{subfigure}
  \begin{subfigure}[b]{0.19\linewidth}
    \centering{
      \includegraphics[width=0.9\linewidth]{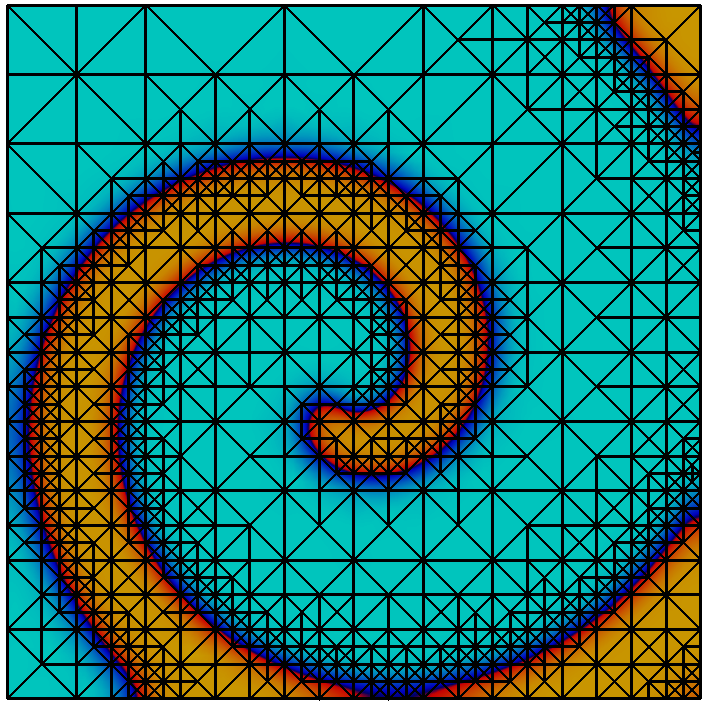}
    }
    \caption{$t=11$}
  \end{subfigure}
  \begin{subfigure}[b]{0.19\linewidth}
    \centering{
      \includegraphics[width=0.9\linewidth]{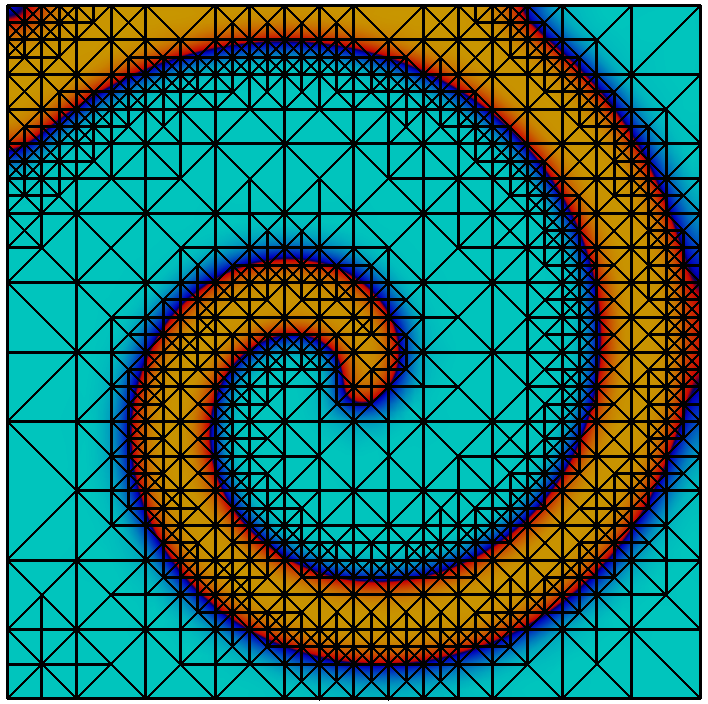}
    }
    \caption{$t=14$}
  \end{subfigure}
  \caption{The evolution of $u$ for different times using conforming bisection
  grid adaptation in 2d.}
  \label{fig:spiral-conf}
\end{figure}

Figure \ref{fig:spiral-conf} shows results using the quadratic Lagrange basis and a 
conforming simplicial grid with bisection refinement. 

\begin{figure}[!ht]
  \begin{subfigure}[b]{0.49\linewidth}
    \centering{
      \includegraphics[width=0.7\linewidth]{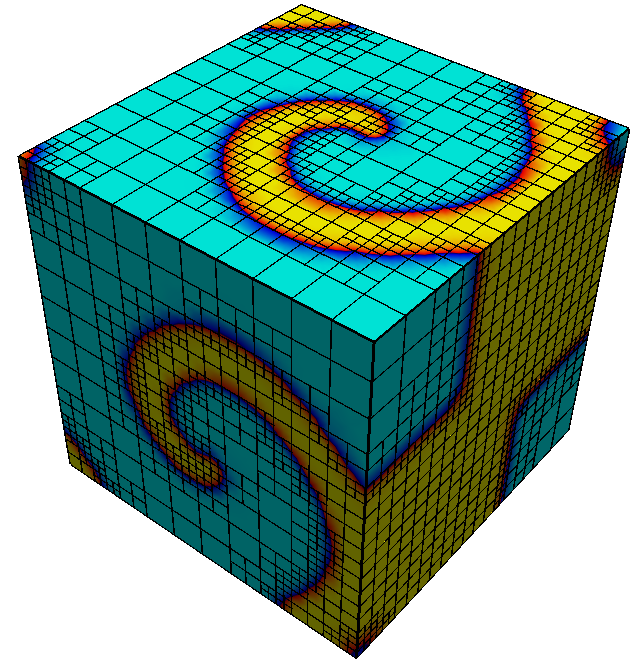}
    }
    \caption{\cpp{ALUGrid}(cube, nonconforming, $Q_1$)}
    \label{fig:spiral-3d-a}
  \end{subfigure}
  \begin{subfigure}[b]{0.49\linewidth}
    \centering{
      \includegraphics[width=0.7\linewidth]{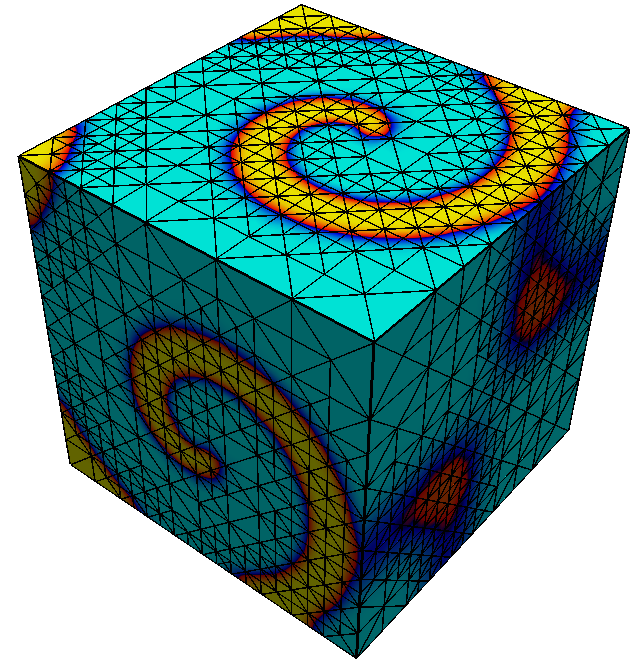}
    }
    \caption{\cpp{ALUGrid}(simplex, conforming, $P_2$)}
    \label{fig:spiral-3d-b}
  \end{subfigure} 
  \caption{The solution $u$ at $t=11$ for (a) a non-conforming cube grid in 3d and (b) conforming bisection
  grid adaptation in 3d. The visual differences between the two solutions
  especially on the right face is caused by the spiral having rotated a
  fraction more on the cube grid compared to the spiral on the simplex
  grid. This is also noticeable on the top left corner. Looking at the
  front left face it is clear that the difference in angle of the spiral is
  quite small.}
  \label{fig:spiral-3d}
\end{figure}

Figure \ref{fig:spiral-3d} shows the same example for 3d grids, 
using a bi-linear Lagrange basis for a non-conforming hexahedral grid in Figure \ref{fig:spiral-3d-a} and 
using a quadratic Lagrange basis on a conforming simplicial grid with bisection refinement in Figure
\ref{fig:spiral-3d-b}.

Details on the available load balancing algorithms and parallel performance studies 
for the \dunealugrid package can be found in \cite{alugrid:16}, \cite{weakcomp:18}, and \cite{dnvb:17}, 
and for UGGrid in \cite{Bastian1997}.

\subsubsection{Moving grids}
\label{sec:moving_grids}

In this section we touch on another important topic 
for modern scientific computing: moving domains. 
Typically this is supported by moving nodes in the computational grid. 
In \dune{} this can be done in a very elegant way. 
The presence of an abstract grid interface allows the construction of meta grids
where only parts of the grid implementation are re-implemented and, in addition,
the original grid implementation stays untouched. 
Thus meta grids provide a very sophisticated way of adding 
features to the complete feature stack and keeping the code base modular. 
In \dunegrid  one can use the meta grid \cpp{GeometryGrid} 
(see also Section \ref{sec:gridimplemenations}) which allows to move nodes of the grid by providing an
overloaded \cpp{Geometry} implementation. 
Another, slightly easier way, is to only overload geometries of 
grid views which is, for example, done in \dunefem.

Both approaches re-implement the reference geometry mapping. In
\cpp{GeometryGrid} an external vector of nodes providing the positions of the computational grid 
is used while for \cpp{GeometryGridView} a grid function, i.e., a function
which is evaluated on each entity given reference coordinates, is used to provide
a mapping for the coordinates. 
The advantage of both approaches is, that the implementation of the numerical
algorithm does not need to change at all. The new grid or grid view follows the
same interface as the original implementation. A moving grid can now be
realized by modifying this grid function.

To demonstrate this feature of \dune{} 
we solve a mean curvature flow problem which 
is a specific example of a geometric evolution
equation where the evolution is governed by the mean curvature $H$.
One real-life example of this is in how soap films change over time,
although it can also be applied to other problems such as image processing.
Assume we are given a reference surface $\bar\Gamma$ such that
we can write the evolving surface in the form $\Gamma_t = X(t,\bar\Gamma)$.
It is now possible to show that the vector valued function $X=X(t,\bar{x})$
with $\bar{x}\in\bar\Gamma$ satisfies
\begin{gather*}
  \frac{\partial}{\partial t}X = - H(X)\nu(X),
\end{gather*}
where $H$ is the mean curvature of $\Gamma_t$ and $\nu$ is its outward pointing normal.

We use the following time discrete approximation as suggest in \cite{MCFDDE:05}
\begin{gather*}
  \int_{\Gamma^n} \big( U^{n+1} - x\big) \cdot \varphi\;d\sigma+
    \Delta t \int_{\Gamma^n} \nabla_{\Gamma^n} U^{n+1}
    \colon\nabla_{\Gamma^n}\varphi\;d\sigma
  =0.
\end{gather*}
Here $U^n$ parametrizes $\Gamma^{n+1}\approx \Gamma_{t^{n+1}}$ over
$\Gamma^n\colonequals\Gamma_{t^{n}}$ and $\Delta t$ is the time step.

In the example used here, the work flow can be set up as follows. 
First one creates a reference grid and a corresponding quadratic Lagrange finite element
space to represent the geometry of the mapped grid. 
\begin{python}
refView = leafGridView("sphere.dgf", dimgrid=2, dimworld=3)
refView.hierarchicalGrid.globalRefine( 2 )
space = solutionSpace(refView, dimRange=refView.dimWorld, order=2)
\end{python}
Then, a deformation function is projected onto this Lagrange space 
\begin{python}
x = ufl.SpatialCoordinate(space)
positions = space.interpolate(
      x*(1 + 0.5*sin(2*pi*x[0]*x[1])*cos(pi*x[2])), name="position")
\end{python}
Using this grid function, a \cpp{GeometryGridView} can be
created that uses these new coordinates to represent the grid geometries. 
This grid view is then used to create the solution space.
\begin{python}
gridView  = geometryGridView(positions)
space = lagrange(gridView, dimRange=gridView.dimWorld, order=2)
\end{python}

In each step of the time loop the coordinate positions can be updated, for
example, by assigning the values from the computed solution of the mean curvature
flow. 
\begin{python}
positions.dofVector.assign(uh.dofVector)
\end{python}

\begin{figure}[!ht]
  \begin{subfigure}[b]{0.19\linewidth}
    \centering{
      \includegraphics[width=0.8\linewidth]{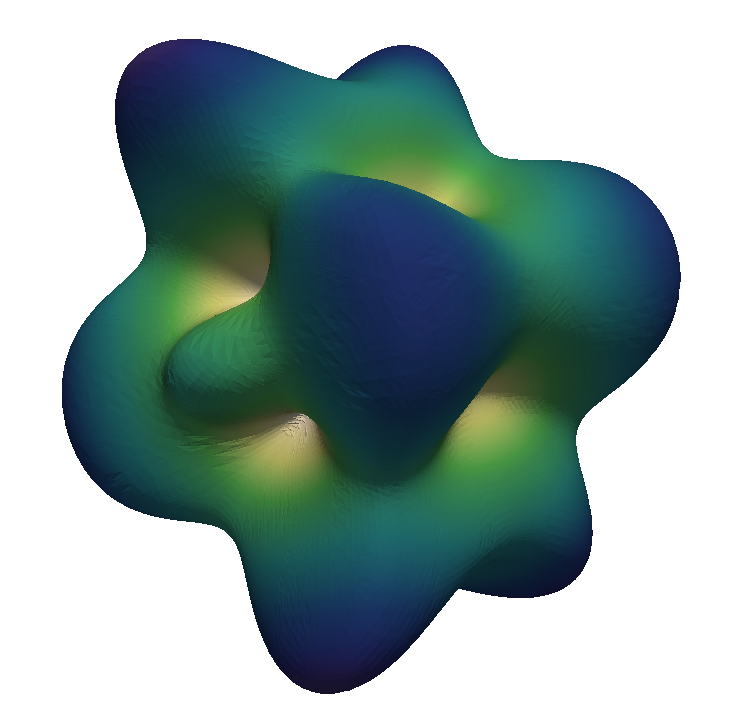}
    }
    \caption{t=0}
  \end{subfigure}
  \begin{subfigure}[b]{0.19\linewidth}
    \centering{
      \includegraphics[width=0.8\linewidth]{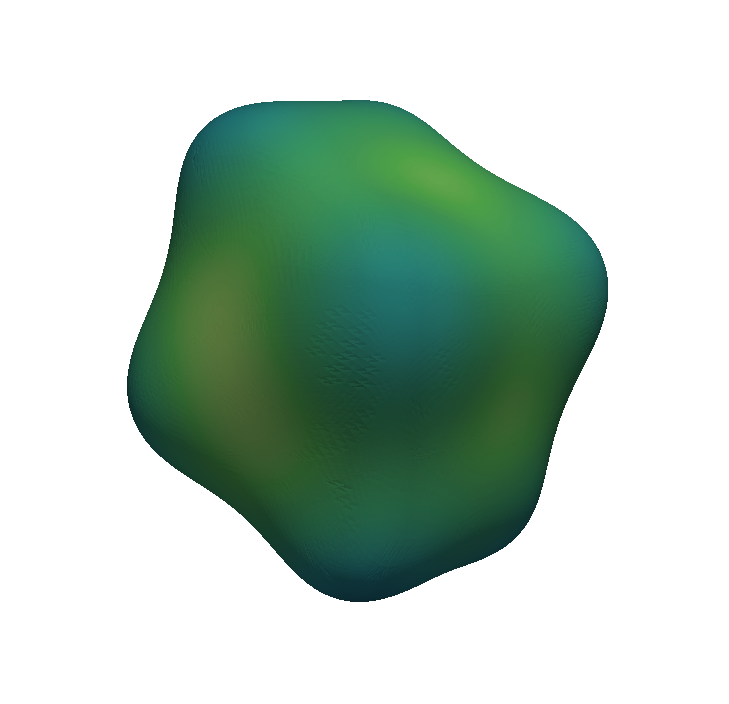}
    }
    \caption{t=0.05}
  \end{subfigure}
  \begin{subfigure}[b]{0.19\linewidth}
    \centering{
      \includegraphics[width=0.8\linewidth]{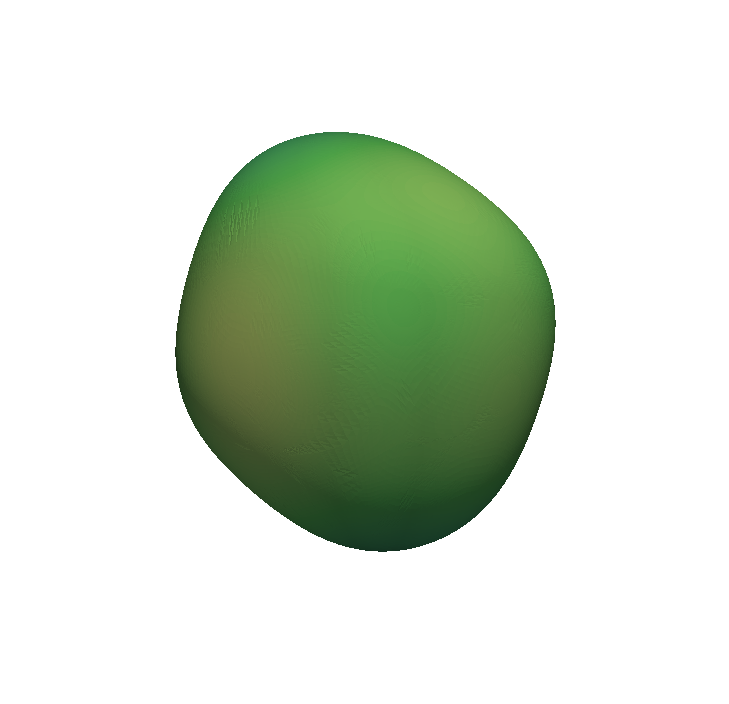}
    }
    \caption{t=0.1}
  \end{subfigure}
  \begin{subfigure}[b]{0.19\linewidth}
    \centering{
      \includegraphics[width=0.8\linewidth]{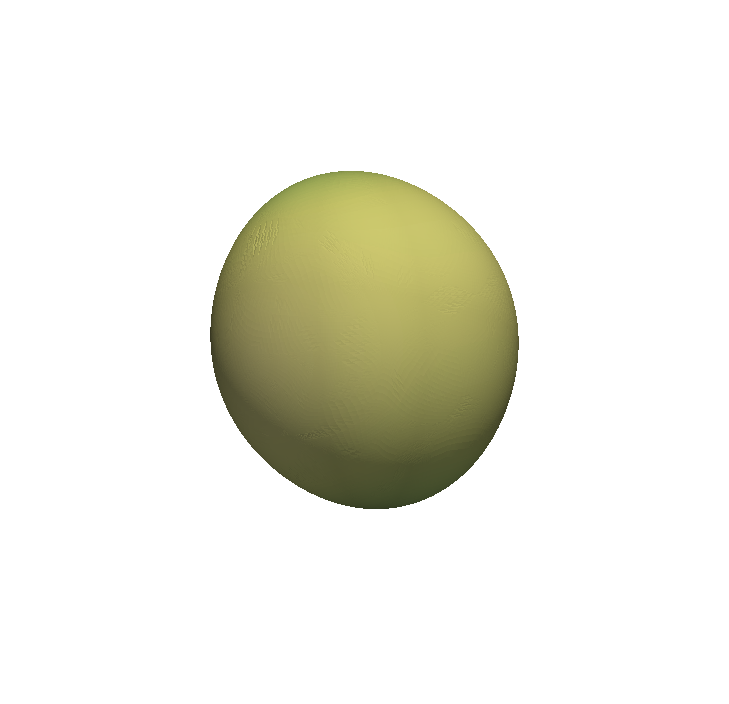}
    }
    \caption{t=0.15}
  \end{subfigure}
  \begin{subfigure}[b]{0.19\linewidth}
    \centering{
      \includegraphics[width=0.8\linewidth]{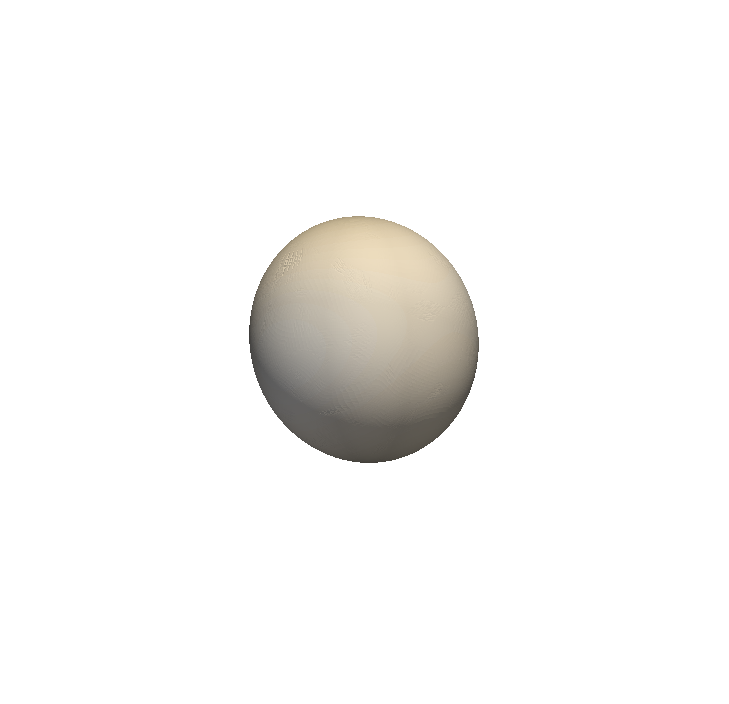}
    }
    \caption{t=0.2}
  \end{subfigure}
  \caption{Surface evolution towards a sphere using the \cpp{ALUGrid<2,3,conforming>} grid
  implementation. The color coding reflects the magnitude of the surface velocity $U$.} 
  \label{fig:mcf}
\end{figure}

In Figure \ref{fig:mcf} the evolution of the surface is presented.

Other successful applications of this meta grid concept for moving domains
can be found, for example, in \cite{nsreac:13} where the compressible Navier-Stokes equations are 
solved in a moving domain and in \cite{dune-prismgrid:12} where free surface
shallow water flow is considered.


%% file: complexdomains.tex
\subsection{Grid coupling and complex domains}

In recent years \dune has gained support for different strategies to
handle couplings of PDEs on different subdomains. One can distinguish three
different approaches to describe and handle such different domains involved in a multi-physics simulation.
As an important side effect, the last approach also provides support
for domains with complex shapes.

\begin{enumerate}
\item \textbf{Coupling of individual grids}:
  \begin{figure}[h]
    \centering
    \includegraphics[height=2cm]{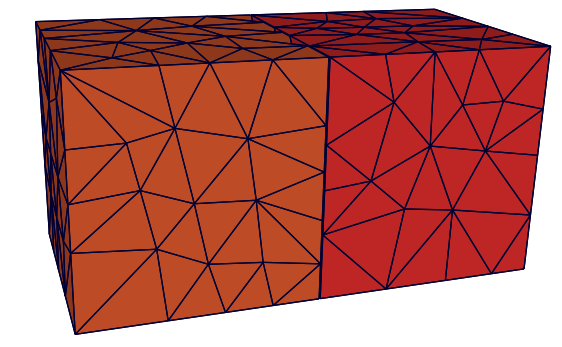}~\raisebox{.85cm}{\large :}\qquad
    \includegraphics[height=2cm]{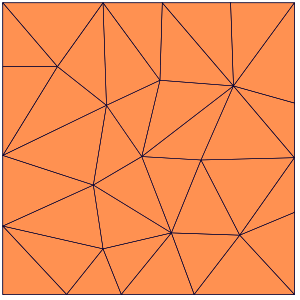}~\raisebox{.85cm}{\large $\cap$}
    \includegraphics[height=2cm]{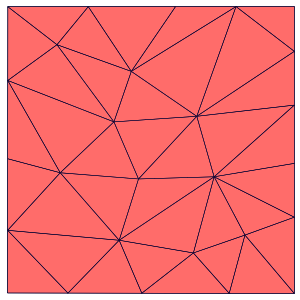}~\raisebox{.85cm}{\large =}
    \includegraphics[height=2cm]{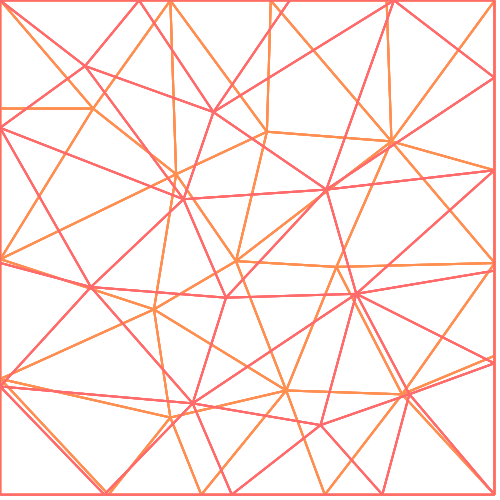}\\
  \caption{Coupling of two unrelated meshes via a merged grid: Intersecting the coupling patches
    yields a set of remote intersections, which can be
    used to evaluate the coupling conditions.}
  \label{fig:grid_glue_coupling}
  \end{figure}
  In the first approach, each subdomain is treated as a separate grid, and meshed individually.
  The challenge is then the construction
  of the coupling interfaces, i.e., the geometrical relationships at common subdomain
  boundaries. As it is natural to construct nonconforming interfaces in this way,
  coupling between the
  subdomains will in general require some type of weak coupling, like
  the mortar method~\cite{bernardi_maday_patera:1993},
  penalty methods~\cite{becker2003finite,lazarov2003almost}, or
  flux-based coupling~\cite{gander2005new}.

\item \textbf{Partition of a single grid}:
  \begin{figure}[h]
    \centering
    \includegraphics[width=0.6\textwidth]{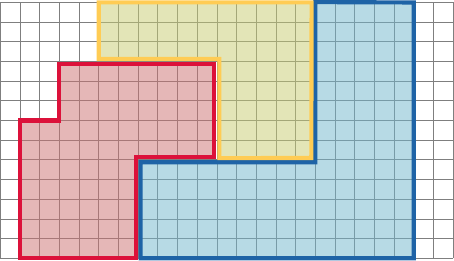}
    \caption{Partition of a given host mesh into subdomains.}
  \end{figure}
  In contrast, one may construct a single host grid that resolves the
  boundaries of all subdomains. The subdomain meshes are then
  defined as subsets of elements of the host grid.
  While the construction of the
  coupling interface is straightforward, generating the
  initial mesh is an involved process, if the subdomains have
  complicated shapes.
  As the coupling interfaces are conforming
  (as long as the host grid is), it is
  possible to enforce coupling conditions in strong form.

\item \textbf{Cut-cell grids}:
\begin{figure}[h]
  \centering
  \includegraphics[width=0.6\textwidth]{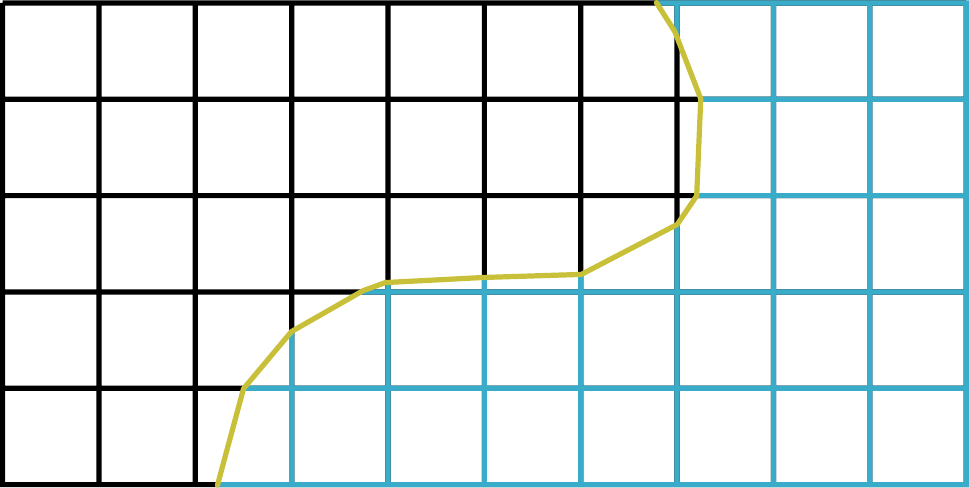}
  \caption{Construction of two cut-cell subdomain grids from a Cartesian
    background grid and a level-set geometry representation: cut-cells
    are constructed by intersecting a background cell with the
    zero-iso-surface of the level-set.}
\end{figure}
  The third approach is similar to the second one, and again involves a host grid.
  However, it is more flexible because this time the host grid can be arbitrary,
  and does not have to resolve the boundaries of the
  subdomains. Instead, subdomain grids are constructed by intersecting
  the elements with the independent subdomain geometry, typically described as the
  0-level set of a given function. This results in so-called cut-cells,
  which are fragments of host grid elements. Correspondingly,
  the coupling interfaces are constructed by intersecting host grid elements
  with the subdomain boundary.

  It is important to note that the shapes
  of the cut-cells can be arbitrary and the resulting cut-cell grids
  are not necessarily shape-regular.  As a consequence,
  it is not possible to employ standard discretization
  techniques. Instead, a range of different methods
  like the unfitted discontinuous
  Galerkin (UDG) method \cite{Bastian_Engwer,Engwer_UDG2012} and the CutFEM method \cite{cutFEM_2015}
  have been developed for cut-cell grids.
\end{enumerate}

All three concepts for handling complex domains are
available as special \dune modules.

\subsubsection{\dunegridglue{} --- Coupling of individual grids}
\label{sec:dunegridglue}
When coupling simulations on separate grids, the main challenge is the
construction of coupling operators, as these require detailed neighborhood information
between cells in different meshes. The \dunegridglue
module \cite{gridglue,engwer2016concepts}, available from \url{https://dune-project.org/modules/dune-grid-glue},
provides infrastructure to construct such relations efficiently.
Neighborhood relationships are described by the concept of \cpp{RemoteIntersection}s,
which are closely related to the \cpp{Intersection}s known from the \dunegrid module
(Section~\ref{sec:dune_grid:geometric_aspects}):
Both keep references to the two elements that make up the intersection, they store
the shape of the intersection in world space, and the local shapes of the intersection
when embedded into one or the other of the two elements. However, a \cpp{RemoteIntersection}
is more general than its \dunegrid cousin: For example, the two elements do not
have to be part of the same grid object, or even the same grid implementation.
Also, there is no requirement for the two elements to have the same dimension.
This allows mixed-dimensional couplings like the one in~\cite{koch_heck_schroeder_class_helmig:2018}.

Constructing the set of remote intersections for a pair of grids first requires
the selection of two coupling patches. These are two sets of entities that are
known to be involved in the coupling, like a contact boundary, or the overlap
between two overlapping grids. Coupling entities can have any codimension.
In principle all entities of a given codimension could always be selected as
coupling entities, but it is usually easy and more efficient to preselect
the relevant ones.

There are several algorithms for constructing the set of remote intersections for a
given pair of coupling patches. Assuming that both patches consist of roughly $N$
coupling entities, the naive algorithm will require $O(N^2)$ operations.
This is too expensive for many situations. Gander and Japhet \cite{gander2013algorithm}
proposed an advancing front algorithm with expected linear complexity,
which, however, slows down considerably when the coupling patches consist of
many connected components, or contain too many entities not actually involved
in the coupling. Both algorithms are available in \dunegridglue.
A third approach using a spatial data structure and a run-time of $O(N \log N)$
still awaits implementation.

A particular challenge arises in the case of
parallel grids, as the partitioning of both grids is also
unrelated. \dunegridglue can also compute the set of
\cpp{RemoteIntersection} in parallel codes, using additional
communication. For details on the algorithm and how to
handle couplings in the parallel case we refer to \cite{engwer2016concepts}.

As an example we implement the assembly of mortar mass matrices using \dunegridglue.
Let $\Omega$ be a domain in $\R^d$, split into two parts $\Omega_1$, $\Omega_2$ by a hypersurface $\Gamma$,
as in Figure~\ref{fig:grid_glue_coupling}.
On $\Omega$ we consider an elliptic PDE for a scalar function~$u$, subject to
the continuity conditions
\begin{align*}
   u_1 & = u_2,
   \qquad
   \langle \nabla u_1, \mathbf{n} \rangle = \langle \nabla u_2, \mathbf{n} \rangle
   \qquad
   \text{on $\Gamma$},
\end{align*}
where $u_1$ and $u_2$ are the restrictions of $u$ to the subdomains $\Omega_1$ and $\Omega_2$,
respectively, and $\mathbf{n}$ is a unit normal of $\Gamma$.

For a variationally consistent discretization, the mortar methods discretizes the weak form of the continuity condition
\begin{equation}
 \label{eq:domain_decomposition:mortar_weak_continuity}
 \int_{\Omega \cap \Gamma} (u_1|_\Gamma - u_2|_\Gamma) w \, ds = 0,
\end{equation}
which has to hold for a space of test functions $w$ defined on the common subdomain boundary.
Let $\Omega_1$ and $\Omega_2$ be discretized by two independent grids,
and let  $\{ \theta_i^1 \}_{i=1}^{n_1}$ and $\{ \theta_i^2 \}_{i=1}^{n_2}$
be nodal basis functions for these grids, respectively.
We use the nonzero restrictions of the $\{ \theta_i^1 \}$ on $\Gamma$ to discretize
the test function space.
Then \eqref{eq:domain_decomposition:mortar_weak_continuity} has the algebraic form
\begin{equation*}
 M_1 \overline{u}_1 - M_2 \overline{u}_2 = 0,
\end{equation*}
with mortar mass matrices
\begin{alignat*}{2}
 M_1 & \in \R^{n_{\Gamma,1} \times n_{\Gamma,1}}, & \qquad & (M_1)_{ij} = \int_{\Omega \cap \Gamma} \theta_i^1 \theta_j^1 \,ds \\
 M_2 & \in \R^{n_{\Gamma,1} \times n_{\Gamma,2}}, & \qquad & (M_2)_{ij} = \int_{\Omega \cap \Gamma} \theta_i^1 \theta_j^2 \,ds.
\end{alignat*}
The numbers $n_{\Gamma,1}$ and $n_{\Gamma_2}$ denote the numbers of degrees of freedom
on the interface $\Omega \cap \Gamma$.
Assembling these matrices is not easy, because $M_2$
involves shape functions from two different grids.

For the implementation, assume that the grids on $\Omega_1$ and $\Omega_2$ are
available as two \dune grid view objects
\cpp{gridView1} and \cpp{gridView2}, of types \cpp{GridView1} and \cpp{GridView2},
respectively.
The code first constructs the coupling patches, i.e.,
those parts of the boundaries of $\Omega_1$, $\Omega_2$ that are on the interface $\Gamma$.
These are represented in \dunegridglue by objects called \cpp{Extractor}s.
Since we are coupling on the
grid boundaries---which have codimension~1---we need two \cpp{Codim1Extractor}s:
\begin{c++}
  using Extractor1 = GridGlue::Codim1Extractor<GridView1>;
  using Extractor2 = GridGlue::Codim1Extractor<GridView2>;

  VerticalFacetPredicate<GridView1> facetPredicate1;
  VerticalFacetPredicate<GridView2> facetPredicate2;

  auto domEx = std::make_shared<Extractor1>(gridView1, facetPredicate1);
  auto tarEx = std::make_shared<Extractor2>(gridView2, facetPredicate2);
\end{c++}
The extractors receive the information on what part of the boundary to use by two predicate objects
\cpp{facetPredicate1} and \cpp{facetPredicate2}.  Both implement a method
\begin{c++}
bool contains(const typename GridView::Traits::template Codim<0>::Entity& element,
              unsigned int facet) const
\end{c++}
that returns true if the \cpp{facet}-th face of the element given in \cpp{element}
is part of the coupling boundary $\Gamma$.  For the example we use the
hyperplane $\Gamma \subset \R^d$ of all points with first coordinate equal to zero.
Then the complete predicate class is
\begin{c++}
template <class GridView>
struct VerticalFacetPredicate
{
  bool operator()(const typename GridView::template Codim<0>::Entity& element,
                  unsigned int facet) const
  {
    const int dim = GridView::dimension;
    const auto& refElement = Dune::ReferenceElements<double, dim>::general(element.type());

    // Return true if all vertices of the facet 
    // have the coordinate (numerically) equal to zero
    for (const auto& c : refElement.subEntities(facet,1,dim))
      if ( std::abs(element.geometry().corner(c)[0] ) > 1e-6 )
        return false;

    return true;
  }
};
\end{c++}

Next, we need to compute the set of remote intersections from the two coupling patches.
The different algorithms for this mentioned above are implemented in objects called
``mergers''. The most appropriate one for the mortar example is called \cpp{ContactMerge},
and it implements the advancing front algorithm of Gander and Japhet.%
\footnote{It is called \cpp{ContactMerge} because it can also handle the case where
the two subdomains are separated by a physical gap, which is common in contact problems.}
The entire code to construct the remote intersections for the two trace grids at the interface $\Gamma$ is

\begin{c++}
  using GlueType = GridGlue::GridGlue<Extractor1,Extractor2>;

  // Backend for the computation of the remote intersections
  auto merger = std::make_shared<GridGlue::ContactMerge<dim,double> >();
  GlueType gridGlue(domEx, tarEx, merger);

  gridGlue.build();
\end{c++}

The \cpp{gridGlue} object is a container for the remote intersections. These can now
be used to compute the two mass matrices $M_1$ and $M_2$.  Let \cpp{mortarMatrix1} and
\cpp{mortarMatrix2} be two objects of some
(deliberately) unspecified matrix type.  We assume that both are initialized and all entries
are set to zero.
The nodal bases $\{ \theta^1_i\}_{i=1}^{n_1}$ and $\{ \theta^2_i\}_{i=1}^{n_2}$
are represented by two \dunefunctions bases.
The mortar assembly loop is much like the loop for a regular mass matrix

\begin{c++}
  auto nonmortarView = nonmortarBasis.localView();
  auto mortarView    = mortarBasis.localView();

  for (const auto& intersection : intersections(gridGlue))
  {
    nonmortarView.bind(intersection.inside());
    mortarView.bind(intersection.outside());

    const auto& nonmortarFiniteElement = nonmortarView.tree().finiteElement();
    const auto& mortarFiniteElement    = mortarView.tree().finiteElement();
    const auto& testFiniteElement      = nonmortarView.tree().finiteElement();

    // Select a quadrature rule:  Use order = 2 just for simplicity
    int quadOrder = 2;
    const auto& quad = QuadratureRules<double, dim-1>::rule(intersection.type(), quadOrder);

    // Loop over all quadrature points
    for (const auto& quadPoint : quad)
    {
      // compute integration element of overlap
      double integrationElement = 
        intersection.geometry().integrationElement(quadPoint.position());

      // quadrature point positions on the reference element
      FieldVector<double,dim> nonmortarQuadPos = intersection.geometryInInside().global(quadPoint.position());
      FieldVector<double,dim> mortarQuadPos    = intersection.geometryInOutside().global(quadPoint.position());

      //evaluate all shapefunctions at the quadrature point
      std::vector<FieldVector<double,1> >
          nonmortarValues, mortarValues, testValues;

      nonmortarFiniteElement.localBasis()
          .evaluateFunction(nonmortarQuadPos,nonmortarValues);
      mortarFiniteElement.localBasis()
          .evaluateFunction(mortarQuadPos,mortarValues);
      testFiniteElement.localBasis()
          .evaluateFunction(nonmortarQuadPos,testValues);

      // Loop over all shape functions of the test space
      for (size_t i=0; i<testFiniteElement.size(); i++)
      {
        auto testIdx = nonmortarView.index(i);

        // loop over all shape functions on the nonmortar side
        for (size_t j=0; j<nonmortarFiniteElement.localBasis().size(); j++)
        {
          auto nonmortarIdx = nonmortarView.index(j);

          mortarMatrix1[testIdx][nonmortarIdx] += integrationElement *
              quadPoint.weight() * testValues[i] * nonmortarValues[j];
        }

        // loop over all shape functions on the mortar side
        for (size_t j=0; j<mortarFiniteElement.size(); j++)
        {
          auto mortarIdx = mortarView.index(j);

          mortarMatrix2[testIdx][mortarIdx] += integrationElement *
              quadPoint.weight() * testValues[i] * mortarValues[j];
        }
      }
    }
  }
\end{c++}
After these loops, the objects \cpp{mortarMatrix1} and \cpp{mortarMatrix2}
contain the matrices $M_1$ and $M_2$, respectively.

The problem gets more complicated when $\Gamma$ is not a hyperplane.
The approximation of a non-planar $\Gamma$
by unrelated grids will lead to ``holes'' at the interface,
and the jump $u_1|_\Gamma - u_2|_\Gamma$ is not well-defined anymore.
This situation is usually dealt with by identifying $\Gamma_1^h$ and $\Gamma_2^h$ with a homeomorphism $\Phi$,
and replacing the second mass matrix by
\begin{equation*}
 M_2 \in \R^{n_{\Gamma,1} \times n_{\Gamma,2}},
 \qquad
 (M_2)_{ij} = \int_{\Gamma_1^h} \theta_i^2 (\theta^1_j \circ \Phi) \, ds.
\end{equation*}

Only few changes have to be done to the code to implement this.  First of all, the vertical predicate class
has to be exchanged for something that correctly finds the curved coupling boundaries.  Then, setting up
extractor and \cpp{GridGlue} objects remains unchanged.  The extra magic needed to  handle the
mapping $\Phi$ is completely concealed in the \cpp{ContactMerge} implementation, which does not rely on $\Gamma_1^h$ and
$\Gamma_2^h$ being identical.  Instead, if there is a gap between them, a projection in normal direction
is computed automatically and used for $\Phi$.

\subsubsection{\dunemultidomaingrid{} --- Using element subsets as subdomains}
\label{sec:multidomaingrid}

The second approach to the coupling of subdomains is implemented in the
\dunemultidomaingrid module, available at \url{https://dune-project.org/modules/dune-multidomaingrid}.
This module allows to structure a given host grid into
  different subdomains.
It is implemented in terms of two cooperating
grid implementations \cpp{MultiDomainGrid} and \cpp{SubDomainGrid}:
\cpp{MultiDomainGrid} is a meta grid that wraps
a given host grid and extends it with an interface for setting up and accessing
subdomains. It also stores all data required to manage the subdomains. The
individual subdomains are exposed as \cpp{SubDomainGrid} instances, which are
lightweight objects that combine the information from the host grid and the
associated \cpp{MultiDomainGrid}. \cpp{SubDomainGrid} objects present a subdomain as a
regular \dune grid.
A \cpp{MultiDomainGrid} inherits all capabilities of the underlying
grid, including features like $h$-adaptivity and MPI parallelism.
Extensions of the official grid interface allow to obtain the
associate entities in the fundamental mesh and the corresponding
indices in both grids.

A fair share of the ideas from \dunemultidomaingrid{} were incorporated in the
coupling capabilities of DuMu$^x$ 3 \cite{dumux3}.



\subsubsection{\dunetpmc{} --- Assembly of cut-cell discretizations}
The main challenge for cut-cell approaches is
the construction of appropriate quadrature rules to evaluate integrals
over the cut-cell and its boundary. We assume that the domain is given
implicitly as a discrete level set function $\Phi_h$, s.t. $\Phi(x) < 0$
if $x \in \Omega^{(i)}$. The goal is now
to compute a polygonal representation of the cut-cell and a
decomposition into sub-elements, such that
standard quadrature can be applied on each sub-element. This allows to
evaluate weak forms on the actual domain, its boundary, and the
internal skeleton (when employing DG methods).

The \dunetpmc library implements a \emph{topology preserving marching
  cubes} (TPMC) algorithm~\cite{tpmc}, assuming that
$\Phi_h$ is given as a piecewise multilinear scalar function (i.e. a
$P^1$ or $Q^1$ function). The fundamental idea in this case is the
same as that of the classical marching cubes algorithm, known from
computer graphics. Given the sign of the vertex values the
library identifies the topology of the cut-cell. In certain ambiguous
cases additional points in the interior of the cell need to be
evaluated. From the topological case the actual decomposition is
retrieved from a lookup table and mapped according to the real function values.

\paragraph{Evaluating integrals over a cut-cell domain using \dunetpmc}
We look at a simple example to learn how to work with cut-cell domains. As
stated, the technical challenge regarding cut-cell methods is the
construction of quadrature rules. We consider a circular domain of
radius $1$ in 2d
and compute the area using numerical quadrature. The scalar function $\Phi :
x \in \mathbb{R}^d \rightarrow |x|_2 - 1$ describes the domain boundary as
the isosurface $\Phi = 0$ and the interior as $\Phi < 0$.
\begin{c++}
using namespace Dune;

double R = 1.0;
auto phi = [R](FieldVector<ctype,dim> x){ return x.two_norm()-R; };
\end{c++}
After having setup a grid, we iterate over a given gridview
\cpp{gv}, compute the $Q_1$ representation of $\Phi$ (or better
to say the vertex values in an element \cpp{e})
\begin{c++}
double volume = 0.0;
std::vector<ctype> values;
for (const auto & e : elements(gv))
{
  const auto & g = e.geometry();
  // fill vertex values
  values.resize(g.corners());
  for (std::size_t i = 0; i < g.corners(); i++)
    values[i] = phi(g.corner(i));
  volume += localVolume(values,g);
}
\end{c++}
We now compute the local volume by quadrature over the cut-cell
\texttt{e}$|_{\Phi < 0}$. In order to evaluate the integral we use the
\cpp{TpmcRefinement} and construct snippets, for which we can
use standard quadrature rules:
\begin{c++}
template<typename Geometry>
double localVolume(std::vector<double> values, const Geometry & g)
{
  double volume = 0.0;
  // calculate tpmc refinement
  TpmcRefinement<ctype,dim> refinement(values);
  // sum over inside domain
  for (const auto & snippet : refinement.volume(tpmc::InteriorDomain))
  {
    // get zero-order quadrature rule
    const QuadratureRule<double,dim>& quad =
      QuadratureRules<double,dim>::rule(snippet.type(),0);
    // sum over snippets
    for (size_t i=0; i<quad.size(); i++)
      volume += quad[i].weight()
        * snippet.integrationElement(quad[i].position())
        * g.integrationElement(snippet.global(quad[i].position()));
  }
}
\end{c++}
This gives us a convergent integral, approximating $\pi$. Unsurprisingly
we obtain an $O(h^2)$ convergence of the quadrature error, as the
geometry is approximated as a polygonal domain.


%% file: solvers.tex
\subsection{Non-smooth multigrid}

Various interesting PDEs from application fields such as computational mechanics
or phase-field modeling can be written as nonsmooth convex minimization problems
with certain separability properties.  For such problems, the module \dunetnnmg offers
an implementation of the Truncated Nonsmooth Newton Multigrid (TNNMG) algorithm~\cite{GraeserKornhuber2009b,graeser_sander:2017}.

\subsubsection{The truncated nonsmooth Newton multigrid algorithm}

TNNMG operates at the algebraic level of PDE problems.
Let $\R^N$ be endowed with a block structure
\begin{equation*}
 \R^N = \prod_{i=1}^m \R^{N_i},
\end{equation*}
and call $R_i : \R^N \to \R^{N_i}$ the canonical restriction to the $i$-th block.
Typically, the factor spaces $\R^{N_i}$ will have small dimension, but the number of factors $m$
is expected to be large.
A strictly convex and coercive objective functional $J:\R^N \to \R \cup \{ \infty \}$ is called
block-separable if it has the form
\begin{align}
    \label{eq:min_problem}
    J(v) = J_0(v) + \sum_{i=1}^m \varphi_i (R_i v),
\end{align}
with a convex $C^2$ functional $J_0 :\R^N \to \R$, and convex, proper, lower semi-continuous functionals
$\varphi_i : \R^{N_i} \to \R \cup \{ \infty \}$.

Given such a functional $J$, the TNNMG method alternates between a nonlinear smoothing step
and a damped inexact Newton correction. The smoother solves local minimization problems
\begin{align}
\label{eq:general_local_problem}
    \tilde{v}^k = \argmin_{\tilde{v} \in \tilde{v}^{k-1} + V_k} J(\tilde{v})
    \qquad
    \text{for all $k=1,\dots, m$},
\end{align}
in the subspaces $V_k \subset \R^N$ of all vectors that have zero
entries everywhere outside of the $k$-th block.
The inexact Newton step typically consists of a single multigrid iteration
for the linearized problem, but other choices are possible as well.

For this method global convergence has been shown even when using only inexact
local smoothers~\cite{graeser_sander:2017}.
In practice it is observed that the method degenerates to a multigrid method
after a finite number of steps, and hence multigrid convergence rates are achieved
asymptotically~\cite{GraeserKornhuber2009b}.

The \dunetnnmg module, available from \url{https://git.imp.fu-berlin.de/agnumpde/dune-tnnmg},
offers an implementation of the TNNMG algorithm in the context of \dune.
The coupling to \dune is very loose---as TNNMG operates on functionals in $\R^N$
only, there is no need for it to know about grids, finite element spaces, etc.%
\footnote{The only exception to this are the multigrid transfer operators.
These require access to finite element bases, but are not central to TNNMG.}
The \dunetnnmg module therefore only depends on \duneistl and \dunesolvers.

\subsubsection{Numerical example: small-strain primal elastoplasticity}

The theory of elastoplasticity describes the behavior of solid objects that can undergo both temporary (elastic)
and permanent (plastic) deformation. In its simplest (primal) form, its variables
are a vector field $\mathbf{u} : \Omega \to \R^d$ of displacements, and a matrix field
$\mathbf{p} : \Omega \to \text{Sym}^{d \times d}_0$ of plastic strains.
These strains are assumed to be symmetric and trace-free~\cite{han_reddy:2013}.
Displacements $\mathbf{u}$ live in the Sobolev space $H^1(\Omega, \R^d)$, and
(in theories without strain gradients) plastic strains
live in the larger space $L^2(\Omega, \text{Sym}^{d\times d}_0)$. Therefore, the easiest
space discretization employs continuous piecewise linear finite elements for
the displacement $\mathbf{u}$, and piecewise constant plastic strains $\mathbf{p}$.

Implicit time discretization of the quasistatic model leads to a sequence of spatial
problems~\cite{sander:2017,han_reddy:2013}.  These can be written as
minimization problems
\begin{align}
 \label{eq:algebraic_plasticity_functional}
 J(u,p)
 & \colonequals
 \frac{1}{2} (u^T \,p^T) A \begin{pmatrix} u \\ p \end{pmatrix} - b^T \begin{pmatrix} u \\ p \end{pmatrix}
 +
 \sigma_c \sum_{i=1}^{n_2} \int_\Omega \theta_i(x)\,dx \cdot \lVert p_i \rVert_F,
\end{align}
which do not involve a time step size because the model is rate-independent.
Here, $u$ and $p$ are the finite element coefficients of the displacement and plastic strains,
respectively, and $A$ is a symmetric positive definite matrix.
The number $\sigma_c$ is the yield stress, and $b$ is the load vector.
The functions $\theta_1,\dots, \theta_{n_2}$ are the canonical basis functions
of the space of piecewise constant functions, and $\lVert \cdot \rVert_F : \text{Sym}^{d \times d}_0 \to \R$
is the Frobenius norm.
In the implementation, trace free symmetric matrices $p_i \in \text{Sym}^{d \times d}_0$
are represented by vectors of length $\frac{1}{2}(d+1)d - 1$.

By comparing~\eqref{eq:algebraic_plasticity_functional} to~\eqref{eq:min_problem},
one can see that the increment functional~\eqref{eq:algebraic_plasticity_functional}
has the required form~\cite{sander:2017}.  By a result of
\citep{alberty_carstensen_zarrabi:1999}, the local nonsmooth minimization
problems~\eqref{eq:general_local_problem} can be solved exactly.

The implementation used in~\cite{sander:2017} employs several of the recent
hybrid features of \dunefunctions and \duneistl.
The pair of finite element spaces for displacements and plastic strains
for a three-dimensional problem forms the tree $(P_1)^3 \times (P_0)^5$
(where we have identified $\text{Sym}^{3 \times 3}_0$ with $\R^5$).
This tree can be constructed by
\begin{c++}
constexpr size_t nPlasticStrainComponents = dim*(dim+1)/2-1;
auto plasticityBasis = makeBasis(gridView, composite(
    power<dim>(lagrange<1>()),                     // Deformation basis
    power<nPlasticStrainComponents>(lagrange<0>()) // Plastic strain basis
  ));
\end{c++}
The corresponding linear algebra data structures must combine block vectors with block size~3
and block vectors with block size~5. Hence the vector data type definition is
\begin{c++}
using DisplacementVector  = BlockVector<FieldVector<double,dim> >;
using PlasticStrainVector = BlockVector<
               FieldVector<double,nPlasticStrainComponents> >;
using Vector = MultiTypeBlockVector<DisplacementVector,
                                    PlasticStrainVector>;
\end{c++}
and this is the type used by \dunetnnmg for the iterates. The corresponding matrix type
combines four \cpp{BCRSMatrix} objects of different block sizes in a single
\cpp{MultiTypeBlockMatrix}, and the multigrid solver operates directly on this type
of matrix.

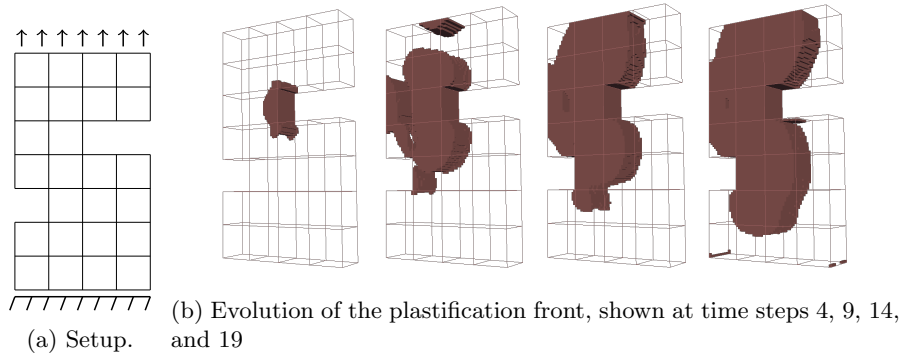
\begin{figure}
 \begin{subfigure}[b]{0.18\linewidth}
 \centering
    \begin{tikzpicture}[scale=0.45]


  \foreach \y in {0,...,7}
    \draw [line width=0.1mm] (0,\y)--(4,\y);

  \draw [line width=0.1mm] (0,0)--(0,2);
  \draw [line width=0.1mm] (0,3)--(0,7);
  \draw [line width=0.1mm] (1,0)--(1,7);
  \draw [line width=0.1mm] (2,0)--(2,7);
  \draw [line width=0.1mm] (3,0)--(3,4);
  \draw [line width=0.1mm] (3,5)--(3,7);
  \draw [line width=0.1mm] (4,0)--(4,4);
  \draw [line width=0.1mm] (4,5)--(4,7);

  \draw [line width=0.2mm] (0,-0.2)--(4,-0.2);

  \foreach \x in {0,...,8}
    \draw [line width=0.2mm] (0+0.5*\x,-0.2)--(-0.15+0.5*\x,-0.6);

  \foreach \x in {0,...,6}
    \draw [->, line width=0.2mm] (0.2 + 0.6*\x,7.2)--(0.2 + 0.6*\x,7.7);

  \end{tikzpicture}
 \caption{Setup.}
 \label{fig:3d_domain}
 \end{subfigure}
\begin{subfigure}[b]{0.8\linewidth}
 \centering
  \includegraphics[width=0.21\linewidth]{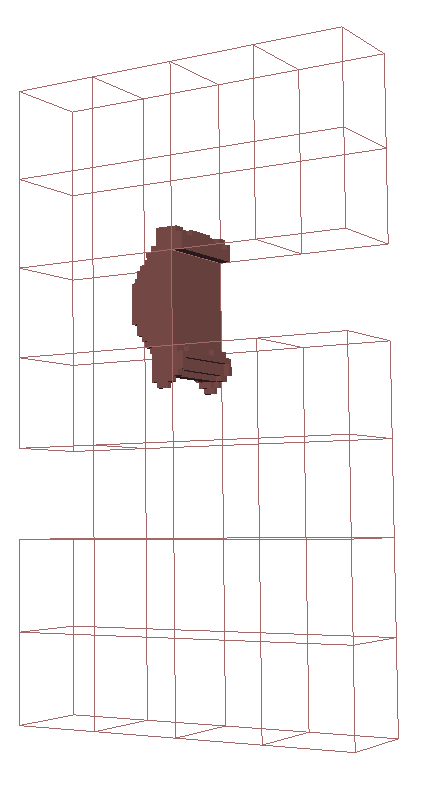}
  \includegraphics[width=0.21\linewidth]{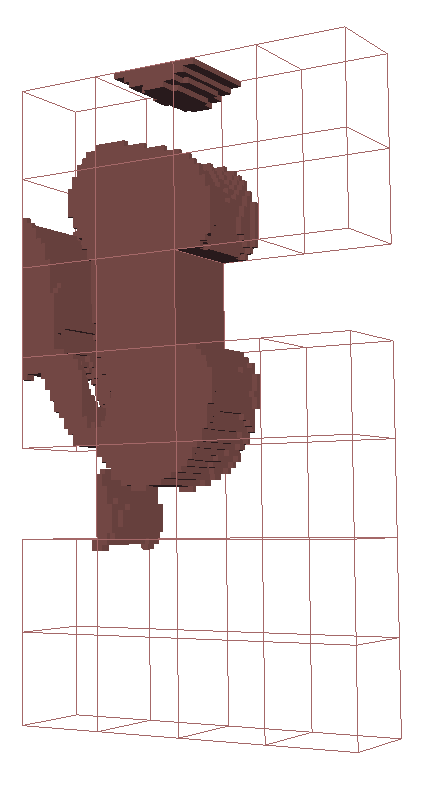}
  \includegraphics[width=0.21\linewidth]{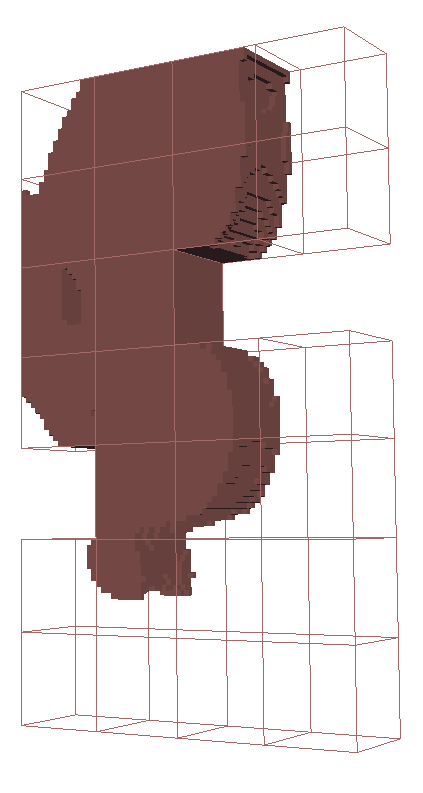}
  \includegraphics[width=0.21\linewidth]{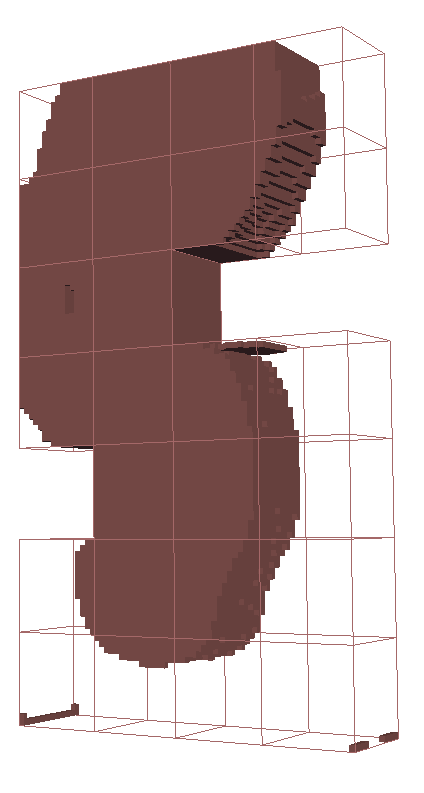}
 \caption{Evolution of the plastification front, shown at time steps 4, 9, 14, and 19}
 \label{fig:plastic_zone_evolution_3d}
\end{subfigure}
\caption{Setup and numerical results for the three-dimensional benchmark from~\cite{neff_sydow_wieners:2009}.}
\end{figure}

We show a numerical example of the TNNMG solver for a three-dimensional test problem.
Note that in this case, the space $\text{Sym}^{d \times d}_0$ is 5-dimensional,
and therefore isomorphic to $\R^5$.  Let $\Omega$ be the domain depicted in
Figure~\ref{fig:3d_domain}, with bounding box $(0,4) \times (0,1) \times (0,7)$.
We clamp the object at $\Gamma_D = (0,4) \times (0,1) \times \{0\}$, and apply a time-dependent
normal load
\begin{equation*}
  \langle l(t), \bu \rangle = 20\,t \int_{\Gamma_N} \bu \cdot \be_3 \, ds
\end{equation*}
on $\Gamma_N = (0,4) \times (0,1) \times \{ 7 \}$.  The material parameters are
taken from~\cite{neff_sydow_wieners:2009}.
Figure~\ref{fig:plastic_zone_evolution_3d} shows the evolution of the plastification front
at several points in time.  See~\cite{sander:2017} for more detailed numerical results
and performance measurements.


%% file: multiscale.tex
\subsection{Multiscale methods}
\label{subsec:ms}

\def\Rec{{\mathcal{R}}}

\newcommand{\Tauh}{\T_h}
\newcommand{\TauH}{\T_H}

There has been a tremendous development of numerical multiscale methods in the last two decades
including the multiscale finite element method (MsFEM)  \cite{HouWu97,Efendiev:Hou:2009,HOS14},
the heterogeneous multiscale method (HMM) \cite{E:Engquist:2003,Ohlberger:2005,Abdulle2005}, the variational multiscale method (VMM) \cite{Hughes:1995,Hughes:et-al:1998,Larson:Malqvist:2005}  or the local orthogonal decomposition (LOD) \cite{MalPet:14,HMP:14,efficientlod}.
More recently, extensions to parameterized multiscale problems have been presented, such as the  localized multiscale reduced basis method (LRBMS) \cite{AlbrechtHaasdonkEtAl2012,OS2015,ORS2017} or the generalized multiscale finite element method (GMsFEM) \cite{Efendiev2013,Chung2014,Chung2014b}.
In \cite{algoritmy12,HO12} we have demonstrated that most of these methods can be cast into a general abstract framework that may then be used for
the design of a common implementation framework for multiscale methods, which has been realized in the \dunemultiscale{} module \cite{dunemultiscale}.
In the following, we concentrate on an efficient parallelized implementation of MsFEM within the DUNE software framework.

\subsubsection{Multiscale model problem}
As a model problem we consider heterogeneous diffusion.
Given a domain $\Omega \subset \R^n$, $n \in \mathbb{N}_{>0}$ with a polygonal boundary,
an elliptic diffusion tensor $A^{\epsilon} \in \left(L^{\infty}(\Omega)\right)^{n \times n}$
with microscopic features, and
an $f \in L^2(\Omega)$  we define our model problem as
\begin{align}
\label{eq:ms:original-problem} \mbox{find} \quad u^{\epsilon} \in \mathring{H}^1(\Omega):
    \quad \int_{\Omega} A^{\epsilon} \nabla u^{\epsilon} \cdot \nabla \Phi
        = \int_{\Omega} f \Phi \quad \forall \Phi \in \mathring{H}^1(\Omega)
\end{align}
with $\mathring{H}^1(\Omega) \colonequals \overline{\mathring{C}^{\infty}(\Omega)}^{\|\cdot\|_{H^1(\Omega)}}$.

For the discretization of Equation~(\ref{eq:ms:original-problem}) we require a regular partition $\T_H$ of $\Omega$ with
elements $T$ and a nested refinement $\Tauh$ of $\T_H$ with elements $t$ and
choose associated piece-wise linear finite element spaces $U_H \colonequals S^1_0(\T_H) \subset U_{h} \colonequals S^1_0(\T_h)\subset \mathring{H}^{1}(\Omega)$.

We assume that $U_h$ is sufficiently accurate.
By $A^{\epsilon}_h$ we denote a suitable piecewise polynomial approximation of $A^{\epsilon}$, and
for $T \in \TauH$, we call $U(T)$ an {\it admissible environment} of $T$, if it is connected,
if $T \subset U(T) \subset \Omega$ and if it is the union of elements of $\Tauh$.
Admissible environments will be used for oversampling. In particular $T$ is an admissible environment of itself.

The MsFEM in Petrov--Galerkin formulation with oversampling is defined in the following. The typical construction of an explicit
multiscale finite element basis is already indirectly incorporated in the method. Also note that for
$U(T) = T$ we obtain the MsFEM without oversampling.

Let now $\mathcal{U}_H = \{ U(T)| \enspace T \in \TauH \}$ denote a set of admissible environments of elements of $\TauH$.
We call $\Rec^{\epsilon}_h(u_H) \in U_h \subset \mathring{H}^1(\Omega)$ the MsFEM-approximation of $u^{\epsilon}$,
if $u_H \in U_H$ solves:
\begin{align*}
 \sum_{T \in \TauH} \int_{T} A^{\epsilon}_h \nabla \Rec^{\epsilon}_h(u_H) \cdot \nabla \Phi_H = \int_{\Omega} f \Phi_H
      \quad \forall \Phi_H \in U_{H}.
\end{align*}

For $\Phi_H \in U_{H}$, the reconstruction $\Rec^{\epsilon}_h(\Phi_H)$ is defined by
$\Rec^{\epsilon}_h(\Phi_H)_{|T}\colonequals \tilde{Q}_{h}^{\epsilon}(\Phi_H) + \Phi_H$,
where $\tilde{Q}_{h}^{\epsilon}(\Phi_H)$ is obtained in the following way: First we solve
for $Q_{h,T}^{\epsilon}(\Phi_H) \in \mathring{U}_h(U(T))$ with
\begin{align}
 \label{eq:msfem:local} \int_{U(T)} A^{\epsilon}_h \left( \nabla \Phi_H + \nabla Q_{h,T}^{\epsilon}(\Phi_H) \right) \cdot \nabla \phi_h = 0
      \quad \forall \phi_h \in \mathring{U}_h(U(T))
\end{align}
for all $T \in \T_H$, where $\mathring{U}_h(U(T))$ is the underlying fine scale finite element space on $U(T)$ with
zero boundary values on $\partial U(T)$.
Since we are interested in a globally continuous approximation, i.e.
$\Rec^{\epsilon}_h(u_H) \in U_h \subset \mathring{H}^1(\Omega)$, we still need a conforming projection
$P_{H,h}$ which maps the discontinuous parts $ Q_{h,T}^{\epsilon}(\Phi_H)_{|T}$ to an element of $U_h$. Therefore, if
\begin{align*}
 P_{H,h} : \{ \phi_h \in L^2(\Omega)| \enspace \phi_h \in U_h(T) \enspace \forall T \in \TauH\} \longrightarrow U_h
\end{align*}
denotes such a projection, we define
\begin{align*}
\tilde{Q}_{h}^{\epsilon}(\Phi_H) \colonequals P_{H,h}( \sum_{T \in \TauH} \chi_T Q_{h,T}^{\epsilon}(\Phi_H))
\end{align*}
with indicator function $\chi_T$.

For a more detailed discussion and analysis of this method we refer to \cite{HOS14}.

\subsubsection{Implementation and parallelization}

Our implementation of the general framework for multiscale methods (\dunems{}, \cite{dunemultiscale})
is an effort birthed from the \exadune{} project \cite{BEGIIOTFKMR14,exa:highlevel,exadune} and is built
using the \dune{} Generic Discretization
Toolbox (\dunegdt{}, \cite{dunegdt}) and \dunext{} \cite{dunext} as well as the
\dune{} core modules described in Section~\ref{sec:coremodules}.

%
To maximize CPU utilization we employ multi-threading to dynamically load balance work items inside one CPU
without expensive memory transfer or cross-node communication. This effectively reduces the communication/overlap
region of the coarse grid in a scenario with a fixed number of available cores.
Within \dune{} we decided to use Intel's Thread Building Blocks (TBB) library as our multithreading abstraction.

Let us now consider an abstract compute cluster that is comprised of
a set of processors $\mathcal{P}$, where a set of cores $C_{P_i} = \{C^j_{P_i}\}$ is associated with each $P_i\in \mathcal{P}$ and
a collection of threads $t_{C_j} = \{t^k_{C_j}\}$. For simplicity, we assume here that $j=k$
across $\mathcal{P}$.

\begin{figure}
 \centering
  \includegraphics[height=0.25\linewidth, clip]{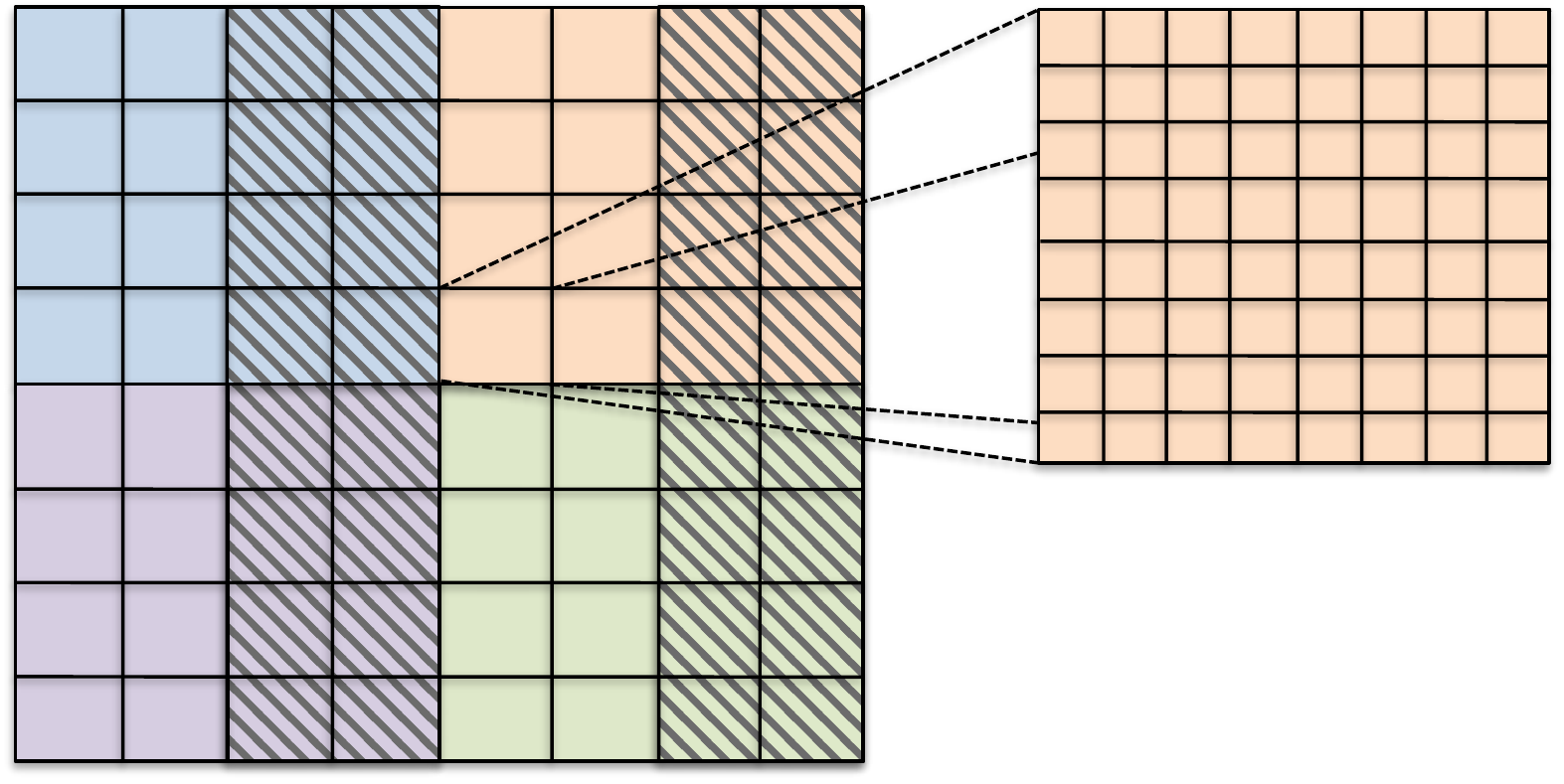}\qquad
  \includegraphics[height=0.25\linewidth, clip]{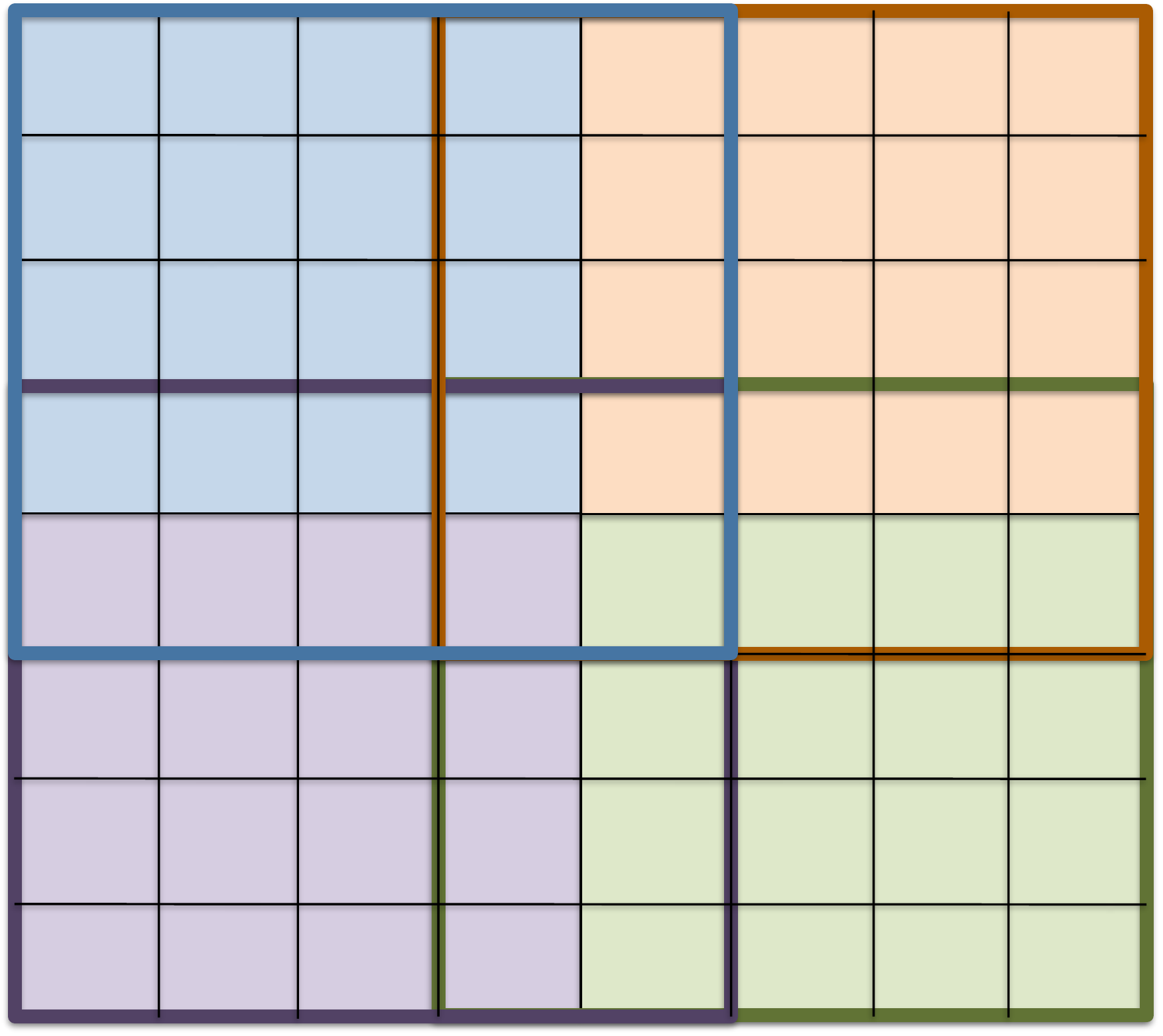}
\caption{\label{fig::msfem::mpigrid} Non-overlapping hybrid macro grid distribution of $\mathcal{T_H}$ for $\mathcal{P} = P_0,\cdots P_3$ with the hatched area symbolizing sub-distribution over $t_{C_j}$
and zoomed fine scale sub-structure of $U_{h,T}$ for $U(T)=T$ (left).
Overlapping macro grid distribution of $\mathcal{T_H}$ for $\mathcal{P} = P_0, \cdots, P_3$ (right).}
\end{figure}
%

Since we are interested in globally continuous solutions in $U_H$, we require an overlapping distribution
$\mathcal{T}_{H,P_i} \subset \mathcal{T}_{H}$ where cells can be present on multiple $P_i$.
Furthermore, we denote by $\mathcal{I}_i \subset \mathcal{T}_{H,P_i}$ the set of inner elements, if for all
$T_H \in \mathcal{I}_i \Rightarrow T_H \notin \mathcal{I}_j$ for all $i,j$ with $i\neq j$.
%
%
%
The first important step in the multiscale algorithm is to solve the cell corrector problems (\ref{eq:msfem:local})
for all $U(T_H), T_H \in \mathcal{I}_i$.
These
are truly locally solvable in the sense of data independence with
respect to neighbouring coarse cells. We build upon
extensions to the \dunegrid{} module made within \exadune{}, presented in \cite{engwer2013:enumath}, that
allow us to partition a given \cpp{GridView} into connected ranges of cells. The assembler was
modified to use TBB such that different threads iterate over
different of these partitions in parallel (Fig. \ref{fig::msfem::mpigrid}).

For each $T_H$ we create a new structured \cpp{Dune::YaspGrid}
to cover $U(T_H)$. Next we need to obtain $Q_{h,T}^{\epsilon}(\Phi_H)$
for all $J$ coarse scale basis function.  After discretization this actually means assembling only one linear system
matrix and $J$ different right hand sides. The assembly handled by \cpp{Dune::GDT::SystemAssembler},
which is parametrized by an 
elliptic operator \cpp{GDT::Operators::EllipticCG}
and corresponding right hand side
functionals
\cpp{GDT::LocalFunctional::Codim0Integral}. The
\cpp{SystemAssembler}
performs loop-fusion by merging
cell-local operations of any number of input functors. This allows to
perform the complete assembly in one single sweep over
the grid, using a configurable amount of
thread-parallelism.

Since the cell problems usually only contain up to about 100,000 elements
it is especially efficient to factorize the assembled system matrix once and then backsolve for all right hand sides.
For this we employ the \texttt{UMFPACK}\cite{umfpack} direct solver from the SuiteSparse library
\footnote{\url{http://faculty.cse.tamu.edu/davis/suitesparse.html}} and its abstraction through \duneistl{}. Another
layer of abstraction on top of that in \dunext{} allows us to switch to an iterative solver at run-time, should
we exceed the suitability constraints of the direct solver.

\begin{figure}
\begin{subfigure}[b]{0.5\linewidth}
  \centering
  \scalebox{0.4}{\input{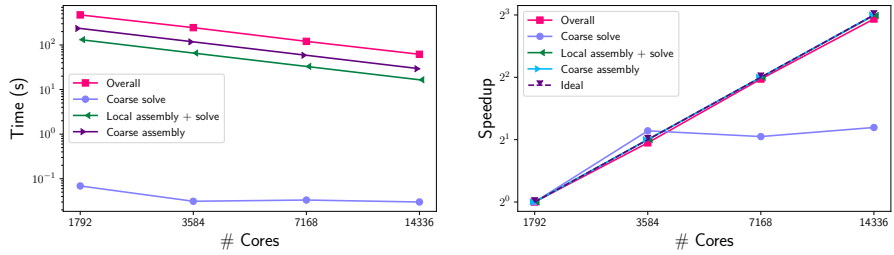}}
\end{subfigure}
\begin{subfigure}[b]{0.5\linewidth}
  \centering
  \scalebox{0.4}{\input{gfx/msfem_speedup.pgf}}
\end{subfigure}
  \caption{\label{fig::msfem::speedup} Strong scaling absolute runtimes (left) and speedup (right) for the MsFEM from $1792$ MPI ranks with roughly $500$ coarse
  cells per rank, up to $14336$ ranks with around $60$ cells per rank. Performed on a full 512-node island
  of the Phase 2 partition of the SuperMUC Petascale System in Garching with 28 ranks per node.}
\end{figure}

After applying the projections $ P_{H,h} $ to get $ \tilde{Q}_{h}^{\epsilon}(\Phi_H) $, we
discretize Eq.~(\ref{eq:msfem:local}) which yields a linear system in the standard way. Since this is
a system with degrees of freedom (DoF) distributed across all $P_i$ we need to select an appropriate iterative solver.
Here we use the implementation of the bi--conjugate gradient stabilized method (BiCGSTAB) in \duneistl{},
preconditioned by an Algebraic Multigrid (AMG)
solver, see Section \ref{sec:duneistl}.
We note that the application of the linear solver for the coarse system is the only step in our algorithm that requires
global
communication. This allows the overall algorithm to scale with high parallel efficiency in setups with few
coarse grid cells per rank, where a distributed iterative solver cannot be expected to scale
with its runtime dominated by communication overhead. 
We demonstrate this case in Figure \ref{fig::msfem::speedup}. 

While the \dunems{} module can function as a standalone application to apply the method to a given problem,
it is also possible to use it as a library in other modules as well (see for example \dunemlmc{}\cite{dunemlmc}).
Run-time parameters like the problem to solve, oversampling size, micro and macro grid
resolution, number of threads per process, etc. are read from a structured INI-style file or
passed as a
\cpp{XT::Common::Configuration} object. New problems with
associated data functions and computational domains can easily be added by defining them in a new header file.
The central library routine to apply the method to a given problem, with nulled solution and prepared grid setup is
very concise as it follows the mathematical abstractions discussed above.
\begin{c++}
void Elliptic_MsFEM_Solver::apply(
    DMP::ProblemContainer& problem,
    const CommonTraits::SpaceType& coarse_space,
    std::unique_ptr<LocalsolutionProxy>& solution,
    LocalGridList& localgrid_list) const
{
  CommonTraits::DiscreteFunctionType coarse_msfem_solution(coarse_space, "Coarse Part MsFEM Solution");
  LocalProblemSolver(problem, coarse_space, localgrid_list).solve_for_all_cells();
  CoarseScaleOperator elliptic_msfem_op(problem, coarse_space, localgrid_list);
  elliptic_msfem_op.apply_inverse(coarse_msfem_solution);
  //! projection and summation
  identify_fine_scale_part(problem, localgrid_list, coarse_msfem_solution, coarse_space, solution);
  solution->add(coarse_msfem_solution);
}
\end{c++}

%% file: gfx/msfem_speedup.pgf
\begingroup%
\makeatletter%
\begin{pgfpicture}%
\pgfpathrectangle{\pgfpointorigin}{\pgfqpoint{5.590641in}{3.396312in}}%
\pgfusepath{use as bounding box, clip}%
\begin{pgfscope}%
\pgfsetbuttcap%
\pgfsetmiterjoin%
\definecolor{currentfill}{rgb}{1.000000,1.000000,1.000000}%
\pgfsetfillcolor{currentfill}%
\pgfsetlinewidth{0.000000pt}%
\definecolor{currentstroke}{rgb}{1.000000,1.000000,1.000000}%
\pgfsetstrokecolor{currentstroke}%
\pgfsetdash{}{0pt}%
\pgfpathmoveto{\pgfqpoint{0.000000in}{0.000000in}}%
\pgfpathlineto{\pgfqpoint{5.590641in}{0.000000in}}%
\pgfpathlineto{\pgfqpoint{5.590641in}{3.396312in}}%
\pgfpathlineto{\pgfqpoint{0.000000in}{3.396312in}}%
\pgfpathclose%
\pgfusepath{fill}%
\end{pgfscope}%
\begin{pgfscope}%
\pgfsetbuttcap%
\pgfsetmiterjoin%
\definecolor{currentfill}{rgb}{1.000000,1.000000,1.000000}%
\pgfsetfillcolor{currentfill}%
\pgfsetlinewidth{0.000000pt}%
\definecolor{currentstroke}{rgb}{0.000000,0.000000,0.000000}%
\pgfsetstrokecolor{currentstroke}%
\pgfsetstrokeopacity{0.000000}%
\pgfsetdash{}{0pt}%
\pgfpathmoveto{\pgfqpoint{0.608141in}{0.601312in}}%
\pgfpathlineto{\pgfqpoint{5.490641in}{0.601312in}}%
\pgfpathlineto{\pgfqpoint{5.490641in}{3.296312in}}%
\pgfpathlineto{\pgfqpoint{0.608141in}{3.296312in}}%
\pgfpathclose%
\pgfusepath{fill}%
\end{pgfscope}%
\begin{pgfscope}%
\pgfsetbuttcap%
\pgfsetroundjoin%
\definecolor{currentfill}{rgb}{0.000000,0.000000,0.000000}%
\pgfsetfillcolor{currentfill}%
\pgfsetlinewidth{0.803000pt}%
\definecolor{currentstroke}{rgb}{0.000000,0.000000,0.000000}%
\pgfsetstrokecolor{currentstroke}%
\pgfsetdash{}{0pt}%
\pgfsys@defobject{currentmarker}{\pgfqpoint{0.000000in}{-0.048611in}}{\pgfqpoint{0.000000in}{0.000000in}}{%
\pgfpathmoveto{\pgfqpoint{0.000000in}{0.000000in}}%
\pgfpathlineto{\pgfqpoint{0.000000in}{-0.048611in}}%
\pgfusepath{stroke,fill}%
}%
\begin{pgfscope}%
\pgfsys@transformshift{0.830073in}{0.601312in}%
\pgfsys@useobject{currentmarker}{}%
\end{pgfscope}%
\end{pgfscope}%
\begin{pgfscope}%
\definecolor{textcolor}{rgb}{0.000000,0.000000,0.000000}%
\pgfsetstrokecolor{textcolor}%
\pgfsetfillcolor{textcolor}%
\pgftext[x=0.830073in,y=0.504089in,,top]{\color{textcolor}\sffamily\fontsize{11.000000}{13.200000}\selectfont 1792}%
\end{pgfscope}%
\begin{pgfscope}%
\pgfsetbuttcap%
\pgfsetroundjoin%
\definecolor{currentfill}{rgb}{0.000000,0.000000,0.000000}%
\pgfsetfillcolor{currentfill}%
\pgfsetlinewidth{0.803000pt}%
\definecolor{currentstroke}{rgb}{0.000000,0.000000,0.000000}%
\pgfsetstrokecolor{currentstroke}%
\pgfsetdash{}{0pt}%
\pgfsys@defobject{currentmarker}{\pgfqpoint{0.000000in}{-0.048611in}}{\pgfqpoint{0.000000in}{0.000000in}}{%
\pgfpathmoveto{\pgfqpoint{0.000000in}{0.000000in}}%
\pgfpathlineto{\pgfqpoint{0.000000in}{-0.048611in}}%
\pgfusepath{stroke,fill}%
}%
\begin{pgfscope}%
\pgfsys@transformshift{2.309619in}{0.601312in}%
\pgfsys@useobject{currentmarker}{}%
\end{pgfscope}%
\end{pgfscope}%
\begin{pgfscope}%
\definecolor{textcolor}{rgb}{0.000000,0.000000,0.000000}%
\pgfsetstrokecolor{textcolor}%
\pgfsetfillcolor{textcolor}%
\pgftext[x=2.309619in,y=0.504089in,,top]{\color{textcolor}\sffamily\fontsize{11.000000}{13.200000}\selectfont 3584}%
\end{pgfscope}%
\begin{pgfscope}%
\pgfsetbuttcap%
\pgfsetroundjoin%
\definecolor{currentfill}{rgb}{0.000000,0.000000,0.000000}%
\pgfsetfillcolor{currentfill}%
\pgfsetlinewidth{0.803000pt}%
\definecolor{currentstroke}{rgb}{0.000000,0.000000,0.000000}%
\pgfsetstrokecolor{currentstroke}%
\pgfsetdash{}{0pt}%
\pgfsys@defobject{currentmarker}{\pgfqpoint{0.000000in}{-0.048611in}}{\pgfqpoint{0.000000in}{0.000000in}}{%
\pgfpathmoveto{\pgfqpoint{0.000000in}{0.000000in}}%
\pgfpathlineto{\pgfqpoint{0.000000in}{-0.048611in}}%
\pgfusepath{stroke,fill}%
}%
\begin{pgfscope}%
\pgfsys@transformshift{3.789164in}{0.601312in}%
\pgfsys@useobject{currentmarker}{}%
\end{pgfscope}%
\end{pgfscope}%
\begin{pgfscope}%
\definecolor{textcolor}{rgb}{0.000000,0.000000,0.000000}%
\pgfsetstrokecolor{textcolor}%
\pgfsetfillcolor{textcolor}%
\pgftext[x=3.789164in,y=0.504089in,,top]{\color{textcolor}\sffamily\fontsize{11.000000}{13.200000}\selectfont 7168}%
\end{pgfscope}%
\begin{pgfscope}%
\pgfsetbuttcap%
\pgfsetroundjoin%
\definecolor{currentfill}{rgb}{0.000000,0.000000,0.000000}%
\pgfsetfillcolor{currentfill}%
\pgfsetlinewidth{0.803000pt}%
\definecolor{currentstroke}{rgb}{0.000000,0.000000,0.000000}%
\pgfsetstrokecolor{currentstroke}%
\pgfsetdash{}{0pt}%
\pgfsys@defobject{currentmarker}{\pgfqpoint{0.000000in}{-0.048611in}}{\pgfqpoint{0.000000in}{0.000000in}}{%
\pgfpathmoveto{\pgfqpoint{0.000000in}{0.000000in}}%
\pgfpathlineto{\pgfqpoint{0.000000in}{-0.048611in}}%
\pgfusepath{stroke,fill}%
}%
\begin{pgfscope}%
\pgfsys@transformshift{5.268709in}{0.601312in}%
\pgfsys@useobject{currentmarker}{}%
\end{pgfscope}%
\end{pgfscope}%
\begin{pgfscope}%
\definecolor{textcolor}{rgb}{0.000000,0.000000,0.000000}%
\pgfsetstrokecolor{textcolor}%
\pgfsetfillcolor{textcolor}%
\pgftext[x=5.268709in,y=0.504089in,,top]{\color{textcolor}\sffamily\fontsize{11.000000}{13.200000}\selectfont 14336}%
\end{pgfscope}%
\begin{pgfscope}%
\definecolor{textcolor}{rgb}{0.000000,0.000000,0.000000}%
\pgfsetstrokecolor{textcolor}%
\pgfsetfillcolor{textcolor}%
\pgftext[x=3.049391in,y=0.313349in,,top]{\color{textcolor}\sffamily\fontsize{16.000000}{19.200000}\selectfont \# Cores}%
\end{pgfscope}%
\begin{pgfscope}%
\pgfsetbuttcap%
\pgfsetroundjoin%
\definecolor{currentfill}{rgb}{0.000000,0.000000,0.000000}%
\pgfsetfillcolor{currentfill}%
\pgfsetlinewidth{0.803000pt}%
\definecolor{currentstroke}{rgb}{0.000000,0.000000,0.000000}%
\pgfsetstrokecolor{currentstroke}%
\pgfsetdash{}{0pt}%
\pgfsys@defobject{currentmarker}{\pgfqpoint{-0.048611in}{0.000000in}}{\pgfqpoint{0.000000in}{0.000000in}}{%
\pgfpathmoveto{\pgfqpoint{0.000000in}{0.000000in}}%
\pgfpathlineto{\pgfqpoint{-0.048611in}{0.000000in}}%
\pgfusepath{stroke,fill}%
}%
\begin{pgfscope}%
\pgfsys@transformshift{0.608141in}{0.723812in}%
\pgfsys@useobject{currentmarker}{}%
\end{pgfscope}%
\end{pgfscope}%
\begin{pgfscope}%
\definecolor{textcolor}{rgb}{0.000000,0.000000,0.000000}%
\pgfsetstrokecolor{textcolor}%
\pgfsetfillcolor{textcolor}%
\pgftext[x=0.368904in, y=0.671005in, left, base]{\color{textcolor}\sffamily\fontsize{11.000000}{13.200000}\selectfont \(\displaystyle 2^{0}\)}%
\end{pgfscope}%
\begin{pgfscope}%
\pgfsetbuttcap%
\pgfsetroundjoin%
\definecolor{currentfill}{rgb}{0.000000,0.000000,0.000000}%
\pgfsetfillcolor{currentfill}%
\pgfsetlinewidth{0.803000pt}%
\definecolor{currentstroke}{rgb}{0.000000,0.000000,0.000000}%
\pgfsetstrokecolor{currentstroke}%
\pgfsetdash{}{0pt}%
\pgfsys@defobject{currentmarker}{\pgfqpoint{-0.048611in}{0.000000in}}{\pgfqpoint{0.000000in}{0.000000in}}{%
\pgfpathmoveto{\pgfqpoint{0.000000in}{0.000000in}}%
\pgfpathlineto{\pgfqpoint{-0.048611in}{0.000000in}}%
\pgfusepath{stroke,fill}%
}%
\begin{pgfscope}%
\pgfsys@transformshift{0.608141in}{1.540135in}%
\pgfsys@useobject{currentmarker}{}%
\end{pgfscope}%
\end{pgfscope}%
\begin{pgfscope}%
\definecolor{textcolor}{rgb}{0.000000,0.000000,0.000000}%
\pgfsetstrokecolor{textcolor}%
\pgfsetfillcolor{textcolor}%
\pgftext[x=0.368904in, y=1.487328in, left, base]{\color{textcolor}\sffamily\fontsize{11.000000}{13.200000}\selectfont \(\displaystyle 2^{1}\)}%
\end{pgfscope}%
\begin{pgfscope}%
\pgfsetbuttcap%
\pgfsetroundjoin%
\definecolor{currentfill}{rgb}{0.000000,0.000000,0.000000}%
\pgfsetfillcolor{currentfill}%
\pgfsetlinewidth{0.803000pt}%
\definecolor{currentstroke}{rgb}{0.000000,0.000000,0.000000}%
\pgfsetstrokecolor{currentstroke}%
\pgfsetdash{}{0pt}%
\pgfsys@defobject{currentmarker}{\pgfqpoint{-0.048611in}{0.000000in}}{\pgfqpoint{0.000000in}{0.000000in}}{%
\pgfpathmoveto{\pgfqpoint{0.000000in}{0.000000in}}%
\pgfpathlineto{\pgfqpoint{-0.048611in}{0.000000in}}%
\pgfusepath{stroke,fill}%
}%
\begin{pgfscope}%
\pgfsys@transformshift{0.608141in}{2.356458in}%
\pgfsys@useobject{currentmarker}{}%
\end{pgfscope}%
\end{pgfscope}%
\begin{pgfscope}%
\definecolor{textcolor}{rgb}{0.000000,0.000000,0.000000}%
\pgfsetstrokecolor{textcolor}%
\pgfsetfillcolor{textcolor}%
\pgftext[x=0.368904in, y=2.303651in, left, base]{\color{textcolor}\sffamily\fontsize{11.000000}{13.200000}\selectfont \(\displaystyle 2^{2}\)}%
\end{pgfscope}%
\begin{pgfscope}%
\pgfsetbuttcap%
\pgfsetroundjoin%
\definecolor{currentfill}{rgb}{0.000000,0.000000,0.000000}%
\pgfsetfillcolor{currentfill}%
\pgfsetlinewidth{0.803000pt}%
\definecolor{currentstroke}{rgb}{0.000000,0.000000,0.000000}%
\pgfsetstrokecolor{currentstroke}%
\pgfsetdash{}{0pt}%
\pgfsys@defobject{currentmarker}{\pgfqpoint{-0.048611in}{0.000000in}}{\pgfqpoint{0.000000in}{0.000000in}}{%
\pgfpathmoveto{\pgfqpoint{0.000000in}{0.000000in}}%
\pgfpathlineto{\pgfqpoint{-0.048611in}{0.000000in}}%
\pgfusepath{stroke,fill}%
}%
\begin{pgfscope}%
\pgfsys@transformshift{0.608141in}{3.172781in}%
\pgfsys@useobject{currentmarker}{}%
\end{pgfscope}%
\end{pgfscope}%
\begin{pgfscope}%
\definecolor{textcolor}{rgb}{0.000000,0.000000,0.000000}%
\pgfsetstrokecolor{textcolor}%
\pgfsetfillcolor{textcolor}%
\pgftext[x=0.368904in, y=3.119975in, left, base]{\color{textcolor}\sffamily\fontsize{11.000000}{13.200000}\selectfont \(\displaystyle 2^{3}\)}%
\end{pgfscope}%
\begin{pgfscope}%
\definecolor{textcolor}{rgb}{0.000000,0.000000,0.000000}%
\pgfsetstrokecolor{textcolor}%
\pgfsetfillcolor{textcolor}%
\pgftext[x=0.313349in,y=1.948812in,,bottom,rotate=90.000000]{\color{textcolor}\sffamily\fontsize{16.000000}{19.200000}\selectfont Speedup}%
\end{pgfscope}%
\begin{pgfscope}%
\pgfpathrectangle{\pgfqpoint{0.608141in}{0.601312in}}{\pgfqpoint{4.882500in}{2.695000in}}%
\pgfusepath{clip}%
\pgfsetrectcap%
\pgfsetroundjoin%
\pgfsetlinewidth{1.505625pt}%
\definecolor{currentstroke}{rgb}{1.000000,0.000000,0.480000}%
\pgfsetstrokecolor{currentstroke}%
\pgfsetdash{}{0pt}%
\pgfpathmoveto{\pgfqpoint{0.830073in}{0.723812in}}%
\pgfpathlineto{\pgfqpoint{2.309619in}{1.498787in}}%
\pgfpathlineto{\pgfqpoint{3.789164in}{2.334167in}}%
\pgfpathlineto{\pgfqpoint{5.268709in}{3.120250in}}%
\pgfusepath{stroke}%
\end{pgfscope}%
\begin{pgfscope}%
\pgfpathrectangle{\pgfqpoint{0.608141in}{0.601312in}}{\pgfqpoint{4.882500in}{2.695000in}}%
\pgfusepath{clip}%
\pgfsetbuttcap%
\pgfsetmiterjoin%
\definecolor{currentfill}{rgb}{1.000000,0.000000,0.480000}%
\pgfsetfillcolor{currentfill}%
\pgfsetlinewidth{1.003750pt}%
\definecolor{currentstroke}{rgb}{1.000000,0.000000,0.480000}%
\pgfsetstrokecolor{currentstroke}%
\pgfsetdash{}{0pt}%
\pgfsys@defobject{currentmarker}{\pgfqpoint{-0.041667in}{-0.041667in}}{\pgfqpoint{0.041667in}{0.041667in}}{%
\pgfpathmoveto{\pgfqpoint{-0.041667in}{-0.041667in}}%
\pgfpathlineto{\pgfqpoint{0.041667in}{-0.041667in}}%
\pgfpathlineto{\pgfqpoint{0.041667in}{0.041667in}}%
\pgfpathlineto{\pgfqpoint{-0.041667in}{0.041667in}}%
\pgfpathclose%
\pgfusepath{stroke,fill}%
}%
\begin{pgfscope}%
\pgfsys@transformshift{0.830073in}{0.723812in}%
\pgfsys@useobject{currentmarker}{}%
\end{pgfscope}%
\begin{pgfscope}%
\pgfsys@transformshift{2.309619in}{1.498787in}%
\pgfsys@useobject{currentmarker}{}%
\end{pgfscope}%
\begin{pgfscope}%
\pgfsys@transformshift{3.789164in}{2.334167in}%
\pgfsys@useobject{currentmarker}{}%
\end{pgfscope}%
\begin{pgfscope}%
\pgfsys@transformshift{5.268709in}{3.120250in}%
\pgfsys@useobject{currentmarker}{}%
\end{pgfscope}%
\end{pgfscope}%
\begin{pgfscope}%
\pgfpathrectangle{\pgfqpoint{0.608141in}{0.601312in}}{\pgfqpoint{4.882500in}{2.695000in}}%
\pgfusepath{clip}%
\pgfsetrectcap%
\pgfsetroundjoin%
\pgfsetlinewidth{1.505625pt}%
\definecolor{currentstroke}{rgb}{0.510000,0.500000,1.000000}%
\pgfsetstrokecolor{currentstroke}%
\pgfsetdash{}{0pt}%
\pgfpathmoveto{\pgfqpoint{0.830073in}{0.723812in}}%
\pgfpathlineto{\pgfqpoint{2.309619in}{1.655590in}}%
\pgfpathlineto{\pgfqpoint{3.789164in}{1.580650in}}%
\pgfpathlineto{\pgfqpoint{5.268709in}{1.699602in}}%
\pgfusepath{stroke}%
\end{pgfscope}%
\begin{pgfscope}%
\pgfpathrectangle{\pgfqpoint{0.608141in}{0.601312in}}{\pgfqpoint{4.882500in}{2.695000in}}%
\pgfusepath{clip}%
\pgfsetbuttcap%
\pgfsetroundjoin%
\definecolor{currentfill}{rgb}{0.510000,0.500000,1.000000}%
\pgfsetfillcolor{currentfill}%
\pgfsetlinewidth{1.003750pt}%
\definecolor{currentstroke}{rgb}{0.510000,0.500000,1.000000}%
\pgfsetstrokecolor{currentstroke}%
\pgfsetdash{}{0pt}%
\pgfsys@defobject{currentmarker}{\pgfqpoint{-0.041667in}{-0.041667in}}{\pgfqpoint{0.041667in}{0.041667in}}{%
\pgfpathmoveto{\pgfqpoint{0.000000in}{-0.041667in}}%
\pgfpathcurveto{\pgfqpoint{0.011050in}{-0.041667in}}{\pgfqpoint{0.021649in}{-0.037276in}}{\pgfqpoint{0.029463in}{-0.029463in}}%
\pgfpathcurveto{\pgfqpoint{0.037276in}{-0.021649in}}{\pgfqpoint{0.041667in}{-0.011050in}}{\pgfqpoint{0.041667in}{0.000000in}}%
\pgfpathcurveto{\pgfqpoint{0.041667in}{0.011050in}}{\pgfqpoint{0.037276in}{0.021649in}}{\pgfqpoint{0.029463in}{0.029463in}}%
\pgfpathcurveto{\pgfqpoint{0.021649in}{0.037276in}}{\pgfqpoint{0.011050in}{0.041667in}}{\pgfqpoint{0.000000in}{0.041667in}}%
\pgfpathcurveto{\pgfqpoint{-0.011050in}{0.041667in}}{\pgfqpoint{-0.021649in}{0.037276in}}{\pgfqpoint{-0.029463in}{0.029463in}}%
\pgfpathcurveto{\pgfqpoint{-0.037276in}{0.021649in}}{\pgfqpoint{-0.041667in}{0.011050in}}{\pgfqpoint{-0.041667in}{0.000000in}}%
\pgfpathcurveto{\pgfqpoint{-0.041667in}{-0.011050in}}{\pgfqpoint{-0.037276in}{-0.021649in}}{\pgfqpoint{-0.029463in}{-0.029463in}}%
\pgfpathcurveto{\pgfqpoint{-0.021649in}{-0.037276in}}{\pgfqpoint{-0.011050in}{-0.041667in}}{\pgfqpoint{0.000000in}{-0.041667in}}%
\pgfpathclose%
\pgfusepath{stroke,fill}%
}%
\begin{pgfscope}%
\pgfsys@transformshift{0.830073in}{0.723812in}%
\pgfsys@useobject{currentmarker}{}%
\end{pgfscope}%
\begin{pgfscope}%
\pgfsys@transformshift{2.309619in}{1.655590in}%
\pgfsys@useobject{currentmarker}{}%
\end{pgfscope}%
\begin{pgfscope}%
\pgfsys@transformshift{3.789164in}{1.580650in}%
\pgfsys@useobject{currentmarker}{}%
\end{pgfscope}%
\begin{pgfscope}%
\pgfsys@transformshift{5.268709in}{1.699602in}%
\pgfsys@useobject{currentmarker}{}%
\end{pgfscope}%
\end{pgfscope}%
\begin{pgfscope}%
\pgfpathrectangle{\pgfqpoint{0.608141in}{0.601312in}}{\pgfqpoint{4.882500in}{2.695000in}}%
\pgfusepath{clip}%
\pgfsetrectcap%
\pgfsetroundjoin%
\pgfsetlinewidth{1.505625pt}%
\definecolor{currentstroke}{rgb}{0.000000,0.500000,0.260000}%
\pgfsetstrokecolor{currentstroke}%
\pgfsetdash{}{0pt}%
\pgfpathmoveto{\pgfqpoint{0.830073in}{0.723812in}}%
\pgfpathlineto{\pgfqpoint{2.309619in}{1.541312in}}%
\pgfpathlineto{\pgfqpoint{3.789164in}{2.354985in}}%
\pgfpathlineto{\pgfqpoint{5.268709in}{3.166582in}}%
\pgfusepath{stroke}%
\end{pgfscope}%
\begin{pgfscope}%
\pgfpathrectangle{\pgfqpoint{0.608141in}{0.601312in}}{\pgfqpoint{4.882500in}{2.695000in}}%
\pgfusepath{clip}%
\pgfsetbuttcap%
\pgfsetmiterjoin%
\definecolor{currentfill}{rgb}{0.000000,0.500000,0.260000}%
\pgfsetfillcolor{currentfill}%
\pgfsetlinewidth{1.003750pt}%
\definecolor{currentstroke}{rgb}{0.000000,0.500000,0.260000}%
\pgfsetstrokecolor{currentstroke}%
\pgfsetdash{}{0pt}%
\pgfsys@defobject{currentmarker}{\pgfqpoint{0.000000in}{-0.041667in}}{\pgfqpoint{0.062500in}{0.041667in}}{%
\pgfpathmoveto{\pgfqpoint{0.062500in}{0.041667in}}%
\pgfpathlineto{\pgfqpoint{0.000000in}{0.000000in}}%
\pgfpathlineto{\pgfqpoint{0.062500in}{-0.041667in}}%
\pgfusepath{stroke,fill}%
}%
\begin{pgfscope}%
\pgfsys@transformshift{0.830073in}{0.723812in}%
\pgfsys@useobject{currentmarker}{}%
\end{pgfscope}%
\begin{pgfscope}%
\pgfsys@transformshift{2.309619in}{1.541312in}%
\pgfsys@useobject{currentmarker}{}%
\end{pgfscope}%
\begin{pgfscope}%
\pgfsys@transformshift{3.789164in}{2.354985in}%
\pgfsys@useobject{currentmarker}{}%
\end{pgfscope}%
\begin{pgfscope}%
\pgfsys@transformshift{5.268709in}{3.166582in}%
\pgfsys@useobject{currentmarker}{}%
\end{pgfscope}%
\end{pgfscope}%
\begin{pgfscope}%
\pgfpathrectangle{\pgfqpoint{0.608141in}{0.601312in}}{\pgfqpoint{4.882500in}{2.695000in}}%
\pgfusepath{clip}%
\pgfsetrectcap%
\pgfsetroundjoin%
\pgfsetlinewidth{1.505625pt}%
\definecolor{currentstroke}{rgb}{0.000000,0.730000,1.000000}%
\pgfsetstrokecolor{currentstroke}%
\pgfsetdash{}{0pt}%
\pgfpathmoveto{\pgfqpoint{0.830073in}{0.723812in}}%
\pgfpathlineto{\pgfqpoint{2.309619in}{1.539546in}}%
\pgfpathlineto{\pgfqpoint{3.789164in}{2.356164in}}%
\pgfpathlineto{\pgfqpoint{5.268709in}{3.173812in}}%
\pgfusepath{stroke}%
\end{pgfscope}%
\begin{pgfscope}%
\pgfpathrectangle{\pgfqpoint{0.608141in}{0.601312in}}{\pgfqpoint{4.882500in}{2.695000in}}%
\pgfusepath{clip}%
\pgfsetbuttcap%
\pgfsetmiterjoin%
\definecolor{currentfill}{rgb}{0.000000,0.730000,1.000000}%
\pgfsetfillcolor{currentfill}%
\pgfsetlinewidth{1.003750pt}%
\definecolor{currentstroke}{rgb}{0.000000,0.730000,1.000000}%
\pgfsetstrokecolor{currentstroke}%
\pgfsetdash{}{0pt}%
\pgfsys@defobject{currentmarker}{\pgfqpoint{-0.062500in}{-0.041667in}}{\pgfqpoint{0.000000in}{0.041667in}}{%
\pgfpathmoveto{\pgfqpoint{-0.062500in}{-0.041667in}}%
\pgfpathlineto{\pgfqpoint{0.000000in}{0.000000in}}%
\pgfpathlineto{\pgfqpoint{-0.062500in}{0.041667in}}%
\pgfusepath{stroke,fill}%
}%
\begin{pgfscope}%
\pgfsys@transformshift{0.830073in}{0.723812in}%
\pgfsys@useobject{currentmarker}{}%
\end{pgfscope}%
\begin{pgfscope}%
\pgfsys@transformshift{2.309619in}{1.539546in}%
\pgfsys@useobject{currentmarker}{}%
\end{pgfscope}%
\begin{pgfscope}%
\pgfsys@transformshift{3.789164in}{2.356164in}%
\pgfsys@useobject{currentmarker}{}%
\end{pgfscope}%
\begin{pgfscope}%
\pgfsys@transformshift{5.268709in}{3.173812in}%
\pgfsys@useobject{currentmarker}{}%
\end{pgfscope}%
\end{pgfscope}%
\begin{pgfscope}%
\pgfpathrectangle{\pgfqpoint{0.608141in}{0.601312in}}{\pgfqpoint{4.882500in}{2.695000in}}%
\pgfusepath{clip}%
\pgfsetbuttcap%
\pgfsetroundjoin%
\pgfsetlinewidth{1.505625pt}%
\definecolor{currentstroke}{rgb}{0.385000,0.000000,0.500000}%
\pgfsetstrokecolor{currentstroke}%
\pgfsetdash{{5.550000pt}{2.400000pt}}{0.000000pt}%
\pgfpathmoveto{\pgfqpoint{0.830073in}{0.723812in}}%
\pgfpathlineto{\pgfqpoint{2.309619in}{1.540135in}}%
\pgfpathlineto{\pgfqpoint{3.789164in}{2.356458in}}%
\pgfpathlineto{\pgfqpoint{5.268709in}{3.172781in}}%
\pgfusepath{stroke}%
\end{pgfscope}%
\begin{pgfscope}%
\pgfpathrectangle{\pgfqpoint{0.608141in}{0.601312in}}{\pgfqpoint{4.882500in}{2.695000in}}%
\pgfusepath{clip}%
\pgfsetbuttcap%
\pgfsetmiterjoin%
\definecolor{currentfill}{rgb}{0.385000,0.000000,0.500000}%
\pgfsetfillcolor{currentfill}%
\pgfsetlinewidth{1.003750pt}%
\definecolor{currentstroke}{rgb}{0.385000,0.000000,0.500000}%
\pgfsetstrokecolor{currentstroke}%
\pgfsetdash{}{0pt}%
\pgfsys@defobject{currentmarker}{\pgfqpoint{-0.041667in}{0.000000in}}{\pgfqpoint{0.041667in}{0.062500in}}{%
\pgfpathmoveto{\pgfqpoint{-0.041667in}{0.062500in}}%
\pgfpathlineto{\pgfqpoint{0.000000in}{0.000000in}}%
\pgfpathlineto{\pgfqpoint{0.041667in}{0.062500in}}%
\pgfusepath{stroke,fill}%
}%
\begin{pgfscope}%
\pgfsys@transformshift{0.830073in}{0.723812in}%
\pgfsys@useobject{currentmarker}{}%
\end{pgfscope}%
\begin{pgfscope}%
\pgfsys@transformshift{2.309619in}{1.540135in}%
\pgfsys@useobject{currentmarker}{}%
\end{pgfscope}%
\begin{pgfscope}%
\pgfsys@transformshift{3.789164in}{2.356458in}%
\pgfsys@useobject{currentmarker}{}%
\end{pgfscope}%
\begin{pgfscope}%
\pgfsys@transformshift{5.268709in}{3.172781in}%
\pgfsys@useobject{currentmarker}{}%
\end{pgfscope}%
\end{pgfscope}%
\begin{pgfscope}%
\pgfsetrectcap%
\pgfsetmiterjoin%
\pgfsetlinewidth{0.803000pt}%
\definecolor{currentstroke}{rgb}{0.000000,0.000000,0.000000}%
\pgfsetstrokecolor{currentstroke}%
\pgfsetdash{}{0pt}%
\pgfpathmoveto{\pgfqpoint{0.608141in}{0.601312in}}%
\pgfpathlineto{\pgfqpoint{0.608141in}{3.296312in}}%
\pgfusepath{stroke}%
\end{pgfscope}%
\begin{pgfscope}%
\pgfsetrectcap%
\pgfsetmiterjoin%
\pgfsetlinewidth{0.803000pt}%
\definecolor{currentstroke}{rgb}{0.000000,0.000000,0.000000}%
\pgfsetstrokecolor{currentstroke}%
\pgfsetdash{}{0pt}%
\pgfpathmoveto{\pgfqpoint{5.490641in}{0.601312in}}%
\pgfpathlineto{\pgfqpoint{5.490641in}{3.296312in}}%
\pgfusepath{stroke}%
\end{pgfscope}%
\begin{pgfscope}%
\pgfsetrectcap%
\pgfsetmiterjoin%
\pgfsetlinewidth{0.803000pt}%
\definecolor{currentstroke}{rgb}{0.000000,0.000000,0.000000}%
\pgfsetstrokecolor{currentstroke}%
\pgfsetdash{}{0pt}%
\pgfpathmoveto{\pgfqpoint{0.608141in}{0.601312in}}%
\pgfpathlineto{\pgfqpoint{5.490641in}{0.601312in}}%
\pgfusepath{stroke}%
\end{pgfscope}%
\begin{pgfscope}%
\pgfsetrectcap%
\pgfsetmiterjoin%
\pgfsetlinewidth{0.803000pt}%
\definecolor{currentstroke}{rgb}{0.000000,0.000000,0.000000}%
\pgfsetstrokecolor{currentstroke}%
\pgfsetdash{}{0pt}%
\pgfpathmoveto{\pgfqpoint{0.608141in}{3.296312in}}%
\pgfpathlineto{\pgfqpoint{5.490641in}{3.296312in}}%
\pgfusepath{stroke}%
\end{pgfscope}%
\begin{pgfscope}%
\pgfsetbuttcap%
\pgfsetmiterjoin%
\definecolor{currentfill}{rgb}{1.000000,1.000000,1.000000}%
\pgfsetfillcolor{currentfill}%
\pgfsetfillopacity{0.800000}%
\pgfsetlinewidth{1.003750pt}%
\definecolor{currentstroke}{rgb}{0.800000,0.800000,0.800000}%
\pgfsetstrokecolor{currentstroke}%
\pgfsetstrokeopacity{0.800000}%
\pgfsetdash{}{0pt}%
\pgfpathmoveto{\pgfqpoint{0.715086in}{2.109564in}}%
\pgfpathlineto{\pgfqpoint{2.685521in}{2.109564in}}%
\pgfpathquadraticcurveto{\pgfqpoint{2.716077in}{2.109564in}}{\pgfqpoint{2.716077in}{2.140119in}}%
\pgfpathlineto{\pgfqpoint{2.716077in}{3.189367in}}%
\pgfpathquadraticcurveto{\pgfqpoint{2.716077in}{3.219923in}}{\pgfqpoint{2.685521in}{3.219923in}}%
\pgfpathlineto{\pgfqpoint{0.715086in}{3.219923in}}%
\pgfpathquadraticcurveto{\pgfqpoint{0.684530in}{3.219923in}}{\pgfqpoint{0.684530in}{3.189367in}}%
\pgfpathlineto{\pgfqpoint{0.684530in}{2.140119in}}%
\pgfpathquadraticcurveto{\pgfqpoint{0.684530in}{2.109564in}}{\pgfqpoint{0.715086in}{2.109564in}}%
\pgfpathclose%
\pgfusepath{stroke,fill}%
\end{pgfscope}%
\begin{pgfscope}%
\pgfsetrectcap%
\pgfsetroundjoin%
\pgfsetlinewidth{1.505625pt}%
\definecolor{currentstroke}{rgb}{1.000000,0.000000,0.480000}%
\pgfsetstrokecolor{currentstroke}%
\pgfsetdash{}{0pt}%
\pgfpathmoveto{\pgfqpoint{0.745641in}{3.105339in}}%
\pgfpathlineto{\pgfqpoint{1.051197in}{3.105339in}}%
\pgfusepath{stroke}%
\end{pgfscope}%
\begin{pgfscope}%
\pgfsetbuttcap%
\pgfsetmiterjoin%
\definecolor{currentfill}{rgb}{1.000000,0.000000,0.480000}%
\pgfsetfillcolor{currentfill}%
\pgfsetlinewidth{1.003750pt}%
\definecolor{currentstroke}{rgb}{1.000000,0.000000,0.480000}%
\pgfsetstrokecolor{currentstroke}%
\pgfsetdash{}{0pt}%
\pgfsys@defobject{currentmarker}{\pgfqpoint{-0.041667in}{-0.041667in}}{\pgfqpoint{0.041667in}{0.041667in}}{%
\pgfpathmoveto{\pgfqpoint{-0.041667in}{-0.041667in}}%
\pgfpathlineto{\pgfqpoint{0.041667in}{-0.041667in}}%
\pgfpathlineto{\pgfqpoint{0.041667in}{0.041667in}}%
\pgfpathlineto{\pgfqpoint{-0.041667in}{0.041667in}}%
\pgfpathclose%
\pgfusepath{stroke,fill}%
}%
\begin{pgfscope}%
\pgfsys@transformshift{0.898419in}{3.105339in}%
\pgfsys@useobject{currentmarker}{}%
\end{pgfscope}%
\end{pgfscope}%
\begin{pgfscope}%
\definecolor{textcolor}{rgb}{0.000000,0.000000,0.000000}%
\pgfsetstrokecolor{textcolor}%
\pgfsetfillcolor{textcolor}%
\pgftext[x=1.173419in,y=3.051867in,left,base]{\color{textcolor}\sffamily\fontsize{11.000000}{13.200000}\selectfont Overall}%
\end{pgfscope}%
\begin{pgfscope}%
\pgfsetrectcap%
\pgfsetroundjoin%
\pgfsetlinewidth{1.505625pt}%
\definecolor{currentstroke}{rgb}{0.510000,0.500000,1.000000}%
\pgfsetstrokecolor{currentstroke}%
\pgfsetdash{}{0pt}%
\pgfpathmoveto{\pgfqpoint{0.745641in}{2.892434in}}%
\pgfpathlineto{\pgfqpoint{1.051197in}{2.892434in}}%
\pgfusepath{stroke}%
\end{pgfscope}%
\begin{pgfscope}%
\pgfsetbuttcap%
\pgfsetroundjoin%
\definecolor{currentfill}{rgb}{0.510000,0.500000,1.000000}%
\pgfsetfillcolor{currentfill}%
\pgfsetlinewidth{1.003750pt}%
\definecolor{currentstroke}{rgb}{0.510000,0.500000,1.000000}%
\pgfsetstrokecolor{currentstroke}%
\pgfsetdash{}{0pt}%
\pgfsys@defobject{currentmarker}{\pgfqpoint{-0.041667in}{-0.041667in}}{\pgfqpoint{0.041667in}{0.041667in}}{%
\pgfpathmoveto{\pgfqpoint{0.000000in}{-0.041667in}}%
\pgfpathcurveto{\pgfqpoint{0.011050in}{-0.041667in}}{\pgfqpoint{0.021649in}{-0.037276in}}{\pgfqpoint{0.029463in}{-0.029463in}}%
\pgfpathcurveto{\pgfqpoint{0.037276in}{-0.021649in}}{\pgfqpoint{0.041667in}{-0.011050in}}{\pgfqpoint{0.041667in}{0.000000in}}%
\pgfpathcurveto{\pgfqpoint{0.041667in}{0.011050in}}{\pgfqpoint{0.037276in}{0.021649in}}{\pgfqpoint{0.029463in}{0.029463in}}%
\pgfpathcurveto{\pgfqpoint{0.021649in}{0.037276in}}{\pgfqpoint{0.011050in}{0.041667in}}{\pgfqpoint{0.000000in}{0.041667in}}%
\pgfpathcurveto{\pgfqpoint{-0.011050in}{0.041667in}}{\pgfqpoint{-0.021649in}{0.037276in}}{\pgfqpoint{-0.029463in}{0.029463in}}%
\pgfpathcurveto{\pgfqpoint{-0.037276in}{0.021649in}}{\pgfqpoint{-0.041667in}{0.011050in}}{\pgfqpoint{-0.041667in}{0.000000in}}%
\pgfpathcurveto{\pgfqpoint{-0.041667in}{-0.011050in}}{\pgfqpoint{-0.037276in}{-0.021649in}}{\pgfqpoint{-0.029463in}{-0.029463in}}%
\pgfpathcurveto{\pgfqpoint{-0.021649in}{-0.037276in}}{\pgfqpoint{-0.011050in}{-0.041667in}}{\pgfqpoint{0.000000in}{-0.041667in}}%
\pgfpathclose%
\pgfusepath{stroke,fill}%
}%
\begin{pgfscope}%
\pgfsys@transformshift{0.898419in}{2.892434in}%
\pgfsys@useobject{currentmarker}{}%
\end{pgfscope}%
\end{pgfscope}%
\begin{pgfscope}%
\definecolor{textcolor}{rgb}{0.000000,0.000000,0.000000}%
\pgfsetstrokecolor{textcolor}%
\pgfsetfillcolor{textcolor}%
\pgftext[x=1.173419in,y=2.838962in,left,base]{\color{textcolor}\sffamily\fontsize{11.000000}{13.200000}\selectfont Coarse solve}%
\end{pgfscope}%
\begin{pgfscope}%
\pgfsetrectcap%
\pgfsetroundjoin%
\pgfsetlinewidth{1.505625pt}%
\definecolor{currentstroke}{rgb}{0.000000,0.500000,0.260000}%
\pgfsetstrokecolor{currentstroke}%
\pgfsetdash{}{0pt}%
\pgfpathmoveto{\pgfqpoint{0.745641in}{2.679529in}}%
\pgfpathlineto{\pgfqpoint{1.051197in}{2.679529in}}%
\pgfusepath{stroke}%
\end{pgfscope}%
\begin{pgfscope}%
\pgfsetbuttcap%
\pgfsetmiterjoin%
\definecolor{currentfill}{rgb}{0.000000,0.500000,0.260000}%
\pgfsetfillcolor{currentfill}%
\pgfsetlinewidth{1.003750pt}%
\definecolor{currentstroke}{rgb}{0.000000,0.500000,0.260000}%
\pgfsetstrokecolor{currentstroke}%
\pgfsetdash{}{0pt}%
\pgfsys@defobject{currentmarker}{\pgfqpoint{0.000000in}{-0.041667in}}{\pgfqpoint{0.062500in}{0.041667in}}{%
\pgfpathmoveto{\pgfqpoint{0.062500in}{0.041667in}}%
\pgfpathlineto{\pgfqpoint{0.000000in}{0.000000in}}%
\pgfpathlineto{\pgfqpoint{0.062500in}{-0.041667in}}%
\pgfusepath{stroke,fill}%
}%
\begin{pgfscope}%
\pgfsys@transformshift{0.898419in}{2.679529in}%
\pgfsys@useobject{currentmarker}{}%
\end{pgfscope}%
\end{pgfscope}%
\begin{pgfscope}%
\definecolor{textcolor}{rgb}{0.000000,0.000000,0.000000}%
\pgfsetstrokecolor{textcolor}%
\pgfsetfillcolor{textcolor}%
\pgftext[x=1.173419in,y=2.626057in,left,base]{\color{textcolor}\sffamily\fontsize{11.000000}{13.200000}\selectfont Local assembly + solve}%
\end{pgfscope}%
\begin{pgfscope}%
\pgfsetrectcap%
\pgfsetroundjoin%
\pgfsetlinewidth{1.505625pt}%
\definecolor{currentstroke}{rgb}{0.000000,0.730000,1.000000}%
\pgfsetstrokecolor{currentstroke}%
\pgfsetdash{}{0pt}%
\pgfpathmoveto{\pgfqpoint{0.745641in}{2.466624in}}%
\pgfpathlineto{\pgfqpoint{1.051197in}{2.466624in}}%
\pgfusepath{stroke}%
\end{pgfscope}%
\begin{pgfscope}%
\pgfsetbuttcap%
\pgfsetmiterjoin%
\definecolor{currentfill}{rgb}{0.000000,0.730000,1.000000}%
\pgfsetfillcolor{currentfill}%
\pgfsetlinewidth{1.003750pt}%
\definecolor{currentstroke}{rgb}{0.000000,0.730000,1.000000}%
\pgfsetstrokecolor{currentstroke}%
\pgfsetdash{}{0pt}%
\pgfsys@defobject{currentmarker}{\pgfqpoint{-0.062500in}{-0.041667in}}{\pgfqpoint{0.000000in}{0.041667in}}{%
\pgfpathmoveto{\pgfqpoint{-0.062500in}{-0.041667in}}%
\pgfpathlineto{\pgfqpoint{0.000000in}{0.000000in}}%
\pgfpathlineto{\pgfqpoint{-0.062500in}{0.041667in}}%
\pgfusepath{stroke,fill}%
}%
\begin{pgfscope}%
\pgfsys@transformshift{0.898419in}{2.466624in}%
\pgfsys@useobject{currentmarker}{}%
\end{pgfscope}%
\end{pgfscope}%
\begin{pgfscope}%
\definecolor{textcolor}{rgb}{0.000000,0.000000,0.000000}%
\pgfsetstrokecolor{textcolor}%
\pgfsetfillcolor{textcolor}%
\pgftext[x=1.173419in,y=2.413152in,left,base]{\color{textcolor}\sffamily\fontsize{11.000000}{13.200000}\selectfont Coarse assembly}%
\end{pgfscope}%
\begin{pgfscope}%
\pgfsetbuttcap%
\pgfsetroundjoin%
\pgfsetlinewidth{1.505625pt}%
\definecolor{currentstroke}{rgb}{0.385000,0.000000,0.500000}%
\pgfsetstrokecolor{currentstroke}%
\pgfsetdash{{5.550000pt}{2.400000pt}}{0.000000pt}%
\pgfpathmoveto{\pgfqpoint{0.745641in}{2.253719in}}%
\pgfpathlineto{\pgfqpoint{1.051197in}{2.253719in}}%
\pgfusepath{stroke}%
\end{pgfscope}%
\begin{pgfscope}%
\pgfsetbuttcap%
\pgfsetmiterjoin%
\definecolor{currentfill}{rgb}{0.385000,0.000000,0.500000}%
\pgfsetfillcolor{currentfill}%
\pgfsetlinewidth{1.003750pt}%
\definecolor{currentstroke}{rgb}{0.385000,0.000000,0.500000}%
\pgfsetstrokecolor{currentstroke}%
\pgfsetdash{}{0pt}%
\pgfsys@defobject{currentmarker}{\pgfqpoint{-0.041667in}{0.000000in}}{\pgfqpoint{0.041667in}{0.062500in}}{%
\pgfpathmoveto{\pgfqpoint{-0.041667in}{0.062500in}}%
\pgfpathlineto{\pgfqpoint{0.000000in}{0.000000in}}%
\pgfpathlineto{\pgfqpoint{0.041667in}{0.062500in}}%
\pgfusepath{stroke,fill}%
}%
\begin{pgfscope}%
\pgfsys@transformshift{0.898419in}{2.253719in}%
\pgfsys@useobject{currentmarker}{}%
\end{pgfscope}%
\end{pgfscope}%
\begin{pgfscope}%
\definecolor{textcolor}{rgb}{0.000000,0.000000,0.000000}%
\pgfsetstrokecolor{textcolor}%
\pgfsetfillcolor{textcolor}%
\pgftext[x=1.173419in,y=2.200247in,left,base]{\color{textcolor}\sffamily\fontsize{11.000000}{13.200000}\selectfont Ideal}%
\end{pgfscope}%
\end{pgfpicture}%
\makeatother%
\endgroup%

%% file: hpc.tex
\subsection{Sum-factorization for high order discretizations to improve node level performance}
\label{sec:hpc}


In this last example we showcase how \dune{} is used to develop HPC simulation code for modern hardware architectures.
We discuss some prevalent trends in hardware development and how they affect finite element software.
Then a matrix-free solution technique for high order discretizations is presented and 
its node level performance on recent architectures is shown.
This work was implemented in \dunepdelab and was originally developed within the \exadune project.
The complexity of the performance engineering efforts have let to a reimplementation and continued development in \dunecodegen .

With the end of frequency scaling, performance increases on current hardware rely on an ever-growing amount of
parallelism in modern architectures.
This includes a drastic increase of CPU floating point performance through instruction level parallelism (SIMD vectorization, superscalar execution, fused multiplication and addition).
However, memory bandwidth has not kept up with these gains, severely restricting the performance of established numerical
codes and leaving them unable to saturate the floating point hardware.
Developers need to both reconsider their choice of algorithms, as well as adapt their implementations in order to overcome this barrier.
E.g. in traditional FEM implementations, the system matrix is assembled in memory and
the sparse linear system is solved with efficient solvers based on sparse matrices.
Optimal complexity solvers scale linearly in the number of unknowns.
Despite their optimal complexity, these schemes cannot leverage the capabilities of modern HPC systems as they rely on sparse matrix
vector products of the assembled system matrix, which have very low arithmetic intensity and are therefore inherently
memory-bound.

One possible approach to leverage the capabilities of current hardware is to directly implement the application of the
sparse matrix on a vector. This direct implementation shifts the arithmetic intensity into the compute-bound regime of
modern CPUs.
Other software projects are pursuing similar ideas for highly performant simulation codes on modern architectures, e.g. libceed \cite{fischer2020scalability} and deal.ii \cite{kronbichler2019matrixfree}.
Given an optimal complexity algorithm on suitable discretizations, it is possible to compute the matrix-vector product faster than the entries of an assembled system matrix can be loaded from main memory.
Such optimal complexity algorithms make use of a technique called sum factorization \cite{orszag1980spectral} which
exploits the tensor product structure of finite element basis functions and quadrature formulae.
Given polynomial degree $k$ and minimal quadrature order, it allows to reduce the computational complexity of one operator application from $\mathcal{O}(k^{2d})$ to $\mathcal{O}(k^{d+1})$ by rewriting the evaluation of finite element functions as a $d$ sequence of tensor contractions.
To compute local contributions of the operator it is necessary to have
access to the 1d shape functions and quadrature rule that was used in
the tensor-product construction of the 2d or 3d variants. Although
this optimal complexity algorithm can not use 3d shape functions, the implementation is still hard-coded, but uses
1d shape functions from
\dunelocalfunctions. By this the implementation can still be fairly
generic and easily switch between different polynomial degrees and
polynomial representation (e.g. Lagrange- or Legendre-Polynomials).

In order to fully exploit the CPU's floating point capabilities, an implementation needs to maximize its use of SIMD instructions.
In our experience, general purpose compilers are not capable to sufficiently autovectorize this type of code, especially
as the loop bounds of tensor contractions depend on the polynomial degree $k$ and are thus not necessarily an integer
multiple of the SIMD width.
Explicit SIMD vectorization is a challenging task that requires both an algorithmic idea of how to group instructions and possibly
rearrange loops as well as a technical realization.
In the following we apply a vectorization strategy developed in \cite{muething2018sumfact}:
Batches of several sum factorization kernels arising from the evaluation of finite element functions and their gradients are parallelized using SIMD instructions.
In order to achieve portability between different instruction sets, code is written using a SIMD abstraction layer \cite{fog2013c++}.
This however requires the innermost loops of finite element assembly to be rewritten using SIMD types.
With \dunepdelab 's abstraction of a local operator, these loops are typically located in user code.
This let to the development of \dunecodegen , which will be further described in Section~\ref{sec:trends}.

%

\begin{figure}
 \includegraphics[width=0.49\textwidth]{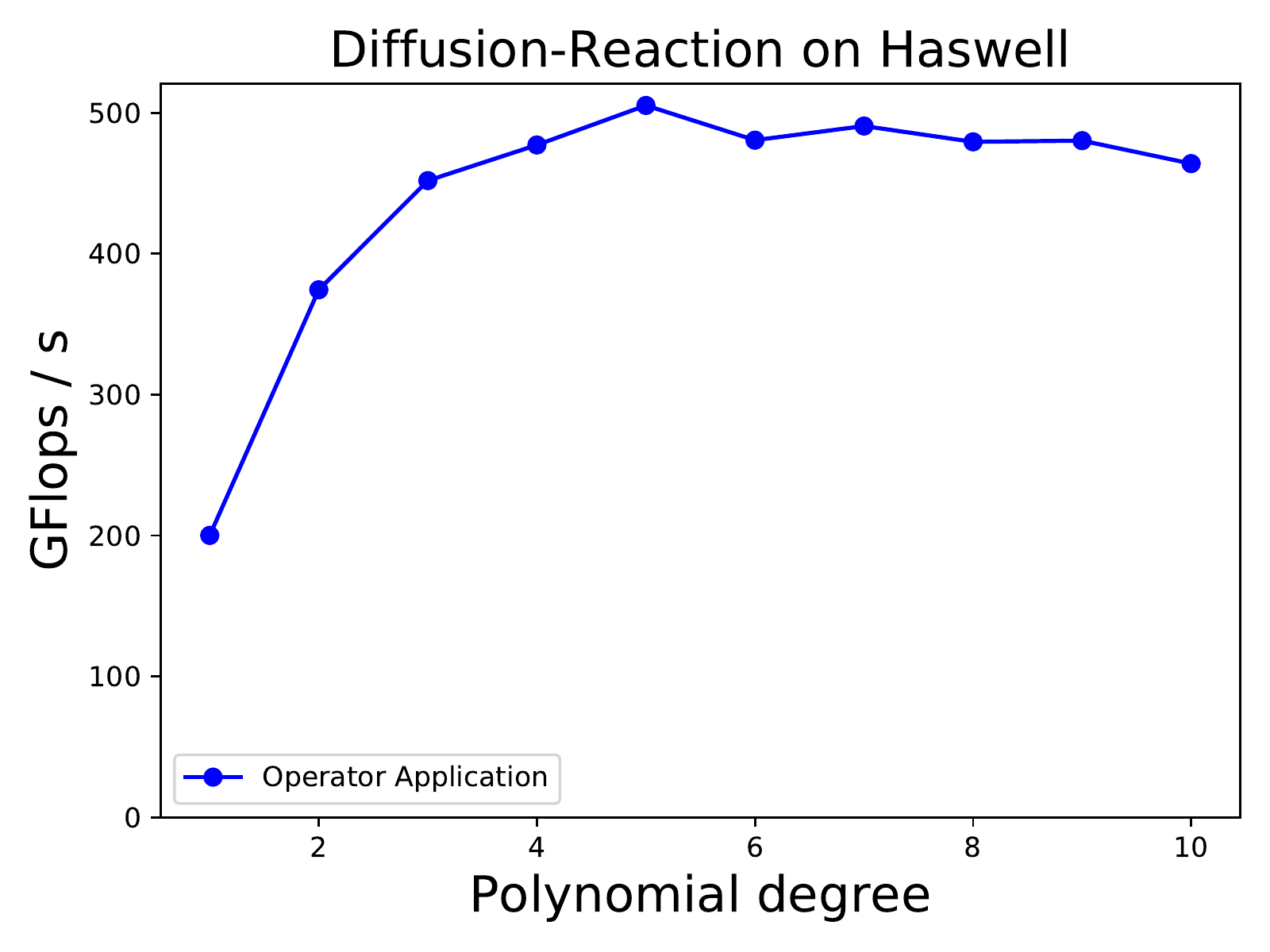}
 \includegraphics[width=0.49\textwidth]{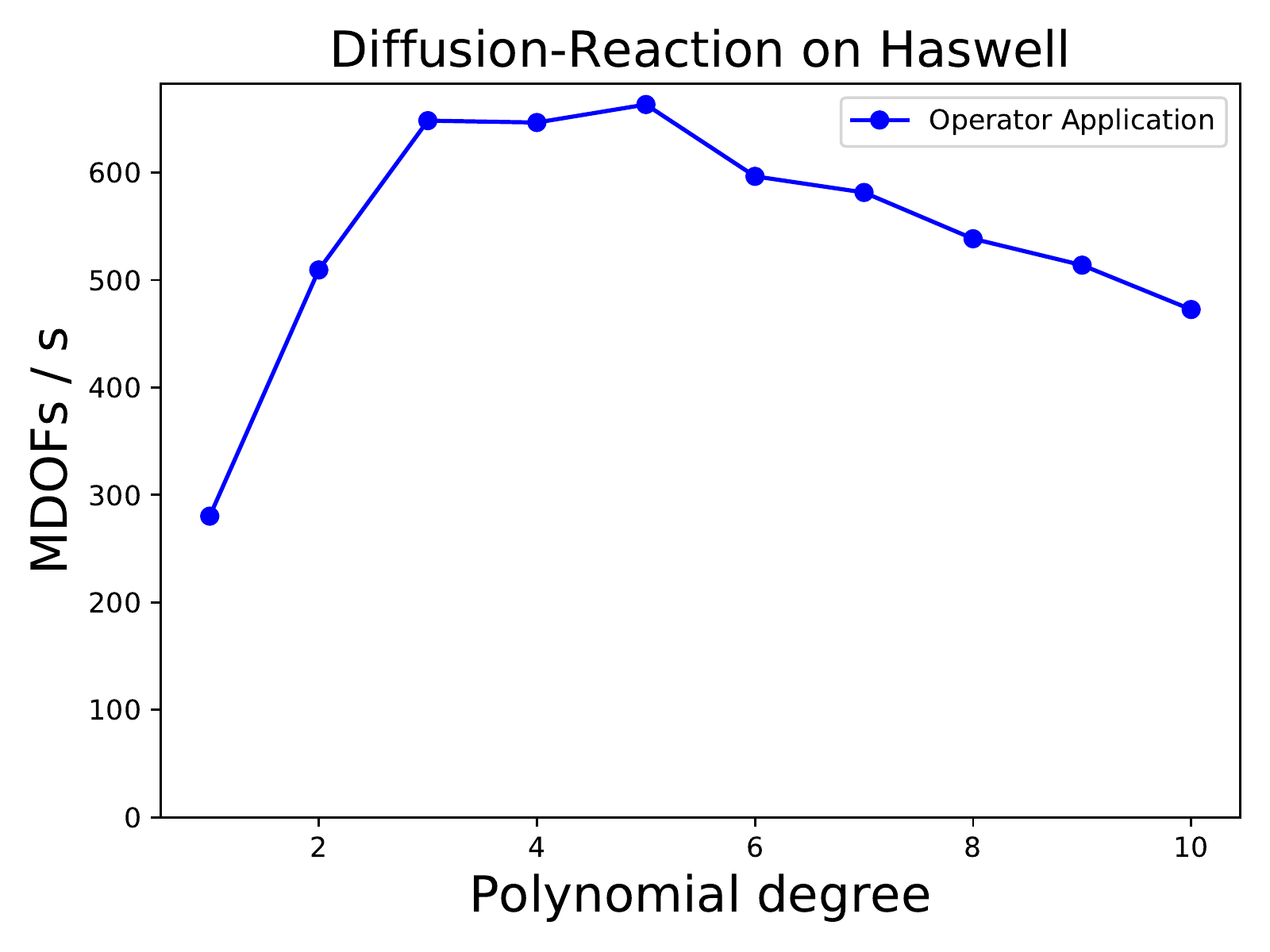}
 \caption{Performance measurements on an Intel Haswell node for a matrix-free application of a convection--diffusion DG operator on an axis-parallel, structured grid:
 On the left side, the machine utilization in GFlops/s ($10^9$ floating point operations per second) is shown. The theoretical peak performance of this Haswell node is 1.17 TFlops/s.
 On the right hand side, the degree of freedom throughput is measured in degrees of freedom per second.
 }
 \label{fig:hpc_numbers}
\end{figure}
Figure~\ref{fig:hpc_numbers} shows node level performance numbers
for a Discontinuous Galerkin finite element operator for the diffusion reaction equation on an Intel Haswell node.
The measurements use MPI to saturate the node and make extensive use of SIMD instructions which lead to a performance of roughly 40\% of the theoretical peak performance of 1.17 TFlops/s ($10^{12}$ floating point operations per second) on this 32 core node.
Discontinuous Galerkin discretizations benefit best from this compute-bound algorithm, as they allow to minimize memory transfers by omitting the costly setup of element-local data structured, operating directly on suitably blocked global data
structures instead.
A dedicated assembler for DG operators,
\cpp{Dune::PDELab::FastDGAssembler}, is
now available in \dunepdelab. It does not gather/scatter data from
global memory into element-local data structures, but just uses views
onto the global data. By this it avoids unnecessary copy operations and
index calculations. This assembler 
is essential to achieve
the presented node level performance, but can also be beneficial for
traditional DG implementations.

It is worth noting that iterative solvers based on this kind of matrix-free operator evaluation require the design of preconditioners that preserve the low memory bandwidth requirements while ensuring good convergence behavior,
as the per-iteration speedup would otherwise be lost to a much higher number of solver iterations.
We studied matrix-free preconditioning techniques for Discontinuous Galerkin problems in \cite{mueller2019matrixfree}.
This matrix-free solution technology have been used for an advanced application with the Navier--Stokes equations in \cite{piatkowski2018stable}.


%% file: trends.tex
\dune, and especially its grid interface, have proven themselves. Use cases range from personal laptops to TOP500 super computers, from mathematical and computer science methodologies to engineering application, and from bachelor thesis to research of Fortune 500 corporations.

As we laid out, \dune's structure and its interfaces remained stable over the time.
The modular structure of \dune is sometimes criticized, as it might lead to different implementations to solve the same problem. We still believe that the decision was right, as it allows to experiment with new features and make them publicly available to the community without compromising the stability of the core. Other projects like FEniCS have taken similar steps. In our experience a high granularity of the software is useful.

\dune remains under constant development and new features are added
regularly. We briefly want to highlight four topics
that are subject of current research and development.

\subsection{Asynchronous communication}
The communication overhead is
expected to be an even greater problem, in future HPC systems, 
as the numbers of processes will increase. Therefore, it is
necessary to use asynchronous communication. A first attempt to establish asynchronous
communication in \dune{} was demonstrated in~\cite{dgimpl:12,alugrid:16}.

With the MPI 3.0 standard, an official interface for asynchronous
communication was established. Based on this standard, 
as part of the \exadune{} project, we are currently developing
high-level abstractions for \dune{} for such asynchronous
communication, following the \textit{future-promise}
concept which is also used in the STL library. An \cpp{MPIFuture} object
encapsulates the \cpp{MPI_Request} as well as the corresponding memory
buffer. Furthermore, it provides methods to check for the state of the
communication and access the result.

Thereby the fault-tolerance with respect to soft- and hard-faults that occur on
remote processes is improved as well. We are following the recommended way of
handling failures by throwing exceptions. Unfortunately, this concept integrates
poorly with MPI. An approach how to propagate exceptions through the entire
system and handle them properly, using the ULFM functionality proposed in
\cite{bosilca2016ulfm,bland2012ulfm}, can be found in
\cite{engwer2018faulttolerant}.

Methods like pipelined CG
\cite{ghysels2014hiding} overlap global communication and operator
application to hide communication costs. Such asynchronous solvers will be
incorporated in \duneistl, along with the described software infrastructure.

\subsection{Thread parallelism}
Modern HPC systems exhibit different levels of concurrency. Many
numerical codes are now adopting the MPI+X paradigm, meaning that they
use inter\-node parallelism via MPI and intranode parallelism,
i.e. threads, via some other interface. While early works where based on
OpenMP and pthreads, for example in~\cite{dgimpl:12}, the upcoming interface changes in
\dune{} will be based on the Intel Thread Building Blocks (TBB) to handle threads. Up to
now the core modules don't use multi-threading directly, but the
consensus on a single library ensures interoperability among different
\dune extension modules.

In the \exadune project several numerical components like assemblers
or specialized linear solvers have been developed using TBB. As many
developments of \exadune are proof of concepts, these can not be
merged into the core modules immediately, but we plan to port
the most promising approaches to mainline \dune. Noteworthy features
include mesh partitioning into entity ranges per thread, as it is used
in the MS-FEM code in Section~\ref{subsec:ms}, the
block-SELL-C-$\sigma$ matrix format~\cite{Muething:2014:IMT} (an extension of the work
of~\cite{kreutzer2014}) and a task-based DG-assembler for \dunepdelab.

\subsection{C++ and Python}
Combining easy to use scripting languages with state-of-the-art
numerical software has been a continuous effort in scientific computing
for a long time. While much of the development of mathematical algorithms
still happens in Matlab, there is increasing use of Python for such efforts,
also in other scientific disciplines. For solution of PDEs
the pioneering work of the FEniCS team~\cite{Logg:2012:ASD:2331176}
inspired many others, e.g.~\cite{petsc4py:11,firedrake} to also provide
Python scripting for high performance PDE solvers usually coded in C++.
As discussed in Section~\ref{sec:dunepython}, \dune{} provides
Python-bindings for central components like meshes, shape functions,
and linear algebra. \dunepython also provides infrastructure for exporting
static polymorphic interfaces to Python using just in time compilation and
without introducing a virtual layer and thus not leading to any performance
losses when objects are passed between different C++ components through the
Python layer. Bindings are now being added to a number of modules like
the \dunegridglue module discussed in Section~\ref{sec:dunegridglue}
and further modules will follow soon.

\subsection{DSLs and code-generation}
Code-generation techniques allow to use scripting
languages, while maintaining high efficiency. Using a
domain-specific language (DSL), the FEniCS project first introduced a
code generator to automatically generate efficient discretization code
in Python. The \emph{Unified Form Language}
UFL\cite{Logg:2012:ASD:2331176,ufl:14} is an embedded Python DSL for
describing a PDE problem in weak form. UFL is now used by several
projects, in particular Firedrake~\cite{firedrake}. 
We also started adopting this input in several places in
\dune. For example, UFL can now be used for the generating model
descriptions for \dunefem~\cite{dunefem:tutorial} as demonstrated
in Section~\ref{sec:gridadaptation}.
Another effort is
the currently developed \dunecodegen module, which tries to make
performance optimization developed in the \exadune project accessible
to the \dune community.

In Section~\ref{sec:hpc} we highlighted how highly tuned matrix-free higher-order
kernels can achieve 40\% peak performance on modern
architectures. While \dune offers the necessary flexibility, this kind
of optimizations is hard to implement for average users.
To overcome this issue and improve  sustainability, we introduced a
code generation toolchain in~\cite{kempf2018codegen}, using UFL as our
input language.
From this DSL, a header file containing a performance-optimized \cpp{LocalOperator} class is generated.
The \cpp{LocalOperator} interface is \dunepdelab{}'s abstraction for local integration kernels.
The design decisions for this code generation toolchain are discussed in detail in~\cite{kempf2019hpc}.
This toolchain achieves near-optimal performance by applying structural transformations to an intermediate
representation based on~\cite{kloeckner2014loopy}.
A search space of SIMD vectorization strategies is explored from within the code generator through an autotuning procedure.
This work now lead to the development of \dunecodegen, which 
also offers other optimizations, like block-structured meshes, similar to the
concepts described in~\cite{hhg2006}, or extruded meshes, like in
\cite{macdonald2011general,Bercea2016}.
This is an ongoing effort and is still in early development.
